\documentclass[journal]{IEEEtran}
\usepackage{amssymb}
\usepackage{color,graphicx}
\usepackage{cite}
\usepackage{algorithm2e}
\usepackage[cmex10]{amsmath}
\usepackage{amsopn}
\usepackage{stfloats}
\usepackage{amsthm}
\usepackage{epstopdf}

\newtheorem{definition}{Definition}
\newtheorem{theorem}{Theorem}
\newtheorem{corollary}{Corollary}

\newtheorem{example}{Example}
\newtheorem{remark}{Remark}

\allowdisplaybreaks

\allowdisplaybreaks

\graphicspath{{fig/}}

\usepackage{stfloats}
\newcounter{mytempeqncnt}

\newtheoremstyle{noparens}%
{}{}%
{\itshape}{}%
{\bfseries}{.}%
{ }%
{\thmname{#1}\thmnumber{ #2}\mdseries\thmnote{ #3}}
\theoremstyle{noparens}

\ifCLASSINFOpdf
\else
\fi
\begin{document}
%
\title{Fundamental Limits of Multiple-Access Integrated Sensing and Communication Systems}
%
%
%

\author{Yao Liu, Min Li, An Liu, Lawrence Ong, and Aylin Yener 
\thanks{Received 10 November 2023; revised 23 January 2025; accepted 8 March 2025. 
The work of Min Li was supported in part by National Natural Science Foundation of China under Grant 62271440, and the Fundamental Research Funds for the Central Universities 226-2022-00195. A preliminary version of this paper has been presented in part at the 2023 IEEE Global Communications (IEEE GLOBECOM) Conference~\cite{liu2023Globecom}. \textit{(Corresponding author: Min Li.)}}
\thanks{Yao Liu, Min Li, and An Liu  are with the College of Information Science and Electronic Engineering and the 
Zhejiang Provincial Key Laboratory of Multi-Modal Communication Networks and Intelligent Information Processing,
Zhejiang University, Hangzhou 310027, China (e-mail: \{yao.liu, min.li, anliu\}@zju.edu.cn). }
\thanks{Lawrence Ong is with the School of Engineering, University of Newcastle, Callaghan, NSW 2308, Australia (e-mail: lawrence.ong@newcastle.edu.au).}
\thanks{Aylin Yener is with the Department of Electrical and Computer Engineering, The Ohio State University, Columbus, OH 43210 USA (e-mail: yener@ece.osu.edu).}
}
\maketitle

\begin{abstract}
A state-dependent discrete memoryless multiple access channel is considered to model an
integrated sensing and communication system, where two transmitters wish to convey messages to a receiver while simultaneously estimating the state parameter sequences through echo signals. In particular, the sensing state parameters are assumed to be correlated with the channel state.
In this setup, improved inner and outer bounds for capacity-distortion region are derived. The inner bound is based on an achievable scheme that combines message cooperation and joint compression of past transmitted codewords and echo signals at each transmitter, resulting in unified cooperative communication and sensing. The outer bound is based on the ideas of dependence balance for communication rate, genie-aided state estimator and rate-limited constraints on sensing distortion. The proposed inner and outer bounds are proved to improve the state-of-the-art bounds. Finally, numerical examples are provided to demonstrate that our new inner and outer bounds strictly improve the existing results. 
\end{abstract}

\begin{IEEEkeywords}
Integrated sensing and communication, multiple access channels, correlated sensing state parameters and channel states, capacity-distortion region.
\end{IEEEkeywords}

%
\IEEEpeerreviewmaketitle

\section{Introduction}
\par~Future 6G mobile networks aim to integrate the functions of communication and sensing to provide advanced intelligent sensing services, such as  smart vehicular networks and smart homes. As mobile networks progress towards millimeter wave (mmWave) bands and embrace massive multi-input multi-output (MIMO) techniques, communication signals tend to have higher resolution in both time and angular domains, and this opens doors for highly accurate sensing via mobile networks. Integrated sensing and communication (ISAC), in which sensing and communication share the same frequency band and hardware, has thus emerged as a pivotal technology for 6G networks \cite{cui2021integrating,liu2022survey}.

\par~A number of previous studies have investigated ISAC in various practical scenarios and system architectures \cite{liu2020joint,zhang2022integrated,mu2021integrated}, demonstrating the advantages of integration. Nonetheless, the optimality of these schemes and the fundamental tradeoff between sensing and communication performance in ISAC systems are worth further studied. To elucidate the tradeoff, Bliss in~\cite{bliss2014cooperative} has introduced the notion of ``estimation information rate'' to quantify the sensing estimation performance and examined the tradeoff between estimation information rate and communication rate for the system considered. Kumari \textit{et~al}.~\cite{kumari2017performance} have instead proposed to convert the communication rate into a mean-square-error (MSE) equivalent quantity, providing a framework for representing the sensing-communication tradeoff under the MSE metric. More recently, Xiong \textit{et~al}.~\cite{10001144,xiong2022fundamental} have utilized the Cramer-Rao bound (CRB), a lower bound of MSE, as the performance metric for sensing, and investigated the sensing-communication tradeoff through CRB-communication rate region. However, the aforementioned studies \cite{bliss2014cooperative,kumari2017performance,10001144,xiong2022fundamental} while insightful, often assume Gaussian parameters for sensing or Gaussian channels, limiting their applicability to more general scenarios.

\par~To address the broader scope of ISAC systems, Kobayashi \textit{et~al}.~\cite{kobayashi2018joint} drew from rate-distortion theory~\cite{el2011network} and introduced a pivotal performance metric known as the capacity-distortion tradeoff. In this framework, the precision of parameter sensing is quantified using general distortion functions, while the effectiveness of communication is evaluated through the classical Shannon communication rate. 
The authors modeled the sensing echo signals as strictly causal feedback and made the assumption that the sensing state parameter coincides with the channel state, with perfect channel state information available at the receiver (CSIR). This allowed them to establish the optimal capacity-distortion tradeoff for ISAC systems employing monostatic sensing over discrete memoryless point-to-point channels. In contrast, our recent work~\cite{liu2022information} has characterized the optimal capacity-distortion tradeoff for point-to-point channels where the sensing state parameters and channel states are correlated with each other, and CSIR is imperfect. While channel state information (CSI) governs how transmitted signals propagate, combine, and are received at their destinations, the sensing operation primarily seeks to detect physical phenomena (represented as sensing state parameters) within the channel. These sensing state parameters are often correlated with but not necessarily identical to the CSI. Furthermore, achieving perfect CSIR in practice is challenging due to channel estimation errors.

\par~References~\cite{kobayashi2019joint,ahmadipour2021joint,ahmadipour2022information,ahmadipour2023information} instead delve deeper into the capacity-distortion tradeoff for multi-terminal ISAC systems. In particular, Kobayashi \textit{et~al.} in \cite{kobayashi2019joint} have considered a multiple-access ISAC model where two transmitters wish to convey messages to a receiver while simultaneously sensing the respective channel states through echo signals. By leveraging the Willem's coding scheme \cite{willems1982informationtheoretical}, they have demonstrated that two transmitters can cooperate by decoding and retransmitting partial messages (called common message) of the other transmitter through echo signals, which can then be leveraged for both state sensing and communication. We refer to this coding scheme~\cite[Theorem~2]{kobayashi2019joint} as the Kobayashi--Hamad--Kramer--Caire scheme for the remainder of this paper. This approach 
enlarged the achievable rate-distortion region compared to the conventional time-sharing approach. They have also established an outer bound for ISAC over multiple access channels (MAC) by combining the principles of dependence balance constraints \cite{hekstra1989dependence,tandon2009outer,tandon2011dependence} on the allowable input distributions and genie-aided side information regarding sensing. We refer to this outer bound~\cite[Theorem~1]{kobayashi2019joint} as the Kobayashi--Hamad--Kramer--Caire outer bound later on. More recently, Ahmadipour \textit{et~al.}\cite{ahmadipour2023information} have proposed a collaborative ISAC scheme, 
where each transmitter conveys information pertaining to the echo signals to the other transmitter. In addition to the message cooperation \cite{kobayashi2019joint}, they have demonstrated that sending compressed information related to the echo signals as part of the common message, decodable by both the other transmitter and the receiver, can further enhance sensing performance. We refer to this scheme~\cite[Theorem~3]{ahmadipour2023information} as the Ahmadipour--Wigger scheme for the remainder of this paper.

\par~Other related studies have also appeared to explore the fundamental limits of ISAC systems, each addressing distinct considerations~\cite{zhang2011joint,choudhuri2013causal,bross2020message,jiao2023rate,joudeh2022joint,wu2022joint,chang2023rate,gunlu2022secure,gunlu2023secure,ahmadipour2023integrated,li2022capacity}. For instance, references~\cite{zhang2011joint,choudhuri2013causal,bross2020message,jiao2023rate} have investigated the capacity-distortion tradeoff in bistatic sensing scenarios where sensing estimation occurs at the receiver. Additionally, references~\cite{joudeh2022joint,wu2022joint,chang2023rate} have considered scenarios where the sensing state parameter remains a fixed parameter correlated with channel states, shedding light on the tradeoff between the classical communication rate and the state detection-error exponent. Furthermore, references~\cite{gunlu2022secure,gunlu2023secure,ahmadipour2023integrated} have addressed security concerns in ISAC systems, establishing the capacity-distortion tradeoff while adhering to secure constraints, with sensing operations carried out at the transmitter and receiver, respectively. Inspired by the application of ISAC in mmWave communication, the authors in~\cite{li2022capacity} have analyzed a binary beam-pointing channel with in-block memory and feedback, deriving its capacity subject to peak transmission cost constraints in closed-form. 
Finally, the problem of joint communication and state amplification, as studied in~\cite{sutivong2005channel,kim2008state,choudhuri2011causal,tian2015gaussian,ramachandran2019joint}, can also be viewed as a special case of ISAC, where the perfect sensing state or a noisy version of the sensing state is acquired at the transmitter and is conveyed to the receiver along with message transmission. The rate-distortion tradeoff has also been investigated in these works.

\subsection{Contributions}
\par~In this paper, we develop improved inner and outer bounds on the capacity-distortion region for ISAC over MACs. The summary of contributions is as follows.

\begin{figure*}[!t]
	\centering
	\includegraphics[width=.8\linewidth]{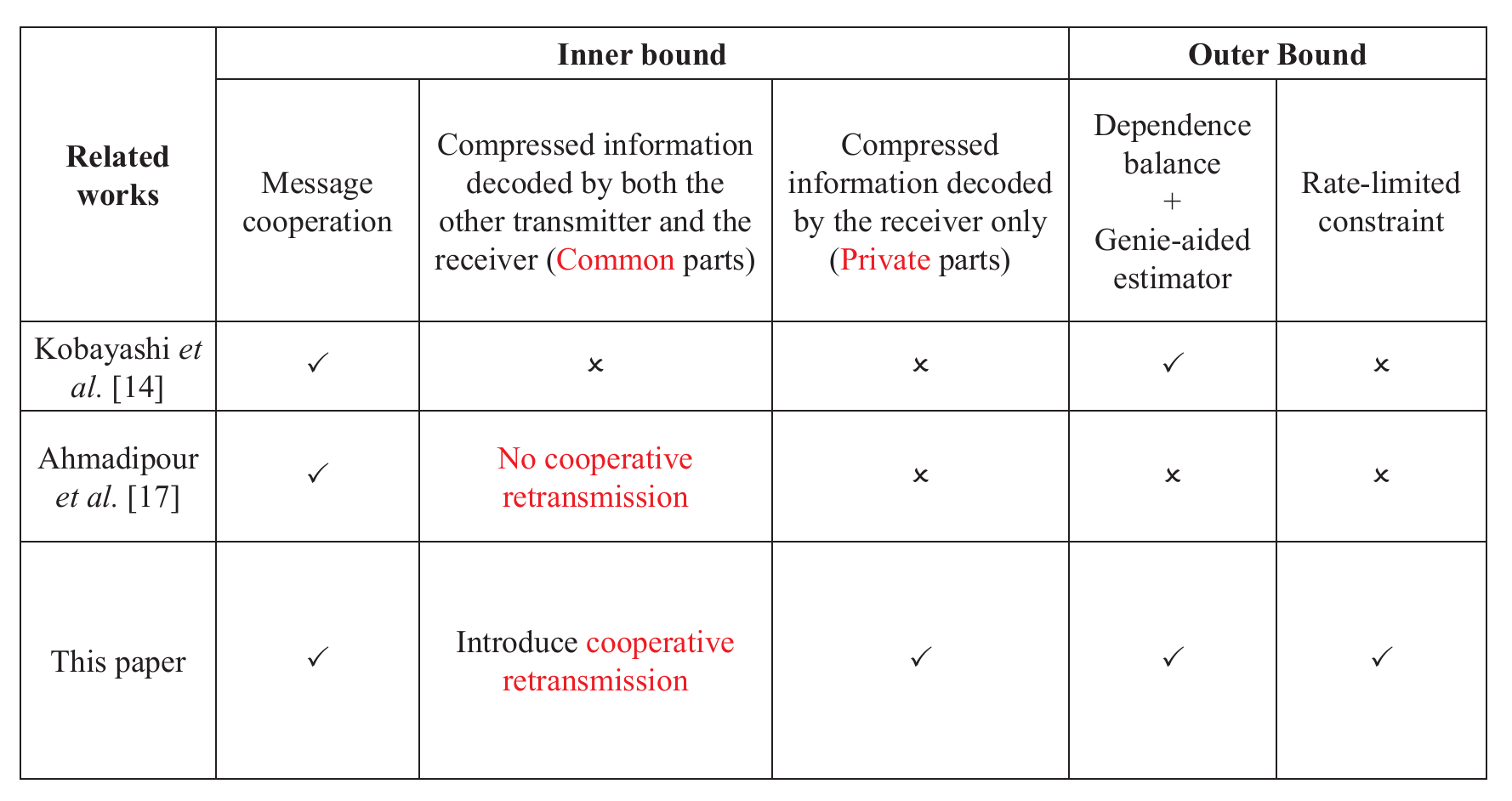}
	\caption{A brief comparison of the key ingredients distinguishing the inner and outer bounds in our work from those in related studies.}
	\label{fig:contributionDiff}
\end{figure*}

\par~For the inner bound, we propose a \textit{new achievable scheme} that combines the concepts of message cooperation \cite{willems1982informationtheoretical,kobayashi2019joint} and joint compression of past transmitted codeword and echo signals \cite{li2012multiple,lapidoth2012multipleA,lapidoth2012multipleB}.
Existing schemes, such as the Kobayashi--Hamad--Kramer--Caire scheme\cite[Theorem~2]{kobayashi2019joint} and the Ahmadipour--Wigger scheme~\cite[Theorem~3]{ahmadipour2023information}, achieve message cooperation by enabling each transmitter to decode the common part of the other transmitter's message and cooperatively retransmit the common messages using block Markov coding. Additionally, the Ahmadipour--Wigger scheme allows the transmitters to send compressed information related to echo signals. While this compressed information is decoded by both the receiver and the other transmitter, it is not retransmitted in subsequent blocks as common messages. Building on these schemes, our proposed scheme introduces two new elements: 
\begin{itemize}
	\item First, we split the compressed information related to echo signals into two parts following the message cooperation mentioned above: (a) the common part to be decoded by both the other transmitter and the receiver (this is the same as the Ahmadipour--Wigger scheme) and (b) the private part to be decoded only by the receiver.
	\item Then, having decoded part (a) of the compressed information, both transmitters can then cooperatively retransmit those common parts to the receiver in the subsequent block.
\end{itemize}
The benefits of our improvement are as follows. First, by enabling cooperative retransmission, our scheme relaxes the rate constraints on the common compressed information and facilitates the decoding at the receiver. Second, incorporating private part of compressed information would allow the receiver to obtain more information about the channel state, akin to the schemes for state-dependent MACs with strictly causal state information at the transmitters~\cite{lapidoth2012multipleA,lapidoth2012multipleB,li2012multiple}. Since the common part is leveraged in the sensing estimator and both the common and private parts are leveraged for message decoding at the receiver, our achievable scheme leads to a unified cooperative communication and sensing approach for ISAC over MACs. The corresponding achievable rate-distortion region is derived and \textit{proved to include the regions achievable by the Kobayashi--Hamad--Kramer--Caire scheme and the Ahmadipour--Wigger scheme in general, and the inclusion is strict for some channels.}

\par~For the outer bound, we refine the Kobayashi--Hamad--Krame--Caire outer bound by incorporating additional rate-limited constraints to more effectively bound the sensing performance. The existing outer bound primarily relies on dependence-balance constraints imposed on the input distribution and assumes a genie-aided sensing estimator at each transmitter. This estimator has perfect knowledge of both the message and the echo signal received from the other transmitter. To tighten this bound, we determine the maximum information rate about the sensing parameter and the message of the other transmitter that can be acquired at a given transmitter, and then construct an additional constraint on sensing distortion by leveraging rate-distortion theory. The resultant outer bound provides a stricter limit on sensing performance, and it includes the Kobayashi--Hamad--Kramer--Caire outer bound in genereal and is strictly tighter for some channels.

\par~Several numerical examples are constructed to show the advantages of our improved inner and outer bounds compared to the existing results. We first provide three examples to intuitively demonstrate the advantages of introducing unified cooperative communication and sensing scheme in inner bound and rate-limited constraints on sensing in outer bound. Then, a general example is presented to show that our proposed scheme can achieve better rate-distortion tradeoff compared to both Ahmadipour--Wigger and Kobayashi--Hamad--Kramer--Caire schemes, and the proposed outer bound is strictly tighter than Kobayashi--Hamad--Kramer--Caire outer bound.

\par~Fig.~\ref{fig:contributionDiff} summarizes the main differences in the key ingredients of the inner and outer bounds between our work and related studies~\cite{ahmadipour2023information,kobayashi2019joint}.

\subsection{Organization and Notation}
\par~The rest of this paper is organized as follows. Section \ref{section:systemModel} describes the general model for ISAC over MAC considered in this work. 
Section \ref{section:mainResults} presents the main results of improved inner and outer bounds, as well as the theoretically comparison with related works. Section \ref{section:example} constructs several numerical examples to demonstrate that our inner and outer bounds strictly improve the existing results. Section \ref{sec:conclusion} concludes the paper. 

\par~\emph{Notation}: Throughout the paper, we use calligraphic letters, uppercase letters, and lowercase letters to denote sets, random variables, and the realizations, respectively, e.g., $\mathcal{X},X,x$. The probability distributions are denoted by $P$ with the subscript indicating the corresponding random variables, e.g., $P_X(x)$ and $P_{Y|X}(y|x)$ are the probability of $X=x$ and conditional probability of $Y=y$ given $X=x$. We use $x^i$ to denote the vector $[x_1,x_2,\cdots,x_i]$, $[1:L]$ to denote the set $\{1,2,\cdots,L\}$ for integer $L$, and $\mathbb{E}(X)$ to denote the expectation of random variable $X$. For $k\in\{1,2\}$, we define $\bar{k} \triangleq 3-k$. For a event $\mathcal{A}$, we use $\mathcal{A}^c$ to denote its complement. Logarithms are taken with respect to base 2.

\section{System Model}\label{section:systemModel}
\par
\begin{figure}[!t]
	\centering
	\includegraphics[width=1\linewidth]{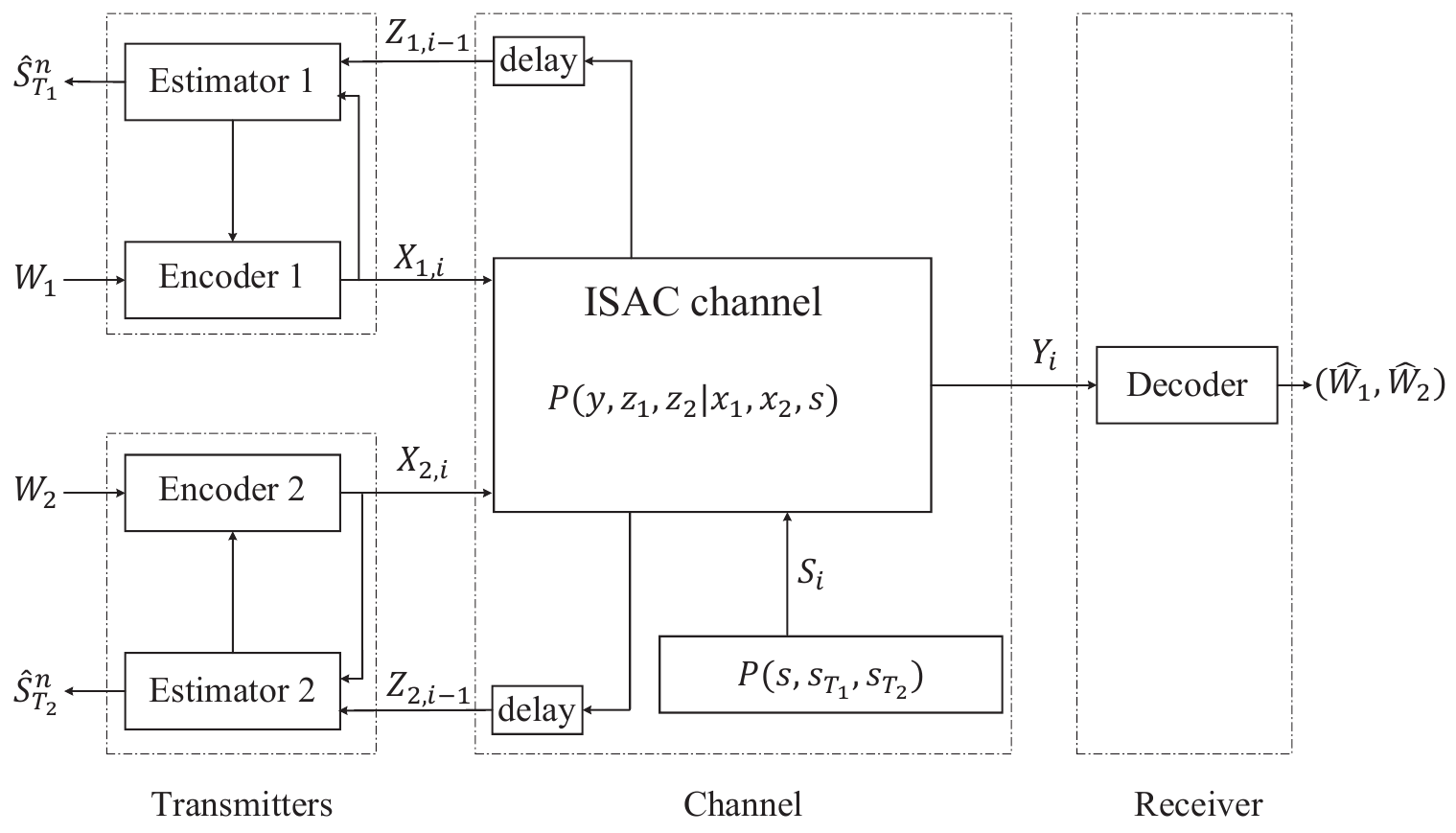}
	\caption{ISAC over SD-DM MAC, where the channel state and sensing state parameters are correlated but not not necessarily the same.}
	\label{fig:channelmodel4mac}
\end{figure}

~Consider a general ISAC over state-dependent discrete memoryless (SD-DM) MAC as shown in Fig.~\ref{fig:channelmodel4mac}. Over $n$ uses of such a channel, transmitter $k\in\{1,2\}$ wishes to convey a message $W_k\in[1:2^{nR_k}]$ to the receiver  while simultaneously estimating the state parameter sequence $S^n_{T_k}$ via output feedback $Z_{k}^n$. Here $S_{T_k,i}$ denotes the estimated state parameter for transmitter~$k$ during channel use $i\in[1:n]$, and output feedback models the communication echo signal reflected back to the transmitter.

\par~The SD-DM MAC considered in Fig. \ref{fig:channelmodel4mac} is denoted by 
\begin{align}
	(\mathcal{X}_1\times\mathcal{X}_2,\mathcal{S},P_{YZ_1Z_2|X_1X_2S},\mathcal{Y}\times\mathcal{Z}_1\times\mathcal{Z}_2,\mathcal{S}_{T_1}\times\mathcal{S}_{T_2})
\end{align}
with input alphabets $\mathcal{X}_1\times\mathcal{X}_2$, channel state alphabet $\mathcal{S}$, a collection of the conditional probability mass functions (pmfs) $P_{YZ_1Z_2|X_1X_2S}$, output and feedback alphabets $\mathcal{Y}\times\mathcal{Z}_1\times\mathcal{Z}_2$, and sensing parameter alphabets $\mathcal{S}_{T_1}\times\mathcal{S}_{T_2}$. The channel is memoryless in the sense that at each time instance $i\in[1:n]$, 
\begin{align}
	P(y_i,z_{1,i},z_{2,i}|&x_1^i,x_2^i,s^i,y^{i-1},z_1^{i-1},z_2^{i-1})\notag\\
	&=P(y_i,z_{1,i},z_{2,i}|x_{1,i},x_{2,i},s_i),
\end{align}
where $x_{1,i},x_{2,i}$ are the realizations of channel inputs during channel use $i$, $s_i$ is the realization of channel state during channel use $i$, $y_i,z_{1,i},z_{2,i}$ are realizations of channel output and feedbacks during channel use $i$. The joint distribution of channel state $S$ and sensing state parameters $S_{T_k}, k\in\{1,2\}$ is given by $P_{SS_{T_1}S_{T_2}}$, which is independent and identically distributed (i.i.d.) according to 
\begin{align}
	P_{S^nS_{T_1}^nS_{T_2}^n}(s^ns_{T_1}^ns_{T_2}^n)=\prod_{i=1}^n P_{SS_{T_1}S_{T_2}}(s_is_{T_1,i}s_{T_2,i}). 
\end{align}
The sensing state parameters $S_{T_k},k\in\{1,2\}$ are usually correlated with channel state $S$ but are not necessarily the same. In general, the state of the channel directly (or say ``physically'') influences the
channel output and feedback, and sensing state parameters are the information that transmitters try to capture. For example, the channel state could be the channel attenuation, and the sensing state parameters may be the relative velocity of the transmitters with respect to the receiver.

\begin{definition}
	For the model considered in Fig.~\ref{fig:channelmodel4mac}, a $(2^{nR_1},2^{nR_2},n)$ code consists of 
	\begin{enumerate}
		\item two message sets $\mathcal{W}_{k}=[1:2^{nR_k}],k\in\{1,2\}$ where the messages $W_1,W_2$ are uniformly distributed;
		\item two encoders where encoder $k\in\{1,2\}$ assigns a symbol $x_{k,i}=f_{k,i}(w_k,z_{k}^{i-1})$ to each message $w_k\in\mathcal{W}_k$ and delayed feedback $z_{k}^{i-1}\in\mathcal{Z}_k^{i-1}$ for $i\in[1:n]$; 
		\item a decoder that produces an estimated message pair $(\hat{w}_1,\hat{w}_2)=h_n(y^n)\in\mathcal{W}_1\times\mathcal{W}_2$ upon observing $y^n$;
		\item two estimators where estimator $k\in\{1,2\}$ assigns an estimated state parameter sequence $\hat{s}_{T_k}^n = g_{k,n}(x_k^n,z_k^n)\in\hat{\mathcal{S}}_{T_k}^n$ based on the codeword $x_k^n\in\mathcal{X}_k^n$ and receiving feedback sequence $z_k^{n}\in\mathcal{Z}_k^n$, where $\hat{\mathcal{S}}_{T_1}^n,\hat{\mathcal{S}}_{T_2}^n$ are given reconstruction alphabets.
	\end{enumerate}
\end{definition}

\par~The sensing performance is measured by the expected distortion of the state parameter estimated, i.e.,
\begin{align}\label{definitionDistortion}
	\mathbb{E}[d_k(S_{T_k}^n,\hat{S}_{T_k}^n)] = \frac{1}{n}\sum_{i=1}^n\mathbb{E}[d_k(S_{T_k,i},\hat{S}_{T_k,i})],\ k\in\{1,2\},
\end{align}
where $d_k:S_{T_k}\times\hat{S}_{T_k}\rightarrow[0,\infty)$ is a bounded distortion function.

\begin{definition}
	A rate-distortion tuple $(R_1,R_2,D_1,D_2)$ is said to be achievable if there exist a sequence of $(2^{nR_1},2^{nR_2},n)$ codes with arbitrarily small error probability for decoding, i.e.,
	\begin{align}
		\lim_{n\rightarrow\infty}P\big((\hat{W}_1,\hat{W}_2)\neq(W_1,W_2)\big)=0
	\end{align}
	that satisfies sensing distortion constraints
	\begin{align}
		{\lim\sup}_{n\rightarrow\infty}\mathbb{E}[d_k(S_{T_k}^n,\hat{S}_{T_k}^n)]\le D_k,\ k\in\{1,2\}.
	\end{align} 
	For any $(D_1,D_2)$, the capacity-distortion region $\mathcal{C}(D_1,D_2)$ is defined as the closure of achievable rate tuple $(R_1,R_2)$ such that $(R_1,R_2,D_1,D_2)$ is achievable.
\end{definition}

\begin{remark}
	\textup{Different from the existing studies \cite{kobayashi2019joint,ahmadipour2023information}, the considered model in this work introduce two random variables $S_{T_1},S_{T_2}$ to denote the two sensing state parameters. The correlation among sensing state parameters and channel state is thus explicitly modeled through $P_{SS_{T_1}S_{T_2}}$, while the sensing state parameters in \cite{kobayashi2019joint,ahmadipour2023information} are denoted as $S_1,S_2$. As claimed in \cite{ahmadipour2023information}, both our model and that in \cite{ahmadipour2023information} can capture the general sensing state parameters. The results in the following sections are thus summarized based on the model in Fig.~\ref{fig:channelmodel4mac}.}
\end{remark}

\section{Main Results}\label{section:mainResults}
\par~In this section, we present the main results of our improved inner and outer bounds for ISAC over SD-DM MAC. For each bound, we first elaborate the key ideas of our proposed scheme. Then, the results and theoretical comparison with the existing ones are provided. 

\subsection{Improved Inner Bound}
\par~Our proposed scheme is based on the ideas of message cooperation~\cite{willems1982informationtheoretical} and transmitting the compressed information related to echo signals~\cite{lapidoth2012multipleA,lapidoth2012multipleB,li2012multiple} via block Markov coding. Specifically, in each block, transmitter~$k$ sends the following six types of message components:
\begin{figure}[!t]
	\centering
	\includegraphics[width=1\linewidth]{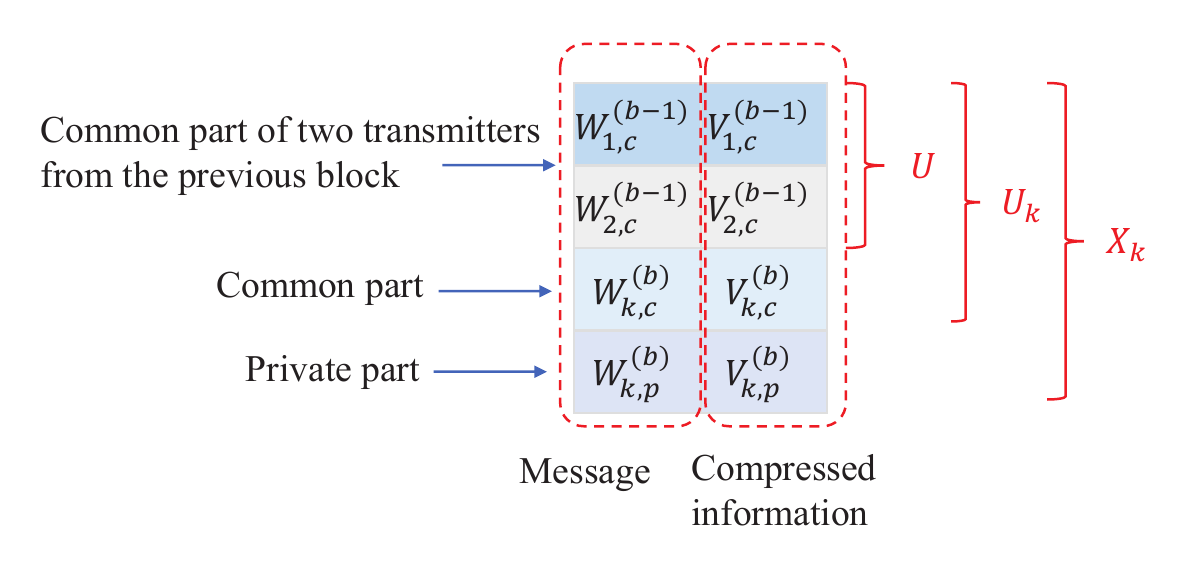}
	\vspace{-1em}
	\caption{A brief illustration of the codeword sent by transmitter~$k$ in block~$b$.}
	\label{fig:innerBoundOurCodeword}
\end{figure}
\begin{itemize}
	\item[(a)] The common part (to be decoded by the other transmitter) of its own message $W_{k,c}^{(b)}$;
	\item[(b)] The private part of its own message $W_{k,p}^{(b)}$;
	\item[(c)] The common part (to be decoded by the other transmitter) of its compressed information $V_{k,c}^{(b)}$ related to the delayed echo signal from the previous block $b-1$;
	\item[(d)] The private part of its compressed information $V_{k,p}^{(b)}$ related to the delayed echo signal from the previous block $b-1$;
	\item[(e)] Cooperative signal about common messages $W_{1,c}^{(b-1)}, W_{2,c}^{(b-1)}$ of both two transmitters from the previous block $b-1$;
	\item[(f)] Cooperative signal about common compressed information $V_{1,c}^{(b-1)}, V_{2,c}^{(b-1)}$ of both two transmitters which are related to the delayed echo signal from the previous block $b-2$.
\end{itemize}

\begin{figure*}[!t]
	\centering
	\includegraphics[width=0.8\linewidth]{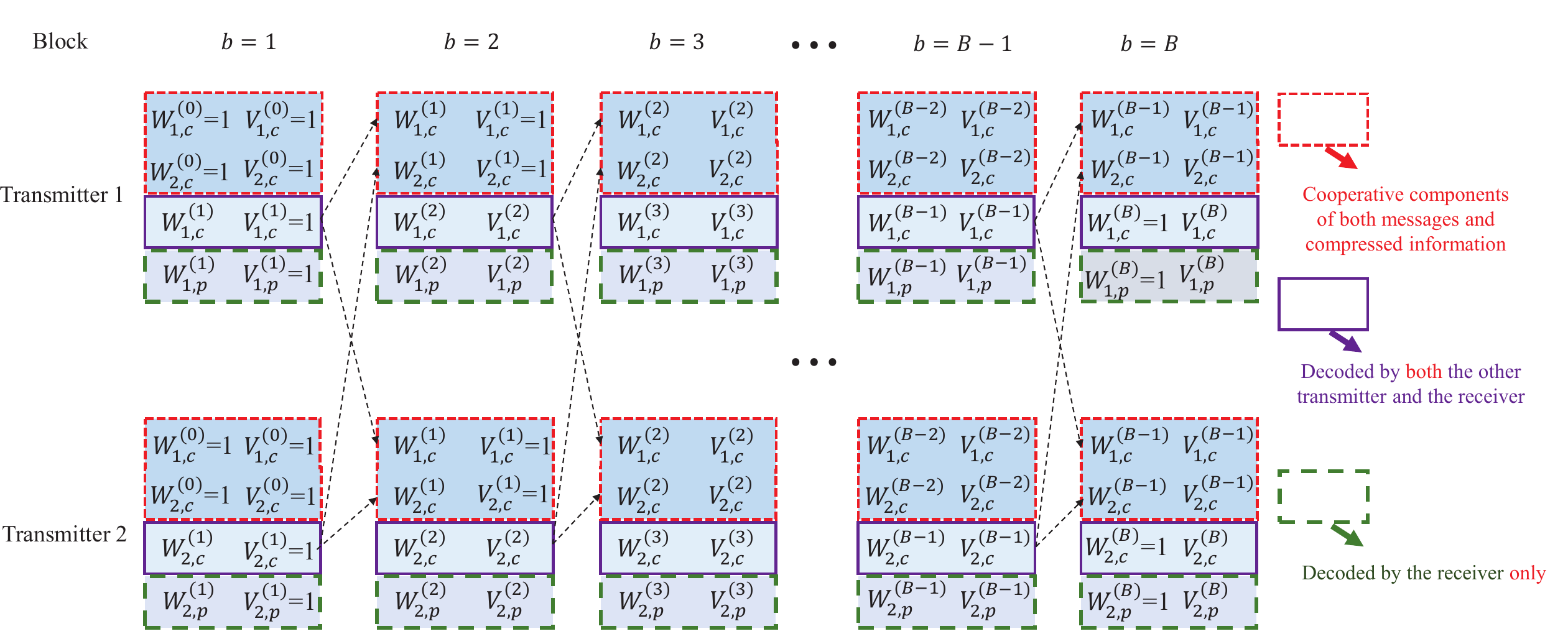}
	\caption{A brief illustration of our coding scheme.}
	\label{fig:innerBoundOur}
\end{figure*}

\par~Fig.~\ref{fig:innerBoundOurCodeword} provides a brief illustration of the codeword constructed. Here, random variable $U$ denotes the cooperative signal of both common messages and common compressed information, i.e., components (e) and (f), and random variable $U_k,k\in\{1,2\}$
denotes both cooperative signal~$U$ and the common part of fresh information for messages and compressed information, i.e., components (a), (c), (e) and (f). 
An illustration of our proposed achievable scheme is shown in Fig.~\ref{fig:innerBoundOur}. We note that in Fig.~\ref{fig:innerBoundOur}, only the details of the first $B$ blocks are provided. There are also $\tilde{B}$ blocks called ``termination blocks'' which are necessary to guarantee the success of backward decoding. The discussion of these termination blocks can be found in Appendix \ref{appendix:proofOfTheorem3}.

\par~The inner bound on the capacity-distortion region achieved by our proposed scheme is given as follows.

\begin{theorem}\label{innerBound:CD}
	Considering the state estimator for transmitter $k\in\{1,2\}$
	\begin{align}\label{equ:estimator}
		\hat{s}_{T_k}(x_k,&u_{\bar{k}},z_k,v_{\bar{k},c}) = \arg\min_{s'_{T_k}\in\hat{\mathcal{S}}_{T_k}}\sum_{s_{T_k}\in\mathcal{S}_{T_k}}\notag\\
		&P_{S_{T_k}|X_kU_{\bar{k}}Z_kV_{\bar{k},c}}(s_{T_k}|x_ku_{\bar{k}}z_kv_{\bar{k},c})d_k(s_{T_k},s'_{T_k}), 
	\end{align}
	an achievable rate-distortion region $\mathcal{I}_{\text{R-D}}^{\text{our}}$ includes any rate-distortion tuple $(R_1,R_2,D_1,D_2)$ that for some choice of variables
	\begin{align}
		UU_1U_2X_1X_2SS_{T_1}S_{T_2}Z_1Z_2YV_{1,c}V_{1,p}V_{2,c}V_{2,p}\hat{S}_{T_1}\hat{S}_{T_2}
	\end{align} 
	with joint distribution 
	\begin{align}\label{jointDistributionOur}
		&P_{U}P_{U_1|U}P_{U_2|U}P_{X_1|UU_1}P_{X_2|UU_2}P_{SS_{T_1}S_{T_2}}P_{YZ_1Z_2|X_1X_2S} \notag\\
		&P_{V_{1,c}V_{1,p}|UU_1U_2X_1Z_1}P_{V_{2,c}V_{2,p}|UU_1U_2X_2Z_2}\notag\\
		&P_{\hat{S}_{T_1}|X_1Z_1U_2V_{2,c}}P_{\hat{S}_{T_2}|X_2Z_2U_1V_{1,c}}
	\end{align}
	satisfying~\eqref{equ:ourRes} and~\eqref{equ:ourResCon} shown in the bottom of the next page 
	as well as the average distortion constraints 
	\begin{align}\label{equ:ourResDistortionCons}
		\setcounter{equation}{11}
		\mathbb{E}[d_k(S_{T_k},\hat{S}_{T_k})] \le D_k, k\in\{1,2\}.
	\end{align}
	It suffices to consider the auxiliary random variables $U$, $U_1$, $U_2$, $V_{1,c}$, $V_{1,p}$,  $V_{2,c}$, $V_{2,p}$ whose alphabets $\mathcal{U}$, $\mathcal{U}_1$, $\mathcal{U}_2$, $\mathcal{V}_{1,c}$, $\mathcal{V}_{1,p}$, $\mathcal{V}_{2,c}$, $\mathcal{V}_{2,p}$ have cardinalities $|\mathcal{U}|\le |\mathcal{Y}|+7$, $|\mathcal{U}_1|\le |\mathcal{U}||\mathcal{X}_{2}||\mathcal{Z}_{2}||\mathcal{Y}|+7$, $|\mathcal{U}_2|\le |\mathcal{U}||\mathcal{X}_{1}||\mathcal{Z}_{1}||\mathcal{Y}|+7$, $|\mathcal{V}_{1,c}|\le |\mathcal{U}||\mathcal{U}_1||\mathcal{U}_2||\mathcal{X}_1||\mathcal{Z}_1|+7$, $|\mathcal{V}_{2,c}|\le |\mathcal{U}||\mathcal{U}_1||\mathcal{U}_2||\mathcal{X}_2||\mathcal{Z}_2|+7$,
	$|\mathcal{V}_{1,p}|\le |\mathcal{U}||\mathcal{U}_1||\mathcal{U}_2||\mathcal{X}_1||\mathcal{Z}_1||\mathcal{V}_{1,c}|+5$, $|\mathcal{V}_{2,p}|\le |\mathcal{U}||\mathcal{U}_1||\mathcal{U}_2||\mathcal{X}_2||\mathcal{Z}_2||\mathcal{V}_{2,c}|+5$.
	\begin{IEEEproof}
		See Appendix \ref{appendix:proofOfTheorem3}.
	\end{IEEEproof}
\end{theorem}

\begin{figure*}[!b]
	\normalsize
	\setcounter{mytempeqncnt}{\value{equation}}
	\setcounter{equation}{9}
	\dotfill
	\begin{subequations}\label{equ:ourRes}
		\begin{align}
			R_k  \le & I(U_k;Z_{\bar{k}}|UU_{\bar{k}}X_{\bar{k}}) - I(V_{k,c};X_kZ_k|UU_1U_2X_{\bar{k}}Z_{\bar{k}}) \notag\\
			& + \min\bigg\{I(X_k;X_{\bar{k}}YV_{1,c}V_{2,c}V_{{\bar{k}},p}|UU_1U_2) - I(V_{k,p};Z_k|UU_1U_2X_1X_{2}YV_{1,c}V_{{2},c}V_{{\bar{k}},p}), \notag\\
			&\qquad\qquad I(X_1X_2;YV_{1,c}V_{2,c}|UU_1U_2)  - I(V_{1,p};Z_1|UU_1U_2X_1X_2YV_{1,c}V_{2,c}) \notag\\
			& \qquad\qquad\qquad\qquad\qquad - I(V_{2,p};Z_2|UU_1U_2X_1X_2YV_{1,c}V_{2,c}V_{1,p})\bigg\}, \ k\in\{1,2\}, \label{equ:ourResA}
		\end{align}
		\begin{align}
			R_1 + R_2 &\le I(U_1;Z_{2}|UU_{2}X_{2}) - I(V_{1,c};X_1Z_1|UU_1U_2X_{2}Z_{2}) + I(U_2;Z_{1}|UU_{1}X_{1}) - I(V_{2,c};X_{2}Z_{2}|UU_1U_2X_{1}Z_{1}) \notag\\
			& + I(X_1X_2;YV_{1,c}V_{2,c}|UU_1U_2)  - I(V_{1,p};Z_1|UU_1U_2X_1X_2YV_{1,c}V_{2,c}) - I(V_{2,p};Z_2|UU_1U_2X_1X_2YV_{1,c}V_{2,c}V_{1,p}), \label{equ:ourResB}
		\end{align}
		\begin{align}
			R_1 + R_2  \le  I(X_1X_2;Y) &- I(V_{1,c};Z_1|UU_1U_2X_1X_2Y)  - I(V_{2,c};Z_2|UU_1U_2X_1X_2YV_{1,c}) \notag\\
			&- I(V_{1,p};Z_1|UU_1U_2X_1X_2YV_{1,c}V_{2,c})-  I(V_{2,p};Z_2|UU_1U_2X_1X_2YV_{1,c}V_{2,c}V_{1,p}),\label{equ:ourResC}
		\end{align}
	\end{subequations}
	\dotfill
	\begin{subequations}\label{equ:ourResCon}
		\begin{align}
			I(U_k;Z_{\bar{k}}|UU_{\bar{k}}X_{\bar{k}}) - I(V_{k,c};X_kZ_k|UU_1U_2X_{\bar{k}}Z_{\bar{k}}) \ge 0, \ k\in\{1,2\}, \label{equ:ourResConsA}
		\end{align}
		\begin{align}
			I(X_k;X_{\bar{k}}YV_{1,c}V_{2,c}V_{{\bar{k}},p}|UU_1U_2) - I(V_{k,p};Z_k|UU_1U_2X_1X_{2}YV_{1,c}V_{{2},c}V_{{\bar{k}},p}) \ge 0 ,\ k\in\{1,2\},\label{equ:ourResConsB}
		\end{align}
		\begin{align}
			I(X_1X_2;&YV_{1,c}V_{2,c}|UU_1U_2)  - I(V_{1,p};Z_1|UU_1U_2X_1X_2YV_{1,c}V_{2,c}) - I(V_{2,p};Z_2|UU_1U_2X_1X_2YV_{1,c}V_{2,c}V_{1,p})\ge 0, \label{equ:ourResConsC}
		\end{align}
		\begin{align}
			I(X_1X_2;Y) &- I(V_{1,c};Z_1|UU_1U_2X_1X_2Y)  - I(V_{2,c};Z_2|UU_1U_2X_1X_2YV_{1,c}) \notag\\
			&- I(V_{1,p};Z_1|UU_1U_2X_1X_2YV_{1,c}V_{2,c})-  I(V_{2,p};Z_2|UU_1U_2X_1X_2YV_{1,c}V_{2,c}V_{1,p})\ge 0 \label{equ:ourResConsD}
		\end{align}
	\end{subequations}
	\setcounter{equation}{12}
\end{figure*}

\begin{remark}
	\textup{In the Ahmadipour--Wigger scheme, the compressed information $V_{k,c}^{(b-1)},V_{k,p}^{(b)},k\in\{1,$ $2\}$ in each block~$b\in[1:B]$ shown in Fig.~\ref{fig:innerBoundOur} are absent. It should be noted that our proposed scheme and the Ahmadipour--Wigger scheme are not special cases of each other. The reason is as follows. By setting $V_{k,c}^{(b-1)}$ to a constant, $V_{k,c}^{(b)}$ will automatically be reduced to a constant. But in Ahmadipour--Wigger scheme, $V_{k,c}^{(b)}$ is present and not a constant. i.e., they are only transmitted once but not retransmitted cooperatively. We show in the following Theorem~\ref{Theorem:CompInner} that our achievable region always includes that by the Ahmadipour--Wigger scheme. Besides, the Kobayashi--Hamad--Kramer--Caire scheme is a special case of both our scheme and the Ahmadipour--Wigger scheme by letting two transmitters send no compressed information.}
\end{remark}

\begin{remark}\label{remark:modelCoincideLI}
	\textup{When the feedback coincides with the channel state, i.e., $Z_1=S_1,Z_2=S_2$, and the sensing state parameters are $S_{T_1}=S_1,S_{T_2}=S_2$, both transmitters can always achieve zero distortion thus the distortion constraints are inactive. In this case, the model considered coincides with that for communication over SD-DM MAC with strictly causal state information at the transmitter~\cite{li2012multiple}.}
\end{remark}

\begin{remark}
	\textup{For the case considered in Remark~\ref{remark:modelCoincideLI}, there is no cooperation for the transmission of message and compressed information. In this case, Theorem~\ref{innerBound:CD} specializes to the results presented in~\cite{li2012multiple} by setting $U=u^*$, $U_k=u^*_k$, $V_{k,c}=v_{k,c}^*$, $k\in\{1,2\}$ almost surely for some specific values $u^*\in\mathcal{U}$, $u^*_k\in\mathcal{U}_k$, $v^*_{k,c}\in\mathcal{V}_{k,c}$, $k\in\{1,2\}$.}
\end{remark}

\begin{remark}
	\textup{The rate constraints in Theorem~\ref{innerBound:CD} can be interpreted as follows. For single-user rate constraints in (\ref{equ:ourResA}), the term $I(U_k;Z_{\bar{k}}|UU_{\bar{k}}X_{\bar{k}})$ represents the achievable rate constraint for common message part of transmitter~$k$, and $I(V_{k,c};X_kZ_k|UU_1U_2X_{\bar{k}}Z_{\bar{k}})$ 
	is the reduction in message rate due to the transmission of common compressed information. 
	The terms in $\text{min}(\cdot,\cdot)$ function represent the achievable rate constraints of private message sent by transmitter~$k$, where the first term $I(X_k;X_{\bar{k}}YV_{1,c}V_{2,c}V_{{\bar{k}},p}|UU_1U_2) - I(V_{k,p};Z_k|UU_1U_2X_1X_{2}YV_{1,c}V_{{2},c}V_{{\bar{k}},p})$ is the rate constraint for private message of transmitter~$k$ when the receiver has already decoded the codeword $X_{\bar{k}}$, and the second term represents the achievable rate constraint for private message when the receiver jointly decodes the private messages of both two transmitters. The sum rate constraints in (\ref{equ:ourResB}) and (\ref{equ:ourResC}) can be explained in a similar manner. 
	}
\end{remark}

\begin{remark}
	\textup{The inequality constraints in (\ref{equ:ourResConsA}), (\ref{equ:ourResConsB}), (\ref{equ:ourResConsC}), and (\ref{equ:ourResConsD}) are imposed to guarantee the successful transmission of compressed information. We emphasize that different from the results in the Ahmadipour--Wigger scheme~\cite[Theorem~3]{ahmadipour2023information}, which contain three inequality constraints on common compressed information, there are only two constraints~\eqref{equ:ourResConsA} and~\eqref{equ:ourResConsD} on common compressed information in Theorem~\ref{innerBound:CD}. This is due to the introduction of cooperative retransmission for common compressed information at two transmitters. As a result, our results in Theorem~\ref{innerBound:CD} supports transmission of a larger amount of common compressed information that can be decoded by both the other transmitter and the receiver, which leads to a larger achievable rate-distortion region in general.
	}
\end{remark}


\begin{remark}
	\textup{We note that our proposed achievable scheme can also address the ISAC scenarios where both the transmitters and the receiver perform sensing estimation. Let $S_R$ denote the sensing state parameter of the receiver, which is assumed to be correlated with the channel state. Our proposed scheme discussed in Appendix~\ref{appendix:proofOfTheorem3} can address this problem with an additional estimator
	\begin{align}
		&\hat{s}_R(x_1,x_2,y,v_{1,c},v_{2,c},v_{1,p},v_{2,p})  =\arg\min_{s'_{R}\in\hat{\mathcal{S}}_{R}}\sum_{s_{R}\in\mathcal{S}_{R}}\notag\\
		&P_{S_{R}|X_1X_2YV_{1,c}V_{2,c}V_{1,p}V_{2,p}}(s_{R}|x_1x_2yv_{1,c}v_{2,c}v_{1,p}v_{2,p})d_R(s_{R},s'_{R}), 
	\end{align}
	where $d_R:S_R\times\hat{S}_R\rightarrow[0,\infty)$ is a bounded distortion function. The corresponding achievable rate-distortion region is given as~\eqref{equ:ourRes}, \eqref{equ:ourResCon}, \eqref{equ:ourResDistortionCons}, and the sensing constraint for the receiver
	\begin{align}
		\mathbb{E}[d_R(S_{R},\hat{S}_{R})] \le D_R.
	\end{align}
	}
\end{remark}

Let $\mathcal{I}_{\text{R-D}}^{\text{aw}}$ denote the achievable rate-distortion region obtained by the Ahmadipour--Wigger scheme. 
Due to the different constructions between our scheme and the Ahmadipour--Wigger scheme as highlighted before,  it is nontrivial to compare
our result $\mathcal{I}_{\text{R-D}}^{\text{our}}$ with the existing one $\mathcal{I}_{\text{R-D}}^{\text{aw}}$. The following theorem indicates that in fact the achievable rate-distortion region $\mathcal{I}_{\text{R-D}}^{\text{our}}$ of our proposed scheme always includes $\mathcal{I}_{\text{R-D}}^{\text{aw}}$.

\begin{theorem}\label{Theorem:CompInner}
	Let $\mathcal{I}_{\text{R-D}}^{\text{our},\text{com}}$ denote the achievable rate-distortion region of Theorem~\ref{innerBound:CD} with $V_{1,p}=v_{1,p}^*,V_{2,p}=v_{2,p}^*$ almost surely for some specific values $v_{1,p}^*\in\mathcal{V}_{1,p}$, $v_{2,p}^*\in\mathcal{V}_{2,p}$, i.e., transmitting no private compressed information. Such a region always includes the achievable region $\mathcal{I}_{\text{R-D}}^{\text{aw}}$, and is a subset of $\mathcal{I}_{\text{R-D}}^{\text{our}}$, i.e., 
	\begin{align}
		\mathcal{I}_{\text{R-D}}^{\text{aw}}  \subseteq \mathcal{I}_{\text{R-D}}^{\text{our},\text{com}}\subseteq\mathcal{I}_{\text{R-D}}^{\text{our}}.
	\end{align}
	\begin{IEEEproof}
		See Appendix \ref{appendix:proofOfTheorem4}.
	\end{IEEEproof}
\end{theorem}

\subsection{Improved Outer Bound}

\begin{figure}[!t]
	\centering
	\includegraphics[width=1\linewidth]{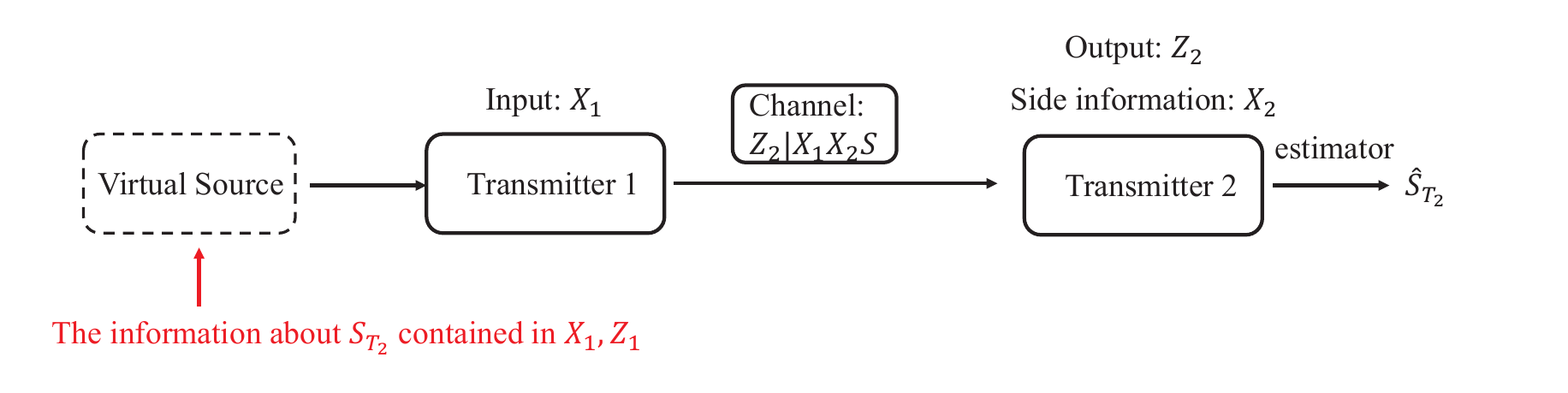}
	\caption{An intuitive illustration for the construction of the constraint on the information rate about $S_{T_2}$ obtained at transmitter 2. In this construction, the link from transmitter~1 to transmitter~2 can be modeled as a channel with transition probability $P_{Z_2|X_1X_2S}$. Transmitter 1 effectively serves as a virtual source, encoding information about $S_{T_{2}}$ and transmitting it to transmitter~2 through the channel $P_{Z_2|X_1X_2S}$. }
	\label{fig:IllustrationOuter}
\end{figure}

The Kobayashi--Hamad--Kramer--Caire outer bound adopts the ideas of dependence balance for communication and genie-aided estimators for parameter sensing. In particular, each transmitter~$k$ is assumed to have perfect knowledge of the transmitted signal $X_{\bar{k}}$ and the received echo signal $Z_{\bar{k}}$ of the other transmitter~$\bar{k}$. This assumption results in the following estimators for $k\in\{1,2\}$:
\begin{align}\label{estimator:sota}
	\hat{s}_{T_k}(&x_1,x_2,z_1,z_2) = \arg\min_{s'_{T_k}\in\hat{\mathcal{S}}_{T_k}}\sum_{s_{T_k}\in\mathcal{S}_{T_k}}\notag\\
	&P_{S_{T_k}|X_1X_2Z_1Z_2}(s_{T_k}|x_1x_2z_1z_2)d_k(s_{T_k},s'_{T_k}).
\end{align} 
Based on these genie-aided estimators, the following constraints on sensing distortion $D_{k},k\in\{1,2\}$ are obtained
\begin{align}\label{estimator:genie-aided}
	\mathbb{E}[d_k(S_{T_k},\hat{S}_{T_k}(X_1,X_2,Z_1,Z_2))]\le D_k,\  k\in\{1,2\}.
\end{align}
Such constraints are generally not tight, since they rely on high signal-to-noise ratio for the received signal $Z_k$ to decode $X_{\bar{k}},Z_{\bar{k}}$ of the other transmitter~$\bar{k}$.
To address this limitation, we propose an improved outer bound by introducing additional rate-limited constraints to further refine the bounds on sensing performance. We first 
introduce the following standard rate-distortion functions~\cite{el2011network} for sensing state parameters:
\begin{align}\label{definitionRateDistortionFunctions}
	f_{k,\text{R-D}}(D_k) &= \min_{P_{A_k|S_{T_k}}: \sum_{s_{T_k}a_k}P_{s_{T_k}}P_{a_k|s_{T_k}}d_k(s_{T_k},a_k)\le D_k}  \notag\\
	&\qquad\qquad\qquad I(S_{T_k}; A_k), \ k\in\{1,2\}.
\end{align}
These rate-distortion functions characterize the minimum rates required by estimators to guarantee the sensing distortion constraints $D_k,k\in\{1,2\}$~\cite{el2011network}. Moreover, as proved in Appendix~\ref{appendix:proofOuterBound}, we can further establish the following upper bound on $f_{k,\text{R-D}}(D_k)$ for $k\in\{1,2\}$: 
\begin{align}
	f_{k,\text{R-D}}(D_k) &\le I(S_{T_k}X_{\bar{k}};Z_k|X_kQ),
\end{align}
where $I(S_{T_k}X_{\bar{k}};Z_k|X_kQ)$ quantifies the maximum amount of information about $S_{T_k}$ that transmitter~$k$ can extract 
from its own echo signal $Z_k$ with side information $X_k$, as well as the information carried in $X_{\bar{k}}$ sent by the other transmitter~$\bar{k}$. Fig.~\ref{fig:IllustrationOuter} provides an intuitive illustration of how the constraint on the information rate about $S_{T_2}$ at transmitter~$2$ is constructed with $k=2$ as an example.

Based on the above discussions, we formally present the improved outer bound in the following theorem. We note that the outer bound involves auxiliary random variables $Q,T$, where $Q$ is the time-sharing random variable, and $T$ is introduced to capture the dependence between the codewords of two transmitters. 
\begin{theorem}\label{ourOuterBound}
	Let $(D_1,D_2)$ be the given distortion constraints for sensing state parameters. An outer bound $\mathcal{O}_{\text{R-D}}^{\text{our}}$ for capacity-distortion region $\mathcal{C}(D_1,D_2)$ is the set of all tuples $(R_1,R_2,D_1,D_2)$ satisfying\begin{subequations}\label{equ:outerDependenceBound}
		\begin{align}
			R_1 \le & I(X_1;YZ_1Z_2|X_2T), \\
			R_2 \le & I(X_2;YZ_1Z_2|X_1T), \\
			R_1 + R_2 \le & I(X_1X_2;YZ_1Z_2|T), \\
			R_1 + R_2 \le & I(X_1X_2;Y),
		\end{align} 
	\end{subequations}	
	with dependence-balanced constraint 
	\begin{align}\label{dependenceBalance}
		I(X_1;X_2|T)\le I(X_1;X_2|Z_1Z_2T),
	\end{align}
	sensing constraints~\eqref{estimator:genie-aided} and
	\begin{align}\label{sensingConsA}
		f_{k,\text{R-D}}(D_k) &\le I(S_{T_k}X_{\bar{k}};Z_k|X_kQ),\ k\in\{1,2\}, 
	\end{align}
	where $\hat{S}_{T_k}(X_1,X_2,Z_1,Z_2)$ in~\eqref{estimator:genie-aided} is the genie-aided estimator defined in~\eqref{estimator:sota},
	the random variables $QTX_1X_2SS_{T_1}S_{T_2}YZ_1Z_2\hat{S}_{T_1}\hat{S}_{T_2}$ have the joint distribution 
	\begin{align}
		P_{QT}P_{X_1X_2|T}P_{SS_{T_1}S_{T_2}}&P_{YZ_1Z_2|X_1X_2S}\notag\\
		&P_{\hat{S}_{T_1}|X_1X_2Z_1Z_2}P_{\hat{S}_{T_2}|X_1X_2Z_1Z_2},
	\end{align}
	and it suffices to consider $Q,T$ whose alphabets $\mathcal{Q},\mathcal{T}$ have cardinalities $|\mathcal{Q}|\le |\mathcal{X}_1||\mathcal{X}_2| + 5$, $|\mathcal{T}|\le (|\mathcal{X}_1||\mathcal{X}_2| + 3)(|\mathcal{X}_1||\mathcal{X}_2| +5)$.
	\begin{IEEEproof}
		See Appendix \ref{appendix:proofOuterBound}.
	\end{IEEEproof}
\end{theorem}


Let $\mathcal{O}_{\text{R-D}}^{\text{khkc}}$ denote Kobayashi--Hamad--Kramer--Caire outer bound, which corresponds to the result of Theorem~\ref{ourOuterBound} without the inclusion of constraints~\eqref{sensingConsA}. We thus conclude that Theorem~\ref{ourOuterBound} provides a tighter outer bound.
\begin{corollary}\label{Theorem:CompOut}
	The proposed outer bound $\mathcal{O}_{\text{R-D}}^{\text{our}}$ is a subset of that $\mathcal{O}_{\text{R-D}}^{\text{khkc}}$ in~\cite{kobayashi2019joint}, i.e., 
	\begin{align}
		\mathcal{O}_{\text{R-D}}^{\text{our}} \subseteq \mathcal{O}_{\text{R-D}}^{\text{khkc}}.
	\end{align}
\end{corollary}

\section{Numerical Examples}\label{section:example}
In this section, we first provide three examples to intuitively demonstrate the improvement of our results. The first two examples are constructed to show the benefits of introducing a unified cooperative communication and sensing scheme in the inner bound. The third example is to demonstrate the improvement of introducing the rate-limited constraints on sensing in the outer bound. Then, an example is constructed to show that the results of our proposed scheme strictly includes that of the Ahmadipour--Wigger scheme, which achieves better performance in both sensing and communication, and the proposed outer bound is strictly tighter than that in \cite{kobayashi2019joint}. In particular, we have assumed that the reconstruction alphabets $\hat{\mathcal{S}}_{T_k}^n,k\in\{1,2\}$ are the same as the sensing parameter alphabets $\mathcal{S}_{T_k}^n$ for the following examples.

\subsection{The Advantages of Unified Cooperative Communication and Sensing Scheme}
\begin{example}\label{example:private}
	\textup{Consider a MAC where the inputs $X_1, X_2$ are binary. The channel output $Y = (Y_1,Y_2)$ is a binary pair with orthogonal components $Y_1 = X_1 \oplus S_1, Y_2 = X_2 \oplus S_2$, where the channel state random variables $S_1,S_2$ are mutually independent binary random variables with $H(S_1)=H(S_2) =0.5$, namely, with $P_{S_1}(1) = P_{S_2}(1) \approx 0.11$. Transmitter~1's feedback is $Z_1 = X_1 \oplus S_2$, and transmitter~2's feedback is $Z_2 = X_2 \oplus S_1$. The sensing state parameters of two transmitters are assumed as $S_{T_1} = S_2,  S_{T_2} = S_1$, and the Hamming distortion $d(s_T,\hat{s}_T)=s_T\oplus\hat{s}_T$ is considered as the distortion metric.}
	
	\textup{In this example, both two transmitters can always achieve zero distortion for sensing state parameters by conducting $\hat{S}_{T_k} = Z_k \oplus X_k, \ k\in\{1,2\}$.
	Moreover, each transmitter can obtain no information sent by the other transmitter, as the feedback $Z_k=X_k\oplus S_{\bar{k}}$ only provides information about channel state $S_{\bar{k}}$. Thus, there is no message cooperation and compressed information gain for the Ahmadipour--Wigger scheme as it requires both the transmitter and the receiver decode common message and common compressed information. However, in our proposed scheme corresponding to Theorem~\ref{innerBound:CD}, transmitter~$k$ can send private compressed information $V_{k,p} = S_{\bar{k}}$ that is related to channel state to the receiver. This private compressed information is beneficial for decoding the message sent by the other transmitter $\bar{k}$ at the receiver. For example, consider the following choice of random variables
	\begin{align}
		X_k = U_k \oplus  \Theta_k = U \oplus \Sigma_k \oplus  \Theta_k , \ k\in\{1,2\}, 
	\end{align}
	where $U,\Sigma_1,\Sigma_2,\Theta_1,\Theta_2$ are independent binary random variables, 
	and let only transmitter~1 send private compressed information about $S_2$, i.e.,
	\begin{align}
		V_{1,p} = S_{2},
	\end{align}
	and 
	\begin{align}
		&V_{1,c}= v^*_{1,c}, V_{2,c}= v^*_{2,c},V_{2,p} = v^*_{2,p},
	\end{align}
	almost surely for some specific values $v^*_{1,c}\in\mathcal{V}_{1,c}$, $v^*_{2,c}\in\mathcal{V}_{2,c}$, $v^*_{2,p}\in\mathcal{V}_{2,p}$. One can find that the rate-distortion tuple $(R_1,R_2,D_1,D_2) = (0,1,0,0)$ lies in $\mathcal{I}_{\text{R-D}}^{\text{our}}$ by letting $U=\Sigma_1=\Sigma_2=0$ almost surely and $P_{\Theta_1}(1)=P_{\Theta_2}(1)=0.5$.
	However, $(R_1,R_2,D_1,D_2) = (0,1,0,0)$ is not in the inner bound $\mathcal{I}_{\text{R-D}}^{\text{aw}}$. In fact, in $\mathcal{I}_{\text{R-D}}^{\text{aw}}$, the achievable rate of transmitter~2 is no more than 0.5. Together with Theorem~\ref{Theorem:CompInner}, we establish $\mathcal{I}_{\text{R-D}}^{\text{aw}} \subsetneq\mathcal{I}_{\text{R-D}}^{\text{our}}$ for this channel.} 
	\begin{IEEEproof}
		See Appendix \ref{appendix:proofExamplePrivate}.
	\end{IEEEproof}
\end{example}

\begin{remark}
	\textup{In Example \ref{example:private}, the echo signals of two transmitters $Z_1 = X_1 \oplus S_2$, $Z_2 = X_2 \oplus S_1$ provide only channel state information. Two transmitters cannot decode any common messages of the other, and thus no direct cooperation through message transmission. However, as shown in Example \ref{example:private}, our scheme allows transmitter~1 to send state information acquired from the echo signals as part of private messages, which can improve the decoding of receiver to achieve a larger transmission rate for transmitter~2. This operation can also be performed by transmitter~2. In this way, two transmitters achieve the cooperation in an implicit manner. 
	}
\end{remark}

\begin{example}\label{example:common}
	\textup{Consider a MAC where the inputs $X_1, X_2$ are binary. The channel output is $Y = (X_1 \oplus S_1 \oplus N, X_2 \oplus S_2)$. Transmitter 1's feedback is $Z_1 = (X_1\oplus S_1, X_1 \oplus S_2)$, and transmitter 2's feedback is $Z_2 = X_1\oplus B$. The channel states $S_1,S_2,B,N$ are mutually independent binary random variables with entropies $H(S_1) =H(S_2)= H(B)=0.5$ and $H(N) = 1$, namely, with $P_{S_1}(1) = P_{S_2}(1) =P_B(1)= 0.11, P_N(1) = 0.5$. The sensing state parameters are assumed as $S_{T_1} = S_2,  S_{T_2} = S_1$, 
	and the Hamming distortion $d(s_T,\hat{s}_T)=s_T\oplus\hat{s}_T$ is considered as the distortion metric.}
	
	\textup{In this example, transmitter 1 can always achieve zero distortion by conducting $\hat{S}_{T_1} = (X_1 \oplus S_2) \oplus X_1$, while transmitter 2 cannot directly get any information about $S_{T_2} = S_1$ as $Z_2 = X_1\oplus B$. In our proposed scheme corresponding to Theorem~\ref{innerBound:CD}, transmitter 1 can send common compressed information related to echo signals to improve the estimation of transmitter 2. One can choose 
	\begin{align}
		X_k = U_k \oplus  \Theta_k = U \oplus \Sigma_k \oplus  \Theta_k , \ k\in\{1,2\},
	\end{align}
	where $U,\Sigma_1,\Sigma_2,\Theta_1,\Theta_2$ are independent binary random variables, and let only transmitter~1 send common compressed information about $S_1$, i.e.,
	\begin{align}
		V_{1,c} = S_1,
	\end{align}
	and
	\begin{align}
		V_{1,p}= v^*_{1,p}, V_{2,c}= v^*_{2,c},V_{2,p} = v^*_{2,p},
	\end{align}
	almost surely for some specific values $v^*_{1,p}\in\mathcal{V}_{1,p}$, $v^*_{2,c}\in\mathcal{V}_{2,c}$, $v^*_{2,p}\in\mathcal{V}_{2,p}$.
	It can be verified that constraints (\ref{equ:ourResConsA}), (\ref{equ:ourResConsB}), (\ref{equ:ourResConsC}), (\ref{equ:ourResConsD}) are satisfied when $U=\Theta_1=\Sigma_2=0$ almost surely and $P_{\Sigma_1}(1)=P_{\Theta_2}(1)=0.5$. As a result, transmitter 2 obtains $V_{1,c} = S_1$ and thus achieves zero distortion. Therefore, the rate-distortion tuple 
	\begin{align}
		(R_1,R_2,D_1,D_2) = (0,0,0,0)
	\end{align} 
	lies in $\mathcal{I}_{\text{R-D}}^{\text{our}}$ (also in $\mathcal{I}_{\text{R-D}}^{\text{our,com}}$) of Theorem \ref{innerBound:CD}. However, $(0,0,0,0)$ is not in the inner bound $\mathcal{I}_{\text{R-D}}^{\text{aw}}$. Together with Theorem~\ref{Theorem:CompInner}, we have $\mathcal{I}_{\text{R-D}}^{\text{aw}} \subsetneq\mathcal{I}_{\text{R-D}}^{\text{our,com}}\subseteq\mathcal{I}_{\text{R-D}}^{\text{our}}$ for this channel.}
	\begin{IEEEproof}
		See Appendix~\ref{appendix:proofExampleCommon}.	
	\end{IEEEproof}
\end{example}

\begin{remark}
	\textup{In Example~\ref{example:common}, transmitter~1 can obtain the perfect sensing state  parameters of both two transmitters, while transmitter~2 can only obtain a noisy version of signal sent by transmitter~1. The channel from transmitter~1 to transmitter~2 with input $X_1$ and output $Z_2$ is with capcity 0.5 as $Z_2 = X_1\oplus B$ and $H(B) =0.5$. The communication channel from the transmitters to the receiver are with capacity $0$ and $0.5$, respectively, as $H(N) = 1$ and $H(S_2) = 0.5$.
	Both our proposed scheme and the Ahmadipour--Wigger scheme enable transmitter~1 to send a description of channel state $S_1$ to improve the sensing of transmitter~2 and the message decoding at the receiver. The amount of common compressed information, i.e., the description of $S_1$, to be transmitted is bounded by the signal-to-noise ratios of both the forward channel output $Y$ and the echo signals $Z_2$. As shown in Example~\ref{example:common}, transmitter cooperation for sending common compressed information in our proposed scheme enables a lossless description $V_{1,c} = S_1$ to be sent to both the receiver and the decoder. We show in Appendix~\ref{appendix:proofExampleCommon} that this is not possible using the Ahmadipour--Wigger scheme.
	}
\end{remark}

\subsection{The Advantage of Introducing Rate-Limited Constraints in Our Outer Bound}
\begin{example}
	\textup{Consider a MAC where the inputs $X_1, X_2$ are binary. The channel output $Y = S_1X_1+S_2X_2$ is ternary, where the channel states $S_1,S_2$ are mutually independent binary random variables with $H(S_1)=H(S_2) =0.5$. Transmitter 1's feedback is $Z_1 = X_1 \oplus S_1$, and transmitter 2's feedback $Z_2 = X_2 \oplus S_2$.  The sensing state parameters are assumed as $S_{T_1} = S_2,  S_{T_2} = S_1$,
	and the Hamming distortion $d(s_T,\hat{s}_T)=s_T\oplus\hat{s}_T$ is considered as the distortion metric.}
	
	\textup{In this example, the rate-distortion tuple
	\begin{align}
		(R_1,R_2,D_1,D_2) = (0,0,0,0)
	\end{align}
	lies in the $\mathcal{O}_{\text{R-D}}^{\text{khkc}}$ but not in our outer bound $\mathcal{O}_{\text{R-D}}^{\text{our}}$ of Theorem \ref{ourOuterBound}. Together with Theorem~\ref{Theorem:CompOut}, we establish $\mathcal{O}_{\text{R-D}}^{\text{our}} \subsetneq \mathcal{O}_{\text{R-D}}^{\text{khkc}}$ for this channel. 
	More specifically, when choosing the genie-aided estimators (\ref{estimator:sota}), both two transmitters can obtain zero distortion for sensing state parameters. While in our outer bound, based on sensing constraints (\ref{sensingConsA}), for $k=1$, we have
	\begin{align}
		f_{1,\text{R-D}}(D_1) &\le I(S_{T_1}X_{2};Z_1|X_1Q) = I(S_2X_2;Z_1|X_1Q) \notag\\
		& = H(Z_1|X_1Q) - H(Z_1|X_1X_2S_2Q)\notag\\
		&\overset{(a)}= H(S_1|Q)- H(S_1|S_2Q)\notag\\
		&\overset{(b)} = 0,
	\end{align}
	where $(a)$ follows from that $Z_1=X_1\oplus S_1$, $(b)$ follows from that $S_1$ is independent of $QS_2$. Similarly, for $k=2$, we have
	\begin{align}
		f_{2,\text{R-D}}(D_2) &\le I(S_{T_2}X_{1};Z_2|X_2Q) = I(S_1X_1;Z_2|X_2Q)=0.
	\end{align} 
	Given $f_{1,\text{R-D}}(D_1)\le0, f_{2,\text{R-D}}(D_2)\le0$, one can obtain that in our outer bound, there is
	\begin{align}
		D_{1,\text{min}} = \text{min}\{P_{S_2}, 1- P_{S_2}\}, 
		D_{2,\text{min}} = \text{min}\{P_{S_1}, 1- P_{S_1}\}. 
	\end{align}
	}
\end{example}

\subsection{The Advantages Can Be Obtained Simultaneously}
\begin{example}\label{example:general}
	\textup{Consider a MAC where the inputs $X_1, X_2$ are binary. The channel output $Y = (Y_1,Y_2)$ is a binary pair with channel law $Y_1 = X_1 \oplus S_1, Y_2 = X_2 \oplus S_2$, where the channel states $S_1,S_2$ are mutually independent binary random variables with $P_{S_1}(1) = 0.24$, $P_{S_2}(1) = 0.05$. Transmitter~1's feedback is $Z_1 = X_1 \oplus N$, and transmitter~2's feedback $Z_2 = (BX_1, X_2 \oplus S_1)$ is a binary pair, where $N, B$ are mutually independent binary random variables with $P_{N}(1) = 0.3$, $P_{B}(1) = 0.5$. The sensing state parameters of two transmitters are assumed as $S_{T_1} = S_{T_2} = N$, and the Hamming distortion $d(s_T,\hat{s}_T)=s_T\oplus\hat{s}_T$ is considered.}

	\subsubsection{Characterization of inner bounds}
	\textup{In this example, transmitter~1 can always achieve zero distortion, and we thus focus on the achievable region $(R_1,R_2,D_2)$. 
	Note that even for the considered binary example, the explictly characterizations of both our results $\mathcal{I}_{\text{R-D}}^{\text{our}},\mathcal{I}_{\text{R-D}}^{\text{our},\text{com}}$ and the existing one $\mathcal{I}_{\text{R-D}}^{\text{aw}}$ of \cite[Theorem~3]{ahmadipour2023information} are intractable due to the presence of auxiliary
	random variables with large alphabet sizes that hinders an exhaustive search.
	To cope with this challenge, 
	we first evaluate $\mathcal{I}_{\text{R-D}}^{\text{aw}}, \mathcal{I}_{\text{R-D}}^{\text{our},\text{com}},\mathcal{I}_{\text{R-D}}^{\text{our}}$
	by considering some particular choice of random variables involved in the rate-distortion characterization. 
	Then, we theoretically prove that $\mathcal{I}_{\text{R-D}}^{\text{aw}}  \subsetneq \mathcal{I}_{\text{R-D}}^{\text{our},\text{com}}\subsetneq\mathcal{I}_{\text{R-D}}^{\text{our}}$, i.e., our results strictly include the existing one, regardless of the choice of the random variables.}
	
	\begin{figure}[!t]
		\centering
		\includegraphics[width=1\linewidth]{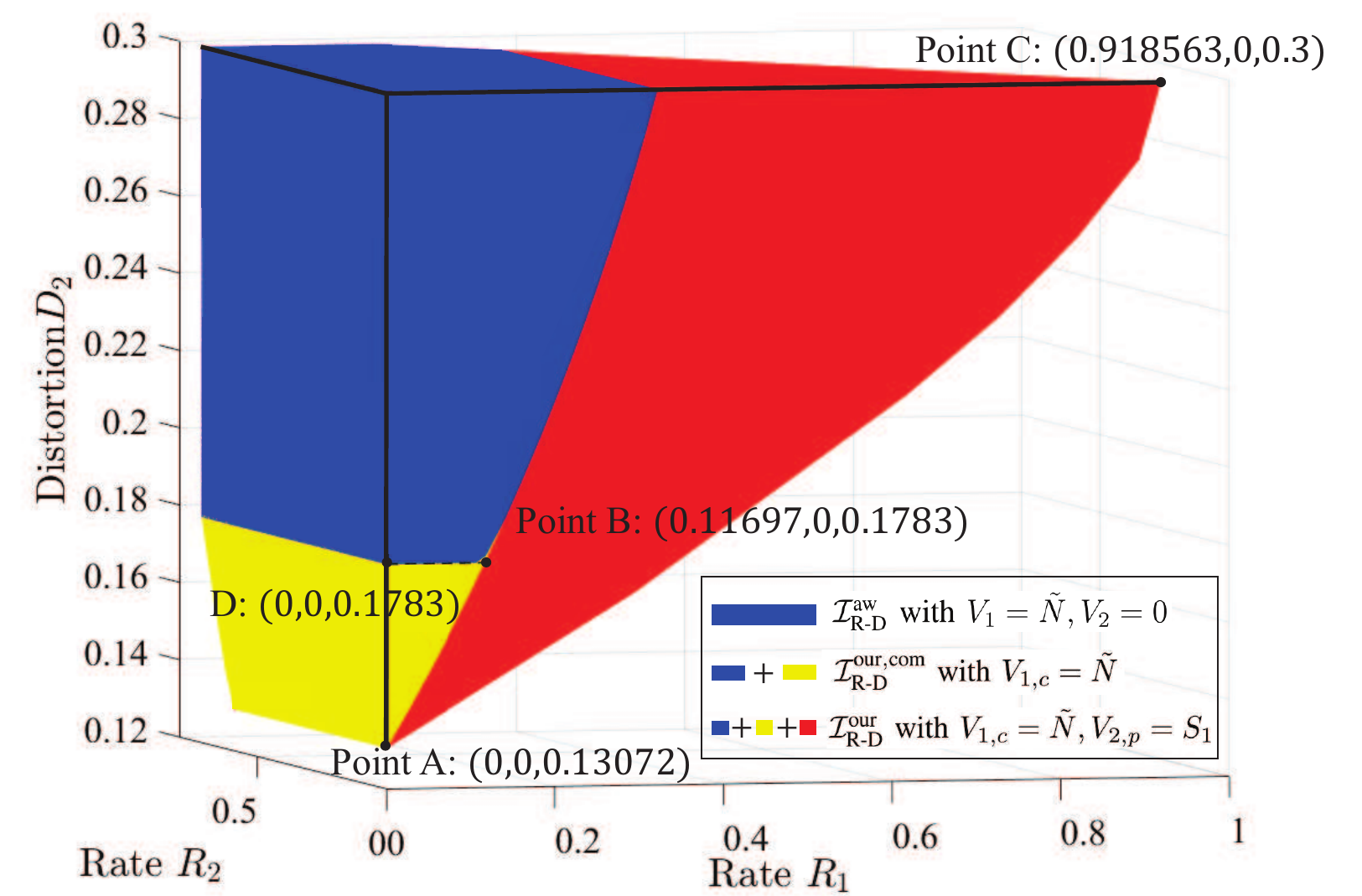}
		\caption{An illustration of $(R_1, R_2, D_2)$ tradeoff for Example \ref{example:general} with the considered coding scheme.}
		\label{fig:binaryCase}
	\end{figure}
	
	\textup{We consider the following choice of random variables to evaluate region $\mathcal{I}_{\text{R-D}}^{\text{our}}$: 
	\begin{subequations}
		\begin{align}
			&X_k = U_k \oplus  \Theta_k = U \oplus \Sigma_k \oplus  \Theta_k , \ k\in\{1,2\}, \label{equ:codingUU1U2}\\
			&V_{1,c} = \tilde{N}, \  V_{2,p} = S_1,  \label{equ:codingCompressA}\\
			&V_{1,p} = v^*_{1,p},\ V_{2,c} =v^*_{2,c}, \label{equ:codingCompressB}
		\end{align}
	\end{subequations}
	where $U,\Sigma_1,\Sigma_2,\Theta_1,\Theta_2$ are independent binary random variables, 
	$N = \tilde{N}\oplus E$, and $E\sim \text{Bern}(P_E)$ is a binary quantization variable with $P_E(1)\in[0,P_N(1)]$, and $V_{1,p} = v^*_{1,p},V_{2,c} =v^*_{2,c}$ almost surely for some specific values $v^*_{1,p}\in\mathcal{V}_{1,p},v^*_{2,c}\in\mathcal{V}_{2,c}$. For $\mathcal{I}_{\text{R-D}}^{\text{our},\text{com}}$, we still consider \eqref{equ:codingUU1U2}, \eqref{equ:codingCompressA}, and \eqref{equ:codingCompressB} except that $V_{2,p} = v^*_{2,p}$ almost surely for some specific value $v^*_{2,p}\in\mathcal{V}_{2,p}$. To characterize region $\mathcal{I}_{\text{R-D}}^{\text{aw}}$, we consider (\ref{equ:codingUU1U2}) and set variables $V_1 = \tilde{N}$ and $V_2 = v^*_2$ almost surely for some specific value $v^*_2\in\mathcal{V}_{2}$ . For the above choices of random variables, the achievable regions of all three schemes are presented in Fig.~\ref{fig:binaryCase}.}

	\textup{It can be shown that tuples $(R_1,R_2,D_2) = (0,0,0.13072)$ and $(0.11697,0,0.1783)$ are in $\mathcal{I}_{\text{R-D}}^{\text{our},\text{com}}$ (as shown in Fig. \ref{fig:binaryCase}) but not in $\mathcal{I}_{\text{R-D}}^{\text{aw}}$, and tuple $(0.918563,0,0.3)$ is in $\mathcal{I}_{\text{R-D}}^{\text{our}}$ (as shown in Fig. \ref{fig:binaryCase}) but not in $\mathcal{I}_{\text{R-D}}^{\text{our},\text{com}}$. Together with Theorem~\ref{Theorem:CompInner}, we have
	\begin{align}
		\mathcal{I}_{\text{R-D}}^{\text{aw}}  \subsetneq \mathcal{I}_{\text{R-D}}^{\text{our},\text{com}} \subsetneq \mathcal{I}_{\text{R-D}}^{\text{our}}
	\end{align}
	for this channel.}
	\begin{IEEEproof}
		The proof of $\mathcal{I}_{\text{R-D}}^{\text{aw}}  \subsetneq \mathcal{I}_{\text{R-D}}^{\text{our},\text{com}} \subsetneq \mathcal{I}_{\text{R-D}}^{\text{our}}$ can be found in  Appendix \ref{appendix:proofExampleGeneral}.	
	\end{IEEEproof}
\subsubsection{Characterization of outer bounds}
\textup{Transmitter~1 can always achieve zero distortion, and we notice that both $\mathcal{O}_{\text{R-D}}^{\text{our}}$ and $\mathcal{O}_{\text{R-D}}^{\text{khkc}}$ are based on the idea of dependence balance and have the same results on communication when ignoring the sensing task. Therefore, we focus on the tradeoff between symmetric rate $R_1=R_2=R$ and distrotion $D_2$.}

\textup{It should be noted that the evaluations of outer bounds $\mathcal{O}_{\text{R-D}}^{\text{our}}$ and $\mathcal{O}_{\text{R-D}}^{\text{khkc}}$ are intractable due to the fact that the large cardinality of involved auxiliary random variables prohibits an exhaustive search \cite{tandon2009outer}. To address these challenges, we apply the technique of \cite{tandon2009outer} by introducing an adaptive parallel channel extension for the dependence balance bound and using a composite function $\phi(t) = \frac{1-\sqrt{|1-2t|}}{2},\ t\in[0,1]$ and its properties to evaluate both our outer bound $\mathcal{O}_{\text{R-D}}^{\text{our}}$ and the existing one $\mathcal{O}_{\text{R-D}}^{\text{khkc}}$, denoted as $\mathcal{O}_{\text{R-D-PC}}^{\text{our},X_2}$ and $\mathcal{O}_{\text{R-D-PC}}^{\text{khkc},X_2}$, respectively. The details are given in Appendix~\ref{appendix:parallelChannelExtension}.}

\begin{figure}[!t]
	\centering
	\includegraphics[width=1\linewidth]{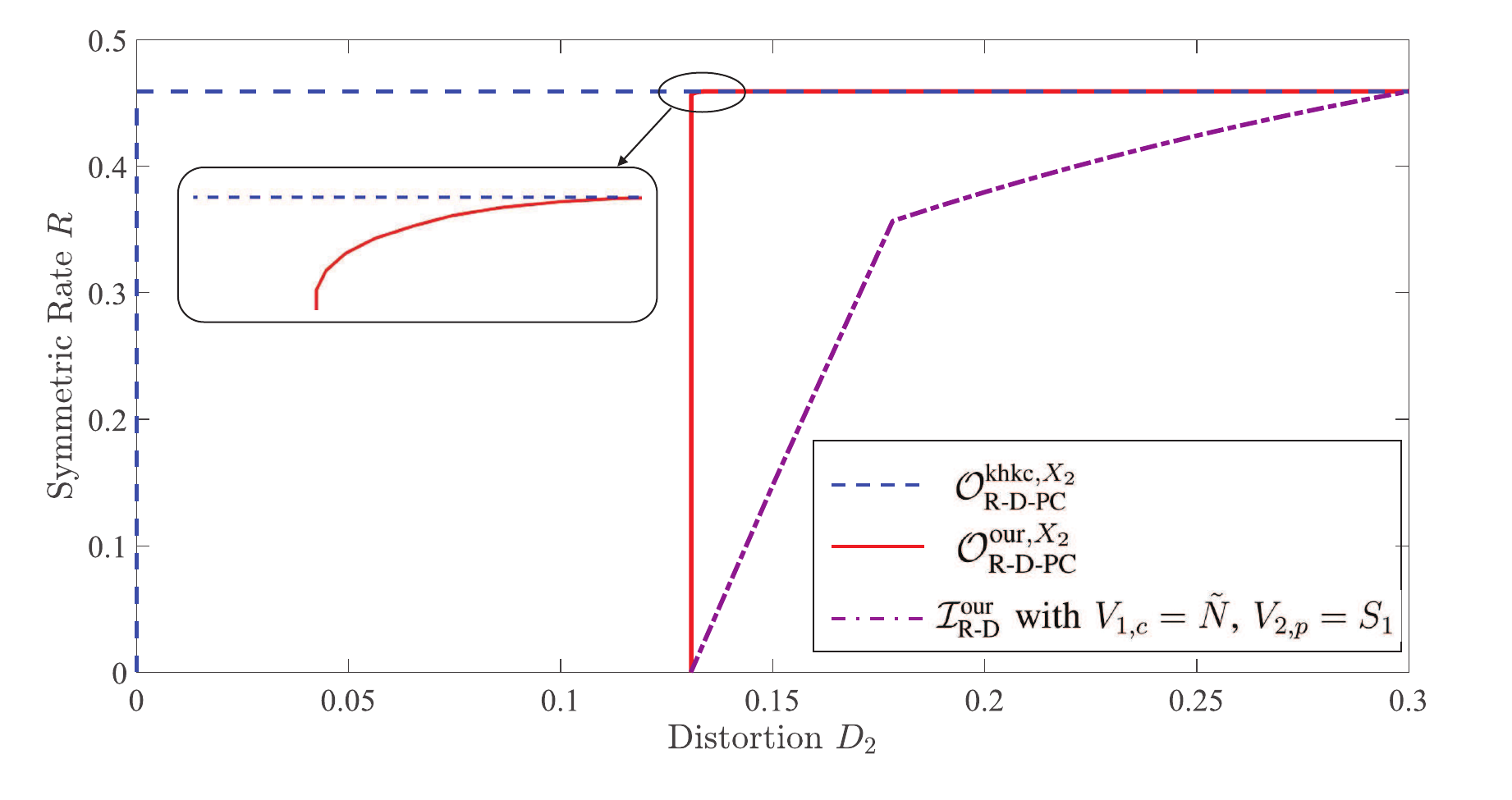}
	\caption{An illustration of the symmetric--rate-distortion tradeoff $(R_1=R_2=R, D_2)$ for Example \ref{example:general}.}
	\label{fig:OuterRes}
\end{figure}

\textup{Based on the parallel channel extension, one can conduct numerical characterization of symmetric--rate-distortion region $(R_1=R_2=R,D_2)$ for $\mathcal{O}_{\text{R-D-PC}}^{\text{our},X_2}$, which is an outer bound of $\mathcal{O}_{\text{R-D}}^{\text{our}}$ in Theorem~\ref{ourOuterBound}. The same technique can be applied to get a valid outer bound $\mathcal{O}_{\text{R-D-PC}}^{\text{khkc},X_2}$ for the results $\mathcal{O}_{\text{R-D}}^{\text{khkc}}$ in~\cite{kobayashi2019joint} and to characterize the symmetric--rate-distortion region $(R_1=R_2=R,D_2)$ in $\mathcal{O}_{\text{R-D-PC}}^{\text{khkc},X_2}$. We note that outer bounds $\mathcal{O}_{\text{R-D}}^{\text{our}}$ and $\mathcal{O}_{\text{R-D}}^{\text{khkc}}$ both use the idea of dependence balance. Thus, such a parallel channel extension results in the same performance on communication. Moreover, since the genie-aided state estimators are considered in $\mathcal{O}_{\text{R-D}}^{\text{khkc}}$, the parallel channel extension $Z_{\text{PC}}=X_2$ does not effect sensing performance for $\mathcal{O}_{\text{R-D}}^{\text{khkc}}$. Therefore, such a characterization can demonstrate the advantages of our improved outer bound on sensing performance. }

\textup{The results are shown in Fig. \ref{fig:OuterRes}, which shows that our outer bound is tighter than state-of-the-art \cite{kobayashi2019joint}. Specifically, one can easily check that tuple $(R=0,D_2=0)$ is in $\mathcal{O}_{\text{R-D}}^{\text{khkc}}$ but not in $\mathcal{O}_{\text{R-D}}^{\text{our}}$, which reveals that 
\begin{align}
	\mathcal{O}_{\text{R-D}}^{\text{our}} \subsetneq \mathcal{O}_{\text{R-D}}^{\text{khkc}}.
\end{align}
Moreover, it can also be seen that our proposed scheme can achieve the sensing-optimal point $(D_{2,\text{min}},0)$ and communication-optimal point $(D_{2,\text{max}},R_{\text{max}})$, while for certain distortion constraint such as $D_2 = 0.2$, how to achieve the optimal performance for ISAC still remains open.}
\end{example}

\begin{remark}
\textup{The results and analysis presented in Example~\ref{example:general} demonstrate the following facts:
\begin{enumerate}
	\item For the inner bound, the performance gain of our proposed scheme is two-fold. First, by allowing transmitters to cooperatively retransmit the common compressed information $(V_{1,c},V_{2,c})$ in a similar manner to messsage coopeartion, our achievable scheme can get a better sensing performance (point A in Fig. \ref{fig:binaryCase}), and enlarge the rate-region (point B) for the same distortion (point D $(0,0,0.1783)$ is in $\mathcal{I}_{\text{R-D}}^{\text{aw}}$) that the existing one can achieve. The reason behind this improvement is that compressed information cooperation between two transmitters leverages the beamforming gain, which results in transmitting a larger amount of common compressed information $(V_{1,c},V_{2,c})$. Second, allowing the transmitter~2 to send private compressed information $V_{2,p}=S_1$ decoded by the receiver only conveys information about channel states, facilitating the message decoding of $W_1$ at the receiver and enlarging the rate region (point C).
	\item For the outer bound, our results introduce additional rate-limited constraints on sensing performance, which provides a tighter bound for sensing performance, and thus a tighter outer bound for capacity-distortion region.
\end{enumerate}
}
\end{remark}

\section{Conclusion}\label{sec:conclusion}
In this paper, we have investigated the fundamental limits of ISAC over SD-DM MACs with correlated sensing state parameters and channel states. 
A new achievable scheme that combines message cooperation and joint compression of past transmitted codewords and echo signals has been proposed, and the corresponding inner bound of the capacity-distortion region has been proved to include that of \cite[Theorem~3]{ahmadipour2023information}. We have also established an improved outer bound of the capacity-distortion region by introducing the rate-limited constraints on sensing. It has been demonstrated, through several numerical examples, that
the improved inner bound can achieve better communication and sensing performance, and the proposed outer bound  strictly improves the existing one. We finally remark that while the inner and outer bounds presented in this paper improve upon the existing results \cite{kobayashi2019joint,ahmadipour2023information}, the optimal capacity-distortion region for the model considered remains open. Future work includes further tightening the bounds for ISAC over MACs and investigation of the fundamental limits of other multi-terminal ISAC building blocks, such as in broadcast and relay channels.


%

\appendices
\section{Proof of Theorem~\ref{innerBound:CD}}\label{appendix:proofOfTheorem3}
The achievable scheme proposed, which combines with implicit binning, block Markov encoding, and backward decoding, consists of $B+\tilde{B}$ transmission blocks. The last $\tilde{B}$ blocks called ``termination blocks'' are necessary to guarantee that the receiver can successfully obtain the compressed information corresponding to blocks $B$ and $B-1$. We first focus on the first $B$ blocks by assuming that the receiver can successfully obtain compressed information corresponding to blocks $B$ and $B-1$. Then, the detailed discussion of termination blocks is presented.

\subsection{The Proposed Coding Scheme}\label{achievabilityProposedScheme}
Each of the first $B$ blocks is of length $N$ channel uses. Transmitter~$k$ transmits $B-1$ i.i.d. messages $\{w_k^{(b)}\}_{b=1}^{B-1}$ over $B+\tilde{B}$ blocks. Each message $w_k^{(b)}$ is partitioned into two messages $\{w_{k,p}^{(b)},w_{k,c}^{(b)}\}$. The subscripts ``p'' and ``c'' here stand for ``private'' and ``common'', respectively. The messages $w_{k,p}^{(b)}\in[1:2^{NR_{k,p}}]$ and $w_{k,c}^{(b)}\in[1:2^{NR_{k,c}}]$ are uniformly distributed and mutually independent for all $k\in\{1,2\}$, $b\in[1:B-1]$, where $R_k=R_{k,p}+R_{k,c}$. Let $X_{1,(b)}^N$, $X_{2,(b)}^N$, $S_{(b)}^N$, $S_{T_1,(b)}^N$, $S_{T_2,(b)}^N$, $Y_{(b)}^N$, $Z_{1,(b)}^N$, $Z_{2,(b)}^N$ denote the inputs, states, sensing state parameters, outputs, and feedbacks in block~$b$.

\subsubsection{Codebook Generation}
\begin{figure}[!t]
	\centering
	\includegraphics[width=1\linewidth]{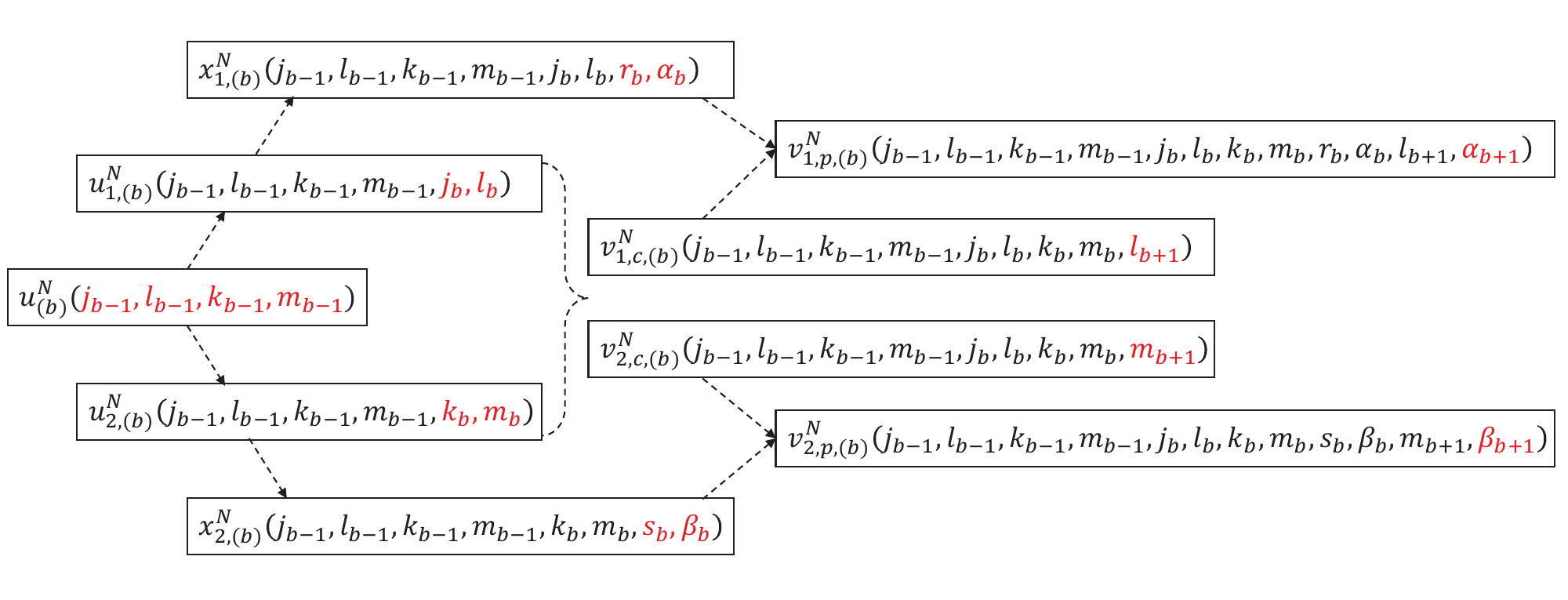}
	\caption{An illustration of codebook generation in block $b$.}
	\label{fig:codebook}
\end{figure}

Fix a joint distribution $P_{U}P_{U_1|U}P_{U_2|U}P_{X_1|UU_1}P_{X_2|UU_2}P_{SS_{T_1}S_{T_2}}P_{YZ_1Z_2|X_1X_2S}$ $P_{V_{1,c}V_{1,p}|UU_1U_2X_1Z_1}P_{V_{2,c}V_{2,p}|UU_1U_2X_2Z_2}P_{\hat{S}_{T_1}|X_1Z_1U_2V_{2,c}}$\\
$P_{\hat{S}_{T_2}|X_2Z_2U_1V_{1,c}}$ as in Theorem~\ref{innerBound:CD}, where
$P_{SS_{T_1}S_{T_2}}$ $P_{YZ_1Z_2|X_1X_2S}$ is defined by the channel. Let $R_k=R_{k,p}+R_{k,c}$, $R_{k,c}\ge0,R_{k,p}\ge0,R_{v_{k,c}}\ge0,R_{v_{k,p}}\ge0$, $k\in\{1,2\}$.
For block $b \in [1:B]$, an illustration of constructed codebook is presented in Fig. \ref{fig:codebook}, and the detailed codebook is as follows. 
\begin{itemize}
	\item Generate $2^{N( R_{1,c} + R_{v_{1,c}} + R_{2,c} + R_{v_{2,c}} )}$ sequences $u_{(b)}^N($ $j_{b-1}, l_{b-1}, k_{b-1}, m_{b-1})$, $j_{b-1} \in [1 : 2^{NR_{1,c}}]$, $l_{b-1} \in$ $ [1 : 2^{NR_{v_{1,c}}}]$, $k_{b-1} \in [1 : 2^{NR_{2,c}}]$, $m_{b-1} \in [1 : 2^{NR_{v_{2,c}}}]$, i.i.d. according to $P_U(\cdot)$.
	\item For each tuple $(j_{b-1},l_{b-1},k_{b-1},m_{b-1})$, generate $2^{N(R_{1,c} + R_{v_{1,c}})}$ sequences $u_{1,(b)}^N(j_{b-1},l_{b-1},k_{b-1},m_{b-1},$ $j_{b},l_{b})$, $j_b\in[1:2^{NR_{1,c}}]$. $l_{b}\in[1:$ $2^{NR_{v_{1,c}}}]$, i.i.d. according to $P_{U_1|U}(\cdot|u)$. Similarly, generate $2^{N(R_{2,c} + R_{v_{2,c}})}$ sequences $u_{2,(b)}^N(j_{b-1},l_{b-1},k_{b-1},m_{b-1},k_b,m_{b})$ i.i.d.
	according to $P_{U_2|U}(\cdot|u)$.
	\item For each tuple $(j_{b-1},l_{b-1},k_{b-1},m_{b-1},j_{b},l_{b})$, generate  $2^{N(R_{1,p}+R_{v_{1,p}})}$ sequences $x_{1,(b)}^N(j_{b-1},l_{b-1},k_{b-1},$ $m_{b-1},j_{b},l_{b},r_{b},\alpha_{b})$, $r_{b} \in [1 : 2^{NR_{1,p}}]$, $\alpha_{b}$ $ \in [1 : 2^{NR_{v_{1,p}}}]$ i.i.d. according to $P_{X_1|UU_1}$ $(\cdot|uu_1)$. 
	Similarly, generate $2^{N(R_{2,p}+R_{v_{2,p}})}$ sequences $x_{2,(b)}^N(j_{b-1},l_{b-1},k_{b-1}$, $m_{b-1},k_{b},m_{b},s_{b},\beta_{b})$
	i.i.d. according to $P_{X_2|UU_2}(\cdot|uu_2)$.
	\item For each tuple $(j_{b-1},l_{b-1},k_{b-1},m_{b-1},j_{b},l_{b},k_{b},m_{b})$, generate $2^{NR_{v_{1,c}}}$ sequences $v^N_{1,c,(b)}(j_{b-1},l_{b-1},k_{b-1},$ $m_{b-1},j_{b},l_{b},k_{b},m_{b},l_{b+1})$, $l_{b+1}\in[1:2^{NR_{v_{1,c}}}]$ i.i.d. according to $P_{V_{1,c}|UU_1U_2}(\cdot|uu_1u_2)$. 
	Similarly, generate $2^{NR_{v_{2,c}}}$ sequences $v^N_{2,c,(b)}(j_{b-1},l_{b-1},k_{b-1},$ $m_{b-1},j_{b},l_{b},k_{b},m_{b},m_{b+1})$, $m_{b+1} \in [1:2^{NR_{v_{2,c}}}]$ i.i.d. according to $P_{V_{2,c}|UU_1U_2}(\cdot|uu_1u_2)$. 
	\item For each tuple $(j_{b-1},l_{b-1},k_{b-1},m_{b-1},j_{b},l_{b},k_{b},m_{b},r_{b},$ $\alpha_{b},l_{b+1})$, generate $2^{NR_{v_{1,p}}}$ sequences $v^N_{1,p,(b)}(j_{b-1},$ $l_{b-1},k_{b-1},m_{b-1},j_{b},l_{b},k_{b},m_{b},r_{b},\alpha_{b},l_{b+1},\alpha_{b+1})$, $\alpha_{b+1}\in $ $[1 : 2^{NR_{v_{1,p}}}]$ i.i.d. according to $P_{V_{1,p}|UU_1U_2X_1V_{1,c}}(\cdot|uu_1u_2x_1v_{1,c})$,
	Similarly, generate $2^{NR_{v_{2,p}}}$ sequences $v^N_{2,p,(b)}($ $j_{b-1},l_{b-1},k_{b-1},m_{b-1},j_{b},l_{b},$ $k_{b},m_{b},s_{b},\beta_{b},m_{b+1},\beta_{b+1})$, $\beta_{b+1} \in [1 : 2^{NR_{v_{2,p}}}]$ i.i.d. according to $P_{V_{2,p}|UU_1U_2X_2V_{2,c}}(\cdot|uu_1u_2x_2v_{2,c})$. 
\end{itemize}

\subsubsection{Encoding}
\begin{figure}[!t]
	\centering
	\includegraphics[width=1\linewidth]{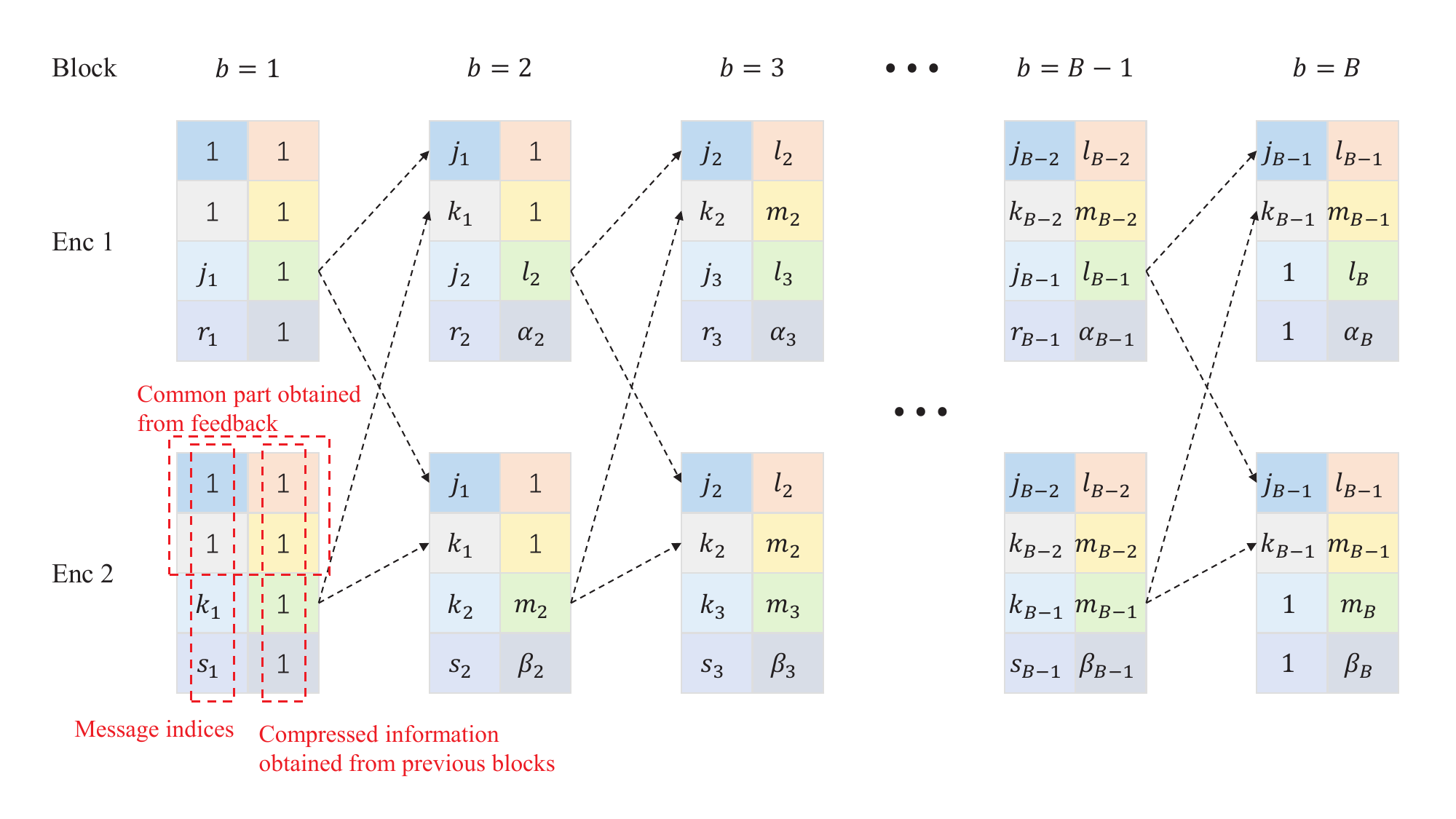}
	\caption{Encoding operations in the first $B$ blocks.}
	\label{fig:encoding}
\end{figure}

We set $j_0=k_0=1$, $l_0=m_0=1$, $l_1=m_1=1$, $\alpha_{1}=\beta_{1}=1$, $j_B=k_B=r_B=s_B=1$. The detailed encoding is as follows, and a brief illustration is provided in Fig.~\ref{fig:encoding}.
\begin{itemize}
	\item {\bf{Block}} $b = 1$. Two transmitters send $x_{1,(1)}^N(1,1,1,1,j_1,1,$ $r_1,1)$ and $x_{2,(1)}^N(1,1,1,1,k_1,1,s_1,1)$, respectively.
	\item {\bf{Block}} $b\in [2:B+1]$. At the beginning of block~$b$, where $b\in[2:B+1]$, transmitter~1 obtains generalized feedback $z^N_{1,(b-1)}$ and indices tuple $(j^*_{b-2},l^*_{b-2},k^*_{b-2},m^*_{b-2},j^*_{b-1},l^*_{b-1},r^*_{b-1},\alpha^*_{b-1})$. If $b\in[3:B]$, it also knows indices tuple $(j^*_{b-3}, l^*_{b-3}, k^*_{b-3}, m^*_{b-3},r^*_{b-2},\alpha^*_{b-2})$ and generalized feedback $z^N_{1,(b-2)}$. Thus, at the beginning of block~$b$, transmitter~1 finds a unique pair $(\hat{k}_{b-1},\hat{m}_{b-1})\in [1 : 2^{NR_{2,c}}] \times [1 : 2^{NR_{v_{2,c}}}]$ satisfying
	\begin{align}\label{equ:Tx1DecodingBlockbMessage}
		\bigg( 
		&u^N_{(b-1)}(j^*_{b-2}, l^*_{b-2}, k^*_{b-2}, m^*_{b-2}), \notag\\
		&u^N_{1,(b-1)}(j^*_{b-2},l^*_{b-2},k^*_{b-2},m^*_{b-2},j^*_{b-1},l^*_{b-1}),\notag\\
		&u^N_{2,(b-1)}(j^*_{b-2},l^*_{b-2},k^*_{b-2},m^*_{b-2},\hat{k}_{b-1},\hat{m}_{b-1}),\notag\\
		& x^N_{1,(b-1)}(j^*_{b-2},l^*_{b-2},k^*_{b-2},m^*_{b-2},j^*_{b-1},l^*_{b-1},r^*_{b-1},\alpha^*_{b-1}),\notag\\
		&z^N_{1,(b-1)}
		\bigg)\in \mathcal{T}_{\epsilon}^N(UU_1U_2X_1Z_1),
	\end{align}
	and 
	\begin{align}\label{equ:Tx1DecodingBlockbCompress}
		\bigg(
		&u^N_{(b-2)}(j^*_{b-3}, l^*_{b-3}, k^*_{b-3}, m^*_{b-3}), \notag\\
		&u^N_{1,(b-2)}(j^*_{b-3},l^*_{b-3},k^*_{b-3},m^*_{b-3},j^*_{b-2},l^*_{b-2}),\notag\\
		&u^N_{2,(b-2)}(j^*_{b-3},l^*_{b-3},k^*_{b-3},m^*_{b-3},k^*_{b-2},m^*_{b-2}),\notag\\
		&x^N_{1,(b-2)}(j^*_{b-3},l^*_{b-3},k^*_{b-3},m^*_{b-3},j^*_{b-2},l^*_{b-2},r^*_{b-2},\alpha^*_{b-2}),\notag\\
		&v^N_{2,c,(b-2)}(j^*_{b-3},l^*_{b-3},k^*_{b-3},m^*_{b-3},j^*_{b-2},l^*_{b-2},k^*_{b-2},m^*_{b-2},\notag\\
		&\hat{m}_{b-1}),z^N_{1,(b-2)}
		\bigg)\in \mathcal{T}_{\epsilon}^N(UU_1U_2X_1V_{2,c}Z_1)
	\end{align}
	simultaneously. For $b=2$, only (\ref{equ:Tx1DecodingBlockbMessage}) needs to be satisfied. If there is exactly one pair $(\hat{k}_{b-1},\hat{m}_{b-1})$ satisfying the above conditions, transmitter~1 sets $k^*_{b-1} = \hat{k}_{b-1}$, $m^*_{b-1} = \hat{m}_{b-1}$. Otherwise, an error is declared. Once obtaining the correct indices pair $(k^*_{b-1}$, $m^*_{b-1})$, transmitter~1 finds a pair $(\hat{l}_{b},\hat{\alpha}_{b})\in[1 : 2^{NR_{v_{1,c}}}] \times [1 : 2^{NR_{v_{1,p}}}]$ satisfying 
	\begin{align}\label{equ:Tx1QuantizationBlockb}
		\bigg(
		&u^N_{(b-1)}(j^*_{b-2}, l^*_{b-2}, k^*_{b-2}, m^*_{b-2}), \notag\\
		&u^N_{1,(b-1)}(j^*_{b-2},l^*_{b-2},k^*_{b-2},m^*_{b-2},j^*_{b-1},l^*_{b-1}),\notag\\
		&u^N_{2,(b-1)}(j^*_{b-2},l^*_{b-2},k^*_{b-2},m^*_{b-2},k^*_{b-1},m^*_{b-1}),\notag\\
		&x^N_{1,(b-1)}(j^*_{b-2},l^*_{b-2},k^*_{b-2},m^*_{b-2},j^*_{b-1},l^*_{b-1},r^*_{b-1},\alpha^*_{b-1}),\notag\\
		&v^N_{1,c,(b-1)}(j^*_{b-2},l^*_{b-2},k^*_{b-2},m^*_{b-2},j^*_{b-1},l^*_{b-1},k^*_{b-1},\notag\\
		&m^*_{b-1},\hat{l}_{b}),v^N_{1,p,(b-1)}(j^*_{b-2},l^*_{b-2},k^*_{b-2},m^*_{b-2},j^*_{b-1},l^*_{b-1},\notag\\
		&k^*_{b-1},m^*_{b-1},r^*_{b-1},\alpha^*_{b-1},\hat{l}_{b},\hat{\alpha}_{b}), z^N_{1,(b-1)}
		\bigg)\notag\\
		&\in \mathcal{T}_{\epsilon}^N(UU_1U_2X_1V_{1,c}V_{1,p}Z_1).
	\end{align}
	and sets $l^*_{b}=\hat{l}_{b},\alpha^*_{b}=\hat{\alpha}_{b}$. If there is no such pair, transmitter~1 sets $l^*_{b}=1,\alpha^*_{b}=1$. Then, transmitter~1 sends $x^N_{1,(b)}(j^*_{b-1},l^*_{b-1},k^*_{b-1},m^*_{b-1},j_{b},l^*_{b},r_{b},\alpha^*_{b})$ in block $b\in[2:B]$. The operations at transmitter~2 are analogous and omitted here.
\end{itemize}

\subsubsection{Decoding}
\begin{figure}[!t]
	\centering
	\includegraphics[width=1\linewidth]{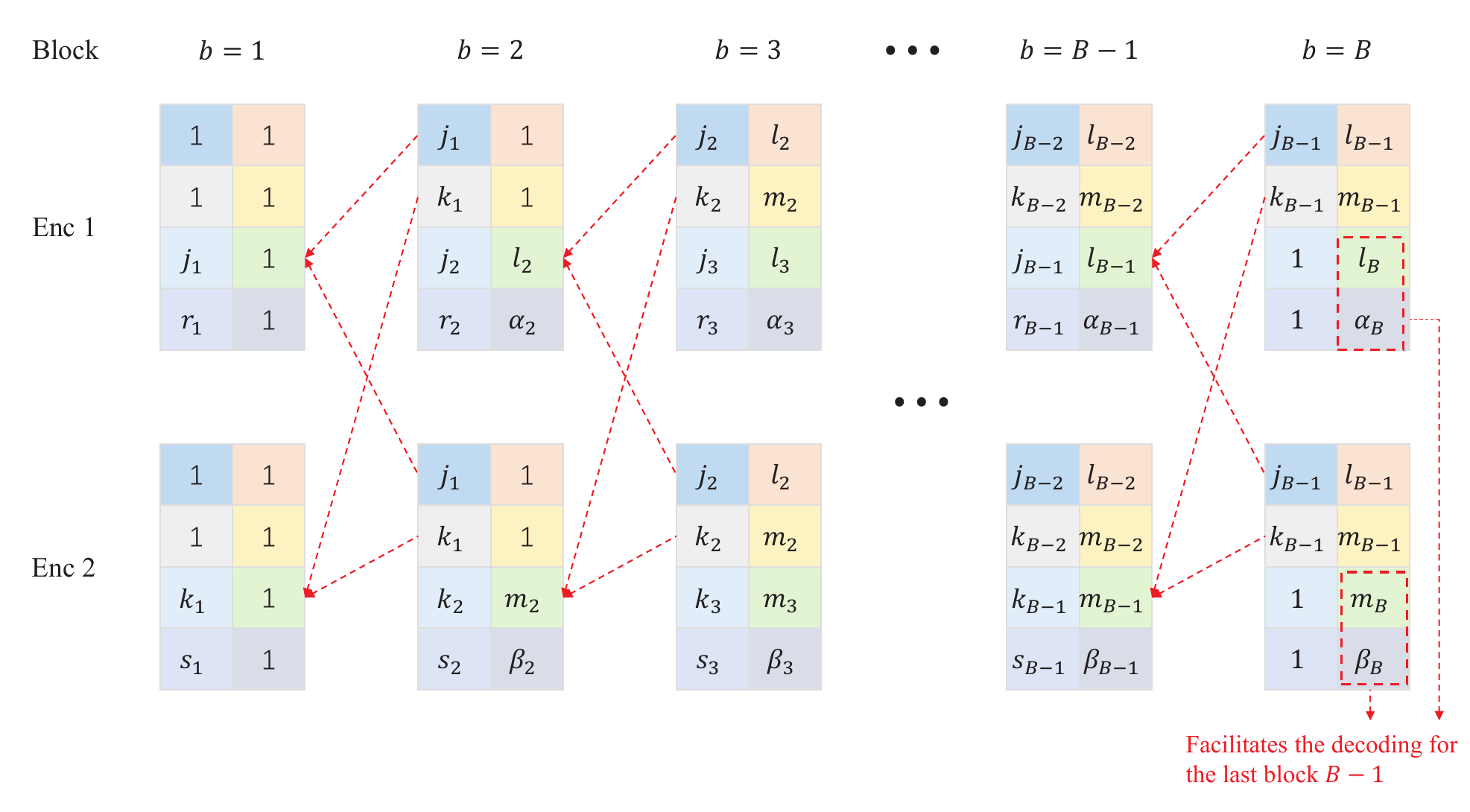}
	\vspace{-1em}
	\caption{Decoding operations in the first $B$ blocks.}
	\label{fig:decoding}
\end{figure}
Decoding begins at the block~$B$ and proceeds backward. A brief illustration of decoding operations is provided in Fig. \ref{fig:decoding}, and the details are as follows. In block~$b\in[B:1]$, the receiver has observed channel output $y^N_{(b)}$. It also has the correct indices tuple $(j^*_b,l^*_b,k^*_b,m^*_b)$ and $(l^*_{b+1},m^*_{b+1},\alpha^*_{b+1},\beta^*_{b+1})$ by the decoding procedures in block~$b+1$\footnote{For block~$b=B$, this is guaranteed by termination blocks.}. The receiver finds a unique tuple $(\hat{j}_{b-1},\hat{l}_{b-1},\hat{k}_{b-1},\hat{m}_{b-1},\hat{r}_b,\hat{\alpha}_{b},\hat{s}_b,\hat{\beta}_{b}) \in [1 : 2^{NR_{1,c}}] \times [1 : 2^{NR_{v_{1,c}}}] \times [1 : 2^{NR_{2,c}}] \times [1 : 2^{NR_{v_{2,c}}}] \times [1 : 2^{NR_{1,p}}] \times [1 : 2^{NR_{v_{1,p}}}] \times [1 : 2^{NR_{2,p}}] \times [1 : 2^{NR_{v_{2,p}}}]$ such that 
\begin{align}\label{equ:RxDecodingBlockb}
	\bigg(
	&u^N_{(b)}(\hat{j}_{b-1},\hat{l}_{b-1},\hat{k}_{b-1},\hat{m}_{b-1}),\notag\\
	&u^N_{1,(b)}(\hat{j}_{b-1},\hat{l}_{b-1},\hat{k}_{b-1},\hat{m}_{b-1},j^*_{b},l^*_{b}),\notag\\
	&u^N_{2,(b)}(\hat{j}_{b-1},\hat{l}_{b-1},\hat{k}_{b-1},\hat{m}_{b-1},k^*_b,m^*_{b}),\notag\\
	&x^N_{1,(b)}(\hat{j}_{b-1},\hat{l}_{b-1},\hat{k}_{b-1},\hat{m}_{b-1},j^*_{b},l^*_{b},\hat{r}_{b},\hat{\alpha}_{b}),\notag\\
	&x^N_{2,(b)}(\hat{j}_{b-1},\hat{l}_{b-1},\hat{k}_{b-1},\hat{m}_{b-1},k^*_b,m^*_{b},\hat{s}_{b},\hat{\beta}_{b}),\notag\\
	& v^N_{1,c,(b)}(\hat{j}_{b-1},\hat{l}_{b-1},\hat{k}_{b-1},\hat{m}_{b-1},j^*_{b},l^*_{b},k^*_{b},m^*_{b},l^*_{b+1}), \notag\\
	&v^N_{2,c,(b)}(\hat{j}_{b-1},\hat{l}_{b-1},\hat{k}_{b-1},\hat{m}_{b-1},j^*_{b},l^*_{b},k^*_{b},m^*_{b},m^*_{b+1})\notag\\
	&v^N_{1,p,(b)}(\hat{j}_{b-1},\hat{l}_{b-1},\hat{k}_{b-1},\hat{m}_{b-1},j^*_{b},l^*_{b},k^*_{b},m^*_{b},\hat{r}_{b},\hat{\alpha}_{b}, l^*_{b+1},\notag\\
	&\alpha^*_{b+1}),v^N_{2,p,(b)}(\hat{j}_{b-1},\hat{l}_{b-1},\hat{k}_{b-1},\hat{m}_{b-1},j^*_{b},l^*_{b},k^*_{b},m^*_{b},\hat{s}_{b},\hat{\beta}_{b},\notag\\
	&m^*_{b+1},\beta^*_{b+1}),y^N_{(b)}
	\bigg)\notag\\
	&\in\mathcal{T}_{\epsilon}^N(UU_1U_2X_1X_2V_{1,c}V_{2,c}V_{1,p}V_{2,p}Y).
\end{align}
If there is exactly one tuple $(\hat{j}_{b-1},\hat{l}_{b-1},\hat{k}_{b-1},\hat{m}_{b-1},$ $\hat{r}_b,\hat{\alpha}_{b},\hat{s}_b,\hat{\beta}_{b})$ satisfying (\ref{equ:RxDecodingBlockb}), the receiver sets $j^*_{b-1} = \hat{j}_{b-1}, l^*_{b-1} = \hat{l}_{b-1}, k^*_{b-1} = \hat{k}_{b-1}, m^*_{b-1} = \hat{m}_{b-1}, r^*_b = \hat{r}_b, \alpha^*_{b} = \hat{\alpha}_{b}, s^*_b = \hat{s}_b, \beta^*_{b} = \hat{\beta}_{b}$. Otherwise, an error is declared. The message output of receiver is $(j^*_{b},r^*_{b},k^*_{b},s^*_{b}),b\in[1:B-1]$.

\subsubsection{State Estimation}
For block~$b\in[1:B-1]$, transmitter~1 has observed the generalized feedback $z^N_{1,(b)}$. By decoding operations corresponding to \eqref{equ:Tx1DecodingBlockbMessage} and \eqref{equ:Tx1DecodingBlockbCompress} at the end of block~$b$, it obtains correct indices pair $(k^*_b,m^*_b)$ and thus knows the codeword $u^N_{2,(b)}(j^*_{b-1},l^*_{b-1},k^*_{b-1},m^*_{b-1},k^*_b,m^*_{b})$. Transmitter~1 can also obtain correct index $m^*_{b+1}$ by the decoding procedures \eqref{equ:Tx1DecodingBlockbMessage} and \eqref{equ:Tx1DecodingBlockbCompress} performed at the end of block~$b+1$ and thus knows the codeword $v^N_{2,c,(b)}(j^*_{b-1},l^*_{b-1},k^*_{b-1},m^*_{b-1},j^*_{b},l^*_{b},k^*_b,$ $m^*_{b},m^*_{b+1})$. Combined with its own transmitted codeword $x^N_{1,(b)}(j^*_{b-1},l^*_{b-1},k^*_{b-1},m^*_{b-1},j^*_{b},l^*_{b},r^*_{b},\alpha^*_{b})$, it produces the estimated parameter sequence 
\begin{align}
	\hat{s}_{T_1,(b)}^N = \hat{s}_{T_1}\bigg(&x^N_{1,(b)}(j^*_{b-1},l^*_{b-1},k^*_{b-1},m^*_{b-1},j^*_{b},l^*_{b},r^*_{b},\alpha^*_{b}),\notag\\
	& u^N_{2,(b)}(j^*_{b-1},l^*_{b-1},k^*_{b-1},m^*_{b-1},k^*_b,m^*_{b}), \notag\\
	&z^N_{1,(b)}, v^N_{2,c,(b)}(j^*_{b-1},l^*_{b-1},k^*_{b-1},m^*_{b-1},j^*_{b},l^*_{b},k^*_b,\notag\\
	&m^*_{b},m^*_{b+1})\bigg) 
\end{align}
by applying the symbol-by-symbol estimator defined in \eqref{equ:estimator} for block~$b \in [1 : B - 1]$. The operations at transmitter~2 are analogous and omitted here.

\subsection{Error Analysis}
We focus on the error analysis of the first $B$ blocks. Denote by $\mathcal{E}_{\text{Tx},k}$ and $\mathcal{E}_{\text{Rx}}$ the ``encoding error'' event of transmitter~$k \in \{1,2\}$ and ``decoding error'' event of the receiver, respectively. By the union bound, we have 
\begin{align}\label{equ:ErrorProWhole}
	P_{\mathcal{E}} \le P(\mathcal{E}_{\text{Tx},1}) + P(\mathcal{E}_{\text{Tx},2}) + P(\mathcal{E}_{\text{Rx}}).
\end{align} 
We proceed to derive an upper bound for each term in the right hand side of (\ref{equ:ErrorProWhole}).

\textbf{Error Analysis at the transmitter~1}: The error probability of transmitter~1 can be bounded as
\begin{align}\label{equ:ErrorProTx1Analysis}
	P(\mathcal{E}_{\text{Tx},1}) \le P(\mathcal{E}_{\text{Tx},1,(2)}) + \sum_{b=3}^B P(\mathcal{E}_{\text{Tx},1,(b)})  + P(\mathcal{E}_{\text{Tx},1,(B+1)}),
\end{align}
where $\mathcal{E}_{\text{Tx},1,(b)}$ represents the error event of transmitter~$1$ in block $b$.

We first bound $P(\mathcal{E}_{\text{Tx},1,(b)})$ for $b\in[3:B]$. Denote by $\mathcal{A}_{1,(b)}$ the error events corresponding to (\ref{equ:Tx1DecodingBlockbMessage}) and (\ref{equ:Tx1DecodingBlockbCompress}), and by 
$\mathcal{B}_{1,(b)}$ the error event corresponding to (\ref{equ:Tx1QuantizationBlockb}). By the union bound, we have
\begin{align}
	P(\mathcal{E}_{\text{Tx},1,(b)}) \le P(\mathcal{A}_{1,(b)}) + P(\mathcal{B}_{1,(b)}).
\end{align}
For term $P(\mathcal{A}_{1,(b)})$, define 
\begin{align}
	\mathcal{A}&_{1,(b)}(\hat{k}_{b-1},\hat{m}_{b-1}) = \bigg\{
	\bigg( 
	u^N_{(b-1)}(j^*_{b-2}, l^*_{b-2}, k^*_{b-2}, m^*_{b-2}),\notag\\ &u^N_{1,(b-1)}(j^*_{b-2},l^*_{b-2},k^*_{b-2},m^*_{b-2},j^*_{b-1},l^*_{b-1}),\notag\\
	&u^N_{2,(b-1)}(j^*_{b-2},l^*_{b-2},k^*_{b-2},m^*_{b-2},\hat{k}_{b-1},\hat{m}_{b-1}),\notag\\ 
	&x^N_{1,(b-1)}(j^*_{b-2},l^*_{b-2},k^*_{b-2},m^*_{b-2},j^*_{b-1},l^*_{b-1},r^*_{b-1},\alpha^*_{b-1}),\notag\\
	&z^N_{1,(b-1)}
	\bigg)\in \mathcal{T}_{\epsilon}^N(UU_1U_2X_1Z_1),\notag\\
	&\text{and} \notag\\
	\bigg(
	&u^N_{(b-2)}(j^*_{b-3}, l^*_{b-3}, k^*_{b-3}, m^*_{b-3}),\notag\\ 
	&u^N_{1,(b-2)}(j^*_{b-3},l^*_{b-3},k^*_{b-3},m^*_{b-3},j^*_{b-2},l^*_{b-2}),\notag\\
	&u^N_{2,(b-2)}(j^*_{b-3},l^*_{b-3},k^*_{b-3},m^*_{b-3},k^*_{b-2},m^*_{b-2}),\notag\\ 
	&x^N_{1,(b-2)}(j^*_{b-3},l^*_{b-3},k^*_{b-3},m^*_{b-3},j^*_{b-2},l^*_{b-2},r^*_{b-2},\alpha^*_{b-2}),\notag\\
	&v^N_{2,c,(b-2)}(j^*_{b-3},l^*_{b-3},k^*_{b-3},m^*_{b-3},j^*_{b-2},l^*_{b-2},k^*_{b-2},m^*_{b-2},\notag\\ 
	&\hat{m}_{b-1}),z^N_{1,(b-2)}
	\bigg)\in \mathcal{T}_{\epsilon}^N(UU_1U_2X_1V_{2,c}Z_1)
	\bigg\}
\end{align}
and we have
\begin{align}
	\mathcal{A}_{1,(b)} =& \mathcal{A}^c_{1,(b)}(k^*_{b-1},m^*_{b-1})\notag\\ 
	&\bigcup
	\big(\cup_{\substack{(\hat{k}_{b-1},\hat{m}_{b-1})\\ \neq (k^*_{b-1},m^*_{b-1})}}\mathcal{A}_{1,(b)}(\hat{k}_{b-1},\hat{m}_{b-1})\big).
\end{align}
By the union bound, we have
\begin{align}\label{equ:Tx1ErrorEventBlockb}
	P(\mathcal{A}_{1,(b)}) \le &P\{\mathcal{A}^c_{1,(b)}(k^*_{b-1},m^*_{b-1})\}\notag\\ 
	& + \sum_{\substack{(\hat{k}_{b-1},\hat{m}_{b-1})\\ \neq (k^*_{b-1},m^*_{b-1})}}P\{\mathcal{A}_{1,(b)}(\hat{k}_{b-1},\hat{m}_{b-1})\}. 
\end{align}
By the law of large numbers, as $N\rightarrow\infty$, there is $P\{\mathcal{A}^c_{1,(b)}(k^*_{b-1},m^*_{b-1})\} \rightarrow 0$. Furthermore, the second term in the right hand side of (\ref{equ:Tx1ErrorEventBlockb}) can be expressed as
\begin{align}\label{equ:TX1ErrorEventBlockbDecompose}
	&\sum_{\substack{(\hat{k}_{b-1},\hat{m}_{b-1})\\ \neq (k^*_{b-1},m^*_{b-1})}}P\{\mathcal{A}_{1,(b)}(\hat{k}_{b-1},\hat{m}_{b-1})\} \notag\\ 
	&\qquad= \sum_{\hat{k}_{b-1}\neq k^*_{b-1}}P\{\mathcal{A}_{1,(b)}(\hat{k}_{b-1},m^*_{b-1})\}\notag\\
	&\qquad +\sum_{\substack{\hat{m}_{b-1} \neq m^*_{b-1}  }}P\{\mathcal{A}_{1,(b)}(k^*_{b-1},\hat{m}_{b-1})\}\notag\\ 
	&\qquad+ \sum_{\substack{\hat{k}_{b-1}\neq k^*_{b-1}\\ \hat{m}_{b-1} \neq m^*_{b-1}  }}P\{\mathcal{A}_{1,(b)}(\hat{k}_{b-1},\hat{m}_{b-1})\}.
\end{align}
According to the codebook generation and standard information-theoretic arguments \cite{el2011network}, the right hand side of (\ref{equ:TX1ErrorEventBlockbDecompose}) tends to zero as $N\rightarrow\infty$ if
\begin{subequations}\label{equ:RateConsForTX1A}
	\begin{align}
		R_{2,c} & < I(U_2;Z_{1}|UU_{1}X_{1}), \\
		R_{2,c} + R_{v_{2,c}}   &< I(U_2;Z_{1}|UU_{1}X_{1}) + I(V_{2,c};X_{1}Z_{1}|UU_1U_2), 
	\end{align}
\end{subequations}
For term $P(\mathcal{B}_{1,(b)})$, define the following events $\mathcal{B}_{1,(b),0}$, $ \mathcal{B}_{1,(b),c}$, and $\mathcal{B}_{1,(b),p}$:
\begin{align}
	\mathcal{B}&_{1,(b),0} =  \bigg\{\big( 
	u^N_{(b-1)}(j^*_{b-2}, l^*_{b-2}, k^*_{b-2}, m^*_{b-2}), \notag\\ 
	&
	u^N_{1,(b-1)}(j^*_{b-2},l^*_{b-2},k^*_{b-2},m^*_{b-2},j^*_{b-1},l^*_{b-1}),\notag\\
	&
	u^N_{2,(b-1)}(j^*_{b-2},l^*_{b-2},k^*_{b-2},m^*_{b-2},k^*_{b-1},m^*_{b-1}),\notag\\ 
	&
	x^N_{1,(b-1)}(j^*_{b-2},l^*_{b-2},k^*_{b-2},m^*_{b-2},j^*_{b-1},l^*_{b-1},r^*_{b-1},\alpha^*_{b-1}),\notag\\
	&z^N_{1,(b-1)} \big) \notin \mathcal{T}_{\epsilon}^N(UU_1U_2X_1Z_1)\bigg\}, \notag\\
	\mathcal{B}&_{1,(b),c} = \bigg\{\big(  
	u^N_{(b-1)}(j^*_{b-2}, l^*_{b-2}, k^*_{b-2}, m^*_{b-2}),\notag\\ 
	&
	u^N_{1,(b-1)}(j^*_{b-2},l^*_{b-2},k^*_{b-2},m^*_{b-2},j^*_{b-1},l^*_{b-1}),\notag\\
	&u^N_{2,(b-1)}(j^*_{b-2},l^*_{b-2},k^*_{b-2},m^*_{b-2},k^*_{b-1},m^*_{b-1}),\notag\\ 
	&
	x^N_{1,(b-1)}(j^*_{b-2},l^*_{b-2},k^*_{b-2},m^*_{b-2},j^*_{b-1},l^*_{b-1},r^*_{b-1},\alpha^*_{b-1}),\notag\\
	&
	v^N_{1,c,(b-1)}(j^*_{b-2},l^*_{b-2},k^*_{b-2},m^*_{b-2},j^*_{b-1},l^*_{b-1},k^*_{b-1},m^*_{b-1},\hat{l}_{b}), \notag\\ 
	&
	z^N_{1,(b-1)}\big) \notin \mathcal{T}_{\epsilon}^N(UU_1U_2X_1V_{1,c}Z_1), \text{for all}\ \hat{l}_{b} \in [1 : 2^{NR_{v_{1,c}}}] \bigg\}, \notag\\
	\mathcal{B}&_{1,(b),p} =  \bigg\{\big(  
	u^N_{(b-1)}(j^*_{b-2}, l^*_{b-2}, k^*_{b-2}, m^*_{b-2}),\notag\\ 
	&
	u^N_{1,(b-1)}(j^*_{b-2},l^*_{b-2},k^*_{b-2},m^*_{b-2},j^*_{b-1},l^*_{b-1}),\notag\\
	& u^N_{2,(b-1)}(j^*_{b-2},l^*_{b-2},k^*_{b-2},m^*_{b-2},k^*_{b-1},m^*_{b-1}),\notag\\ 
	&
	x^N_{1,(b-1)}(j^*_{b-2},l^*_{b-2},k^*_{b-2},m^*_{b-2},j^*_{b-1},l^*_{b-1},r^*_{b-1},\alpha^*_{b-1}),\notag\\
	&
	v^N_{1,c,(b-1)}(j^*_{b-2},l^*_{b-2},k^*_{b-2},m^*_{b-2},j^*_{b-1},l^*_{b-1},k^*_{b-1},m^*_{b-1},l^*_{b}),\notag\\
	& v^N_{1,p,(b-1)}(j^*_{b-2},l^*_{b-2},k^*_{b-2},m^*_{b-2},j^*_{b-1},l^*_{b-1},k^*_{b-1},m^*_{b-1},r^*_{b-1},\notag\\ 
	&\alpha^*_{b-1},l^*_{b},\hat{\alpha}_{b}),z^N_{1,(b-1)} \big) \notin \mathcal{T}_{\epsilon}^N(UU_1U_2X_1V_{1,c}V_{1,p}Z_1),\notag\\ 
	&\text{for all}\ \hat{\alpha}_{b} \in [1 : 2^{NR_{v_{1,p}}}] \bigg\}.
\end{align}
By the union bound, there is 
\begin{align}
	P(\mathcal{B}_{1,(b)}) \le P(\mathcal{B}_{1,(b),0}) &+ P(\mathcal{B}^c_{1,(b),0}\cap \mathcal{B}_{1,(b),c}) \notag\\ 
	&+ P(\mathcal{B}^c_{1,(b),0}\cap\mathcal{B}^c_{1,(b),c}\cap \mathcal{B}_{1,(b),p}).
\end{align}
According to the codebook generation and standard information-theoretic arguments \cite{el2011network}, we can obtain that $P(\mathcal{B}_{1,(b)})$ tends to zero as $N\rightarrow\infty$ if
\begin{subequations}\label{equ:RateConsForTX1B}
	\begin{align}
		R_{v_{1,c}} &> I(V_{1,c};X_1Z_1|UU_1U_2),\\
		R_{v_{1,p}} &> I(V_{1,p};Z_1|UU_1U_2X_1V_{1,c}).
	\end{align}
\end{subequations} 
Following the similar analysis, one can find that the error events in blocks~2 and $B+1$ are special cases those are contained in block~$b\in[3:B]$. Thus, when \eqref{equ:RateConsForTX1A} and \eqref{equ:RateConsForTX1B} hold, $P(\mathcal{E}_{\text{Tx},1,(2)})$ and $P(\mathcal{E}_{\text{Tx},1,(B+1)})$ tend to zero as $N\rightarrow\infty$.

\textbf{Error Analysis at the transmitter~2}: The analysis of error probability for transmitter~2 is analogous to that for transmitter~1. One can prove that $\lim_{N\rightarrow\infty}P(\mathcal{E}_{\text{Tx},2}) = 0$ , 
whenever 
\begin{subequations}\label{equ:RateConsForTX2}
	\begin{align}
		&R_{1,c}  < I(U_1;Z_{2}|UU_{2}X_{2}), \\
		&R_{1,c} + R_{v_{1,c}}   < I(U_1;Z_{2}|UU_{2}X_{2}) + I(V_{1,c};X_{2}Z_{2}|UU_1U_2), \\
		&R_{v_{2,c}} > I(V_{2,c};X_2Z_2|UU_1U_2),\\
		&R_{v_{2,p}} > I(V_{2,p};Z_2|UU_1U_2X_2V_{2,c}).
	\end{align}
\end{subequations}

\par~\textbf{Error Analysis at the receiver}: The error probability of receiver can be bounded as
\begin{align}\label{equ:ErrorProTx1Analysis}
	P(\mathcal{E}_{\text{Rx}}) \le P(\mathcal{E}_{\text{Rx},(B)}) &+  \sum_{b=3}^{B-1} P(\mathcal{E}_{\text{Rx},(b)}) \notag\\ 
	&+ P(\mathcal{E}_{\text{Rx},(2)}) + P(\mathcal{E}_{\text{Rx},(1)}),
\end{align}
where $\mathcal{E}_{\text{Rx},(b)}$ represents the error event of receiver in block~$b$.

We first bound $P(\mathcal{E}_{\text{Rx},(b)})$ for $b \in [3:B-1]$. Define the error event $\mathcal{H}_{(b)}(\hat{j}_{b-1},\hat{l}_{b-1},\hat{k}_{b-1},\hat{m}_{b-1},\hat{r}_b,$ $\hat{\alpha}_{b},\hat{s}_b,\hat{\beta}_{b})$ as
\begin{align}
	\mathcal{H}&_{(b)}(\hat{j}_{b-1},\hat{l}_{b-1},\hat{k}_{b-1},\hat{m}_{b-1},\hat{r}_b,\hat{\alpha}_{b},\hat{s}_b,\hat{\beta}_{b})  
	=\notag\\\bigg\{ \bigg(
	&u^N_{(b)}(\hat{j}_{b-1},\hat{l}_{b-1},\hat{k}_{b-1},\hat{m}_{b-1}),\notag\\
	&u^N_{1,(b)}(\hat{j}_{b-1},\hat{l}_{b-1},\hat{k}_{b-1},\hat{m}_{b-1},j^*_{b},l^*_{b}),\notag\\ 
	&u^N_{2,(b)}(\hat{j}_{b-1},\hat{l}_{b-1},\hat{k}_{b-1},\hat{m}_{b-1},k^*_b,m^*_{b}),\notag\\
	&x^N_{1,(b)}(\hat{j}_{b-1},\hat{l}_{b-1},\hat{k}_{b-1},\hat{m}_{b-1},j^*_{b},l^*_{b},\hat{r}_{b},\hat{\alpha}_{b}),\notag\\ 
	&x^N_{2,(b)}(\hat{j}_{b-1},\hat{l}_{b-1},\hat{k}_{b-1},\hat{m}_{b-1},k^*_b,m^*_{b},\hat{s}_{b},\hat{\beta}_{b}),\notag\\
	& v^N_{1,c,(b)}(\hat{j}_{b-1},\hat{l}_{b-1},\hat{k}_{b-1},\hat{m}_{b-1},j^*_{b},l^*_{b},k^*_{b},m^*_{b},l^*_{b+1}),\notag\\
	& v^N_{2,c,(b)}(\hat{j}_{b-1},\hat{l}_{b-1},\hat{k}_{b-1},\hat{m}_{b-1},j^*_{b},l^*_{b},k^*_{b},m^*_{b},m^*_{b+1})\notag\\
	&v^N_{1,p,(b)}(\hat{j}_{b-1},\hat{l}_{b-1},\hat{k}_{b-1},\hat{m}_{b-1},j^*_{b},l^*_{b},k^*_{b},m^*_{b},\hat{r}_{b},\hat{\alpha}_{b}, l^*_{b+1},\notag\\ 
	&\alpha^*_{b+1}),v^N_{2,p,(b)}(\hat{j}_{b-1},\hat{l}_{b-1},\hat{k}_{b-1},\hat{m}_{b-1},j^*_{b},l^*_{b},k^*_{b},m^*_{b},\hat{s}_{b},\notag\\ 
	&\hat{\beta}_{b},m^*_{b+1},\beta^*_{b+1}),y^N_{(b)}
	\bigg)\notag\\ 
	&\in\mathcal{T}_{\epsilon}^N(UU_1U_2X_1X_2V_{1,c}V_{2,c}V_{1,p}V_{2,p}Y)
	\bigg\}.
\end{align}
Then, we have
\begin{align}
	\mathcal{E}_{\text{Rx},(b)} =& \mathcal{H}^c_{(b)}(j^*_{b-1},l^*_{b-1},k^*_{b-1},m^*_{b-1},r^*_b,\alpha^*_{b},s^*_b,\beta^*_{b}) \notag\\ 
	&\bigcup \bigg(
	\cup_{\substack{(\hat{j}_{b-1},\hat{l}_{b-1},\hat{k}_{b-1},\hat{m}_{b-1},\hat{r}_b,\hat{\alpha}_{b},\hat{s}_b,\hat{\beta}_{b})\\\neq(j^*_{b-1},l^*_{b-1},k^*_{b-1},m^*_{b-1},r^*_b,\alpha^*_{b},s^*_b,\beta^*_{b}) } }\notag\\
	&
	\mathcal{H}_{(b)}(\hat{j}_{b-1},\hat{l}_{b-1},\hat{k}_{b-1},\hat{m}_{b-1},\hat{r}_b,\hat{\alpha}_{b},\hat{s}_b,\hat{\beta}_{b})
	\bigg).
\end{align}
By the union bound, we have 
\begin{align}\label{equ:errorProbRxAnaBlockb}
	&P(\mathcal{E}_{\text{Rx},(b)}) \le P\{\mathcal{H}^c_{(b)}(j^*_{b-1},l^*_{b-1},k^*_{b-1},m^*_{b-1},r^*_b,\alpha^*_{b},s^*_b,\beta^*_{b}) \} \notag\\
	&+ 	\sum_{\substack{(\hat{j}_{b-1},\hat{l}_{b-1},\hat{k}_{b-1},\hat{m}_{b-1},\hat{r}_b,\hat{\alpha}_{b},\hat{s}_b,\hat{\beta}_{b})\\\neq(j^*_{b-1},l^*_{b-1},k^*_{b-1},m^*_{b-1},r^*_b,\alpha^*_{b},s^*_b,\beta^*_{b}) } } \notag\\ 
	&\qquad\qquad P\{ \mathcal{H}_{(b)}(\hat{j}_{b-1},\hat{l}_{b-1},\hat{k}_{b-1},\hat{m}_{b-1},\hat{r}_b,\hat{\alpha}_{b},\hat{s}_b,\hat{\beta}_{b})\}.
\end{align}
By the law of large numbers, we have $P\{\mathcal{H}^c_{(b)}(j^*_{b-1},l^*_{b-1},k^*_{b-1},m^*_{b-1},r^*_b,\alpha^*_{b},s^*_b,\beta^*_{b})\}\rightarrow 0$ as $N\rightarrow\infty$. Moreover, the the second term in the right hand side of (\ref{equ:errorProbRxAnaBlockb}) can be expressed as
\begin{align}\label{equ:RxDecodingErrorFinal}
	&\sum_{\substack{(\hat{j}_{b-1},\hat{l}_{b-1},\hat{k}_{b-1},\\\hat{m}_{b-1},\hat{r}_b,\hat{\alpha}_{b},\hat{s}_b,\hat{\beta}_{b})\\\neq(j^*_{b-1},l^*_{b-1},k^*_{b-1},\\m^*_{b-1},r^*_b,\alpha^*_{b},s^*_b,\beta^*_{b}) } } P\{ \mathcal{H}_{(b)}(\hat{j}_{b-1},\hat{l}_{b-1},\hat{k}_{b-1},\hat{m}_{b-1},\hat{r}_b,\hat{\alpha}_{b},\hat{s}_b,\hat{\beta}_{b}\} \notag\\
	&= \sum_{\substack{(\hat{j}_{b-1},\hat{l}_{b-1},\\\hat{k}_{b-1},\hat{m}_{b-1})\\\neq(j^*_{b-1},l^*_{b-1},\\k^*_{b-1},m^*_{b-1})\\ \forall\ \hat{r}_b,\hat{\alpha}_{b},\hat{s}_b,\hat{\beta}_{b} } } P\{ \mathcal{H}_{(b)}(\hat{j}_{b-1},\hat{l}_{b-1},\hat{k}_{b-1},\hat{m}_{b-1},\hat{r}_b,\hat{\alpha}_{b},\hat{s}_b,\hat{\beta}_{b}\} \notag\\
	& + \sum_{\substack{(\hat{r}_b,\hat{\alpha}_{b})\neq(r^*_b,\alpha^*_{b}) } } P\{ \mathcal{H}_{(b)}(j^*_{b-1},l^*_{b-1},k^*_{b-1},m^*_{b-1},\hat{r}_b,\hat{\alpha}_{b},s^*_b,\beta^*_{b}\} \notag\\
	& + \sum_{\substack{(\hat{s}_b,\hat{\beta}_{b})\neq(s^*_b,\beta^*_{b}) } } P\{ \mathcal{H}_{(b)}(j^*_{b-1},l^*_{b-1},k^*_{b-1},m^*_{b-1},r^*_b,\alpha^*_{b},\hat{s}_b,\hat{\beta}_{b}\} \notag\\
	& + \sum_{\substack{(\hat{r}_b,\hat{\alpha}_{b})\neq(r^*_b,\alpha^*_{b}) \\ (\hat{s}_b,\hat{\beta}_{b})\neq(s^*_b,\beta^*_{b}) } } P\{ \mathcal{H}_{(b)}(j^*_{b-1},l^*_{b-1},k^*_{b-1},m^*_{b-1},\hat{r}_b,\hat{\alpha}_{b},\hat{s}_b,\hat{\beta}_{b}\}.
\end{align}
According to the codebook generation and standard information-theoretic arguements \cite{el2011network}, the right hand side of \eqref{equ:RxDecodingErrorFinal} tends to zero as $N\rightarrow\infty$ if
\begin{subequations}\label{equ:RateConsForRx}
	\begin{align}
		R_{1,p} + R_{v_{1,p}} & < I(X_1;X_{2}YV_{1,c}V_{2,c}V_{{2},p}|UU_1U_2) \notag\\ 
		&+ I(V_{1,p};X_{2}YV_{{2},c}V_{{2},p}|UU_1U_2X_1V_{1,c}),\\
		R_{2,p} + R_{v_{2,p}}&  < I(X_2;X_{1}YV_{1,c}V_{2,c}V_{{1},p}|UU_1U_2)\notag\\ 
		& + I(V_{2,p};X_{1}YV_{{1},c}V_{{1},p}|UU_1U_2X_2V_{2,c}),\\
		R_{1,p} + R_{v_{1,p}}& + R_{2,p} + R_{v_{2,p}} < I(X_1X_2;YV_{1,c}V_{2,c}|UU_1U_2) \notag\\
		& + I(V_{1,p};X_2YV_{2,c}|UU_1U_2X_1V_{1,c}) \notag\\ 
		&+ I(V_{2,p};X_1YV_{1,c}V_{1,p}|UU_1U_2X_2V_{2,c}),\\
		R_{1,c} + R_{v_{1,c}}& + R_{2,c} + R_{v_{2,c}} 
		+ R_{1,p} + R_{v_{1,p}} + R_{2,p} + R_{v_{2,p}} \notag\\
		&<  I(X_1X_2;Y) + I(V_{1,c};X_1X_2Y|UU_1U_2) \notag\\ 
		&+ I(V_{2,c};X_1X_2YV_{1,c}|UU_1U_2) \notag\\
		&+ I(V_{1,p};X_2YV_{2,c}|UU_1U_2X_1V_{1,c}) \notag\\ 
		&+ I(V_{2,p};X_1YV_{1,c}V_{1,p}|UU_1U_2X_2V_{2,c}).
	\end{align}
\end{subequations}
Following the similar analysis, one can find that the error events in blocks~1,~2, and $B$ are special cases those are contained in block~$b\in[3:B-1]$. Thus, when \eqref{equ:RateConsForRx} holds, $P(\mathcal{E}_{\text{Rx},(1)})$, $P(\mathcal{E}_{\text{Rx},(2)})$, $P(\mathcal{E}_{\text{Rx},(B)})$ tend to zero as $N\rightarrow\infty$.

Combining all above error analysis, we conclude that the error probability of the proposed scheme $P_{\mathcal{E}}$ tends to zero as $N\rightarrow\infty$ if conditions 
\begin{subequations}\label{equ:OurRateBeforFME}
	\begin{align}
		&R_{v_{k,c}} > I(V_{k,c};X_kZ_k|UU_1U_2),\\
		&R_{v_{k,p}} > I(V_{k,p};Z_k|UU_1U_2X_kV_{k,c}),\\
		&R_{k,c} < I(U_k;Z_{\bar{k}}|UU_{\bar{k}}X_{\bar{k}}), \\
		&R_{k,c} + R_{v_{k,c}}   < I(U_k;Z_{\bar{k}}|UU_{\bar{k}}X_{\bar{k}}) \notag\\ 
		&\qquad\quad+ I(V_{k,c};X_{\bar{k}}Z_{\bar{k}}|UU_1U_2), \\
		&R_{k,p} + R_{v_{k,p}}  < I(X_k;X_{\bar{k}}YV_{1,c}V_{2,c}V_{{\bar{k}},p}|UU_1U_2) \notag\\ 
		&\qquad\quad+ I(V_{k,p};X_{\bar{k}}YV_{{\bar{k}},c}V_{{\bar{k}},p}|UU_1U_2X_kV_{k,c}),\\
		&R_{1,p} + R_{v_{1,p}} + R_{2,p} + R_{v_{2,p}} < I(X_1X_2;YV_{1,c}V_{2,c}|UU_1U_2) \notag\\
		&\qquad\quad + I(V_{1,p};X_2YV_{2,c}|UU_1U_2X_1V_{1,c}) \notag\\ 
		&\qquad\quad+ I(V_{2,p};X_1YV_{1,c}V_{1,p}|UU_1U_2X_2V_{2,c}),\\
		&R_{1,c} + R_{v_{1,c}} + R_{2,c} + R_{v_{2,c}} 
		+ R_{1,p} + R_{v_{1,p}} + R_{2,p} + R_{v_{2,p}} \notag\\
		&<  I(X_1X_2;Y) + I(V_{1,c};X_1X_2Y|UU_1U_2) \notag\\ 
		&\qquad\quad+ I(V_{2,c};X_1X_2YV_{1,c}|UU_1U_2) \notag\\
		&\qquad\quad+ I(V_{1,p};X_2YV_{2,c}|UU_1U_2X_1V_{1,c}) \notag\\
		&\qquad\quad+ I(V_{2,p};X_1YV_{1,c}V_{1,p}|UU_1U_2X_2V_{2,c}),
	\end{align}
\end{subequations}
hold for $k\in\{1,2\}$.

\subsection{Fourier-Motzkin Elimination}\label{appendix:fme}
Based on~\eqref{equ:OurRateBeforFME} with $R_1=R_{1,c}+R_{1,p}$, $R_2=R_{2,c}+R_{2,p}$, one can apply Fourier-Motzkin elimination 
to obtain the achievable rate region as stated in Theorem~\ref{innerBound:CD}.

\subsection{Analysis of Expected Distortion}
The number of channel uses is $n=NB+N_{\tilde{B}}$, where $N_{\tilde{B}}$ is the number of channel uses of termination blocks and $\frac{N_{\tilde{B}}}{N}$ is a finite number that will be discussed later. Without loss in performance as $N,B\rightarrow\infty$, we focus on the 
average distortion in the first $B-1$ blocks.

Define $w_k=\{w_k^{(b)}\}_{b=1}^{B-1}$ with $|\mathcal{W}_k|=2^{N(B-1)(R_{k,p}+R_{k,c})}$ where $w_k^{(b)}=\{w_{k,p}^{(b)},w_{k,c}^{(b)}\}$ and $k\in\{1,2\}$. For any given message pair $(w_1,w_2)$, define $n'\triangleq N(B-1)$, and the expected distortion of transmitter~1 in first $B-1$ blocks is 
\begin{align}
	\limsup_{n'\rightarrow\infty}&d_1^{(n')}(w_1,w_2)  \mathbb{E}\bigg[\frac{1}{n'}\sum_{i=1}^{n'}d_1(S_{T_1,i},\hat{S}_{T_1,i})\bigg] \notag\\
	&\overset{(a)}\le \limsup_{N,B\rightarrow\infty}\bigg( P_{\mathcal{E}}d_{\text{max}} \notag\\
	&\qquad+ (1-P_{\mathcal{E}})(1+\epsilon)\cdot\mathbb{E}\bigg[\frac{1}{n'}\sum_{i=1}^{n'}d_1(S_{T_1,i},\hat{S}_{T_1,i})\bigg]\bigg)\notag\\
	&\overset{(b)}\le\limsup_{N,B\rightarrow\infty}\bigg( P_{\mathcal{E}}d_{\text{max}} + (1-P_{\mathcal{E}})(1+\epsilon)D_1\bigg)\notag\\
	&\overset{(c)}=D_1,
\end{align}
where $(a)$ follows by applying the upper bound of the distortion function to the decoding error event and the typical average lemma \cite{el2011network} to the successful decoding event; $(b)$ follows from the random codebook generation and state estimating function that achieves $D_1$; $(c)$ follows because $P_{\mathcal{E}}$ tends to zeros as $N,B\rightarrow\infty$ if the rate constraints (\ref{equ:ourRes}) and (\ref{equ:ourResCon}) in Theorem~\ref{innerBound:CD} holds. Since the uniformly distributed messages are considered, it is easy to show that 
\begin{align}
	\limsup_{n'\rightarrow\infty}d_1^{(n')} \le D_1,
\end{align} 
and as $N,B\rightarrow\infty$, there is 
\begin{align}
	\limsup_{n\rightarrow\infty}d_1^{(n)} \le D_1.
\end{align}
The similar analysis can be conducted for transmitter~2 to verify that 
\begin{align}
	\limsup_{n\rightarrow\infty}d_2^{(n)} \le D_2.
\end{align}

\subsection{Discussion on Termination Blocks}
The termination blocks are used to guarantee that the receiver obtains correct indices $(l^*_{B},m^*_{B},$ $l^*_{B+1},m^*_{B+1},\alpha^*_{B+1},\beta^*_{B+1})$ to perform the backward decoding, i.e.,~\eqref{equ:RxDecodingBlockb}. Based on the~\eqref{equ:ourResC} in Theorem~\ref{innerBound:CD}, we have 
\begin{equation}\label{terminationSumConstraint}
	R_1 + R_2 \le I(X_1X_2;Y)
\end{equation}
for any achievable rate-distortion tuples $(R_1,R_2,D_1,D_2)$.

If $I(X_1X_2;Y) = 0$, there must be $R_1=R_2=0$. In this case, there is no need to consider the decoding operations and thus termination blocks. If $I(X_1X_2;Y) > 0$. Since 
\begin{align}
	I(X_1X_2;Y) = I(X_1;Y) + I(X_2;Y|X_1),
\end{align}
we know that $I(X_1;Y)$ and $I(X_2;Y|X_1)$ cannot be both zero. This reveals that at least one transmitter can achieve a positive communication rate to the receiver\footnote{If $I(X_1;Y)>0$, one can treat the MAC as a point-to-point channel from transmitter~1 to the receiver. If $I(X_2;Y|X_1)>0$, transmitter~2 can achieve a positive rate by letting transmitter~1 send a deterministic sequence.}. Without loss of generality, we consider that $I(X_1;Y)>0$ since that $I(X_2;Y|X_1)>0$ can be addressed in a similar manner. 
The inequality $I(X_1;Y)>0$ means that transmitter~1 can achieve a positive rate to the receiver, which guarantees that the compressed information $(l^*_{B},l^*_{B+1},\alpha^*_{B+1})$ can be sent to the receiver within $N_1$ channel uses, where $N_1$ is roughly $\frac{N(2R_{v_{1,c}}+ R_{v_{1,p}})}{I(X_1;Y)}$ and $\frac{N_1}{N} = \frac{2R_{v_{1,c}}+ R_{v_{1,p}}}{I(X_1;Y)}$ is finite. We thus focus on the transmission of compressed information of transmitter~2, i.e., $(m^*_{B},m^*_{B+1},\beta^*_{B+1})$.

\textbf{Case A}: {{$I(U_2;Z_1|UU_1X_1) > 0$ or $I(X_2;UU_1U_2X_1Z_1$ $Y)>0$}}. The inequality $I(U_2;Z_1|UU_1X_1)$ $>0$ means that transmitter~2 can achieve a positive rate to transmitter~1 through feedback~$Z_1$. In this case, compressed information $(m^*_{B},m^*_{B+1},\beta^*_{B+1})$  can be conveyed to transmitter~1 first, which then is sent to the receiver through the channel from transmitter~1 to the receiver. The number of channel uses for the transmission $N_2$ is roughly $\frac{N(2R_{v_{2,c}}+ R_{v_{2,p}})}{I(U_2;Z_1|UU_1X_1)}+\frac{N(2R_{v_{2,c}}+ R_{v_{2,p}})}{I(X_1;Y)}$, where $\frac{N_2}{N}$ is finite. The inequality $I(X_2;UU_1U_2X_1Z_1Y)>0$ means that transmitter~2 can achieve a positive rate to the receiver if the receiver has known $UU_1U_2X_1Z_1$. Thanks to $I(X_1;Y)>0$, transmitter~1 can send a lossless description of codewords $U,U_1,U_2,X_1$ and echo signal $Z_1$ to the receiver, then the compressed information $(m^*_{B},m^*_{B+1},\beta^*_{B+1})$ can be decoded by the receiver as $I(X_2;UU_1U_2X_1Z_1Y)>0$. In this case, the number of channel uses is $N_3$ roughly $\frac{N(2R_{v_{2,c}}+ R_{v_{2,p}})}{I(X_2;UU_1U_2X_1Z_1Y)}+\frac{N(2R_{v_{2,c}}+ R_{v_{2,p}})(H(UU_1U_2X_1Z_1) + \delta)}{I(X_2;UU_1U_2X_1Z_1Y)I(X_1;Y)}$, where $\frac{N_3}{N}$ is finite. In this case, the number of channel uses for termination blocks is $N_{\tilde{B}}\le N_1 + \max\{N_2,N_3\}$ and $\frac{N_{\tilde{B}}}{N}$ is a finite number.

\textbf{Case B}: $I(U_2;Z_1|UU_1X_1) = 0$ and $I(X_2;UU_1U_2X_1Z_1Y) $ $= 0$: In this case, we show that the achievability of Theorem~\ref{innerBound:CD} can be guaranteed by the proposed scheme without private compressed information, i.e., $V_{1,p}=v_{1,p}^*,V_{2,p}=v_{2,p}^*$ almost surely for some specific values $v_{1,p}^*\in\mathcal{V}_{1,p}$, $v_{2,p}^*\in\mathcal{V}_{2,p}$. We first show that $R_2$ in Theorem~\ref{innerBound:CD} must be zero in this case. Based on the (\ref{equ:ourResConsA}) for $k=2$, we have
\begin{align}
	0 = I(U_2;Z_{1}|UU_{1}X_{1}) \ge I(V_{2,c};X_2Z_2|UU_1U_2X_{1}Z_{1}).
\end{align}
Next, based on the (\ref{equ:ourResA}) for $k=2$, we have
\begin{align}
	R_2 &\le I(U_2;Z_1|UU_1X_1) - I(V_{2,c};X_2Z_2|UU_1U_2X_1Z_1) \notag\\
	&\quad+ I(X_2;YV_{1,c}V_{2,c}V_{1,p}|UU_1U_2X_{1}) \notag\\
	&\quad- I(V_{2,p};Z_2|UU_1U_2X_1X_2YV_{1,c}V_{2,c}V_{1,p})\notag\\
	&\overset{(a)}\le I(X_2;YV_{1,c}V_{2,c}V_{1,p}|UU_1U_2X_{1}), \notag\\
	& \le I(X_2;UU_1U_2X_{1}Z_1YV_{1,c}V_{2,c}V_{1,p}) \notag\\
	&\overset{(b)} =  I(X_2;UU_1U_2X_{1}Z_1YV_{2,c}) \notag\\
	& \overset{(c)}= I(X_2;V_{2,c}|UU_1U_2X_{1}Z_1Y)\notag\\
	& = H(V_{2,c}|UU_1U_2X_{1}Z_1Y) - H(V_{2,c}|UU_1U_2X_{1}Z_1YX_2)\notag\\
	&\overset{(d)}\le H(V_{2,c}|UU_1U_2X_{1}Z_1)- H(V_{2,c}|UU_1U_2X_{1}Z_1YX_2Z_2)\notag\\
	&\overset{(e)}=  H(V_{2,c}|UU_1U_2X_{1}Z_1)- H(V_{2,c}|UU_1U_2X_1Z_1X_2Z_2)\notag\\
	& = I(V_{2,c};X_2Z_2|UU_1U_2X_{1}Z_1)\notag\\
	& \overset{(f)}= 0,
\end{align} 
where $(a)$ follows from $I(U_2;Z_{1}|UU_{1}X_{1}) = I(V_{2,c};X_2Z_2|U$ $U_1U_2X_{1}Z_{1}) = 0$ and mutual information is nonnegative, $(b)$ follows from the Markov chain $X_2-UU_1U_2X_1Z_1YV_{2,c}-V_{1,c}V_{1,p}$, $(c)$ follows from that $I(X_2;UU_1U_2X_1Z_1Y) = 0$, $(d)$ follows from  that conditioning reduces entropy, $(e)$ follows from the Markov chain $X_{1}Z_1Y-UU_1U_2X_2Z_2-V_{2,c}$, and $(f)$ follows from $I(V_{2,c};X_2Z_2|UU_1U_2X_{1}Z_{1}) = 0$. Such a result means that when $I(U_2;Z_1|UU_1X_1) = 0$ and $I(X_2;UU_1U_2X_1Z_1YS_R) = 0$, the achievable distortion-rate region is reduced to the convex hull of tuple $(R_1,R_2=0,D_1,D_2)$. We now focus on rate $R_1$. Based on (\ref{equ:ourResA}), (\ref{equ:ourResC}), we have
\begin{align}\label{R1SINGLEBOUNDR2equal0case1}
	R_1&\le  I(X_1X_2;Y) - I(V_{1,c};Z_1|UU_1U_2X_1X_2Y)  \notag\\
	&\quad- I(V_{2,c};Z_2|UU_1U_2X_1X_2YV_{1,c}) \notag\\
	&\quad- I(V_{1,p};Z_1|UU_1U_2X_1X_2YV_{1,c}V_{2,c})\notag\\
	&\quad-  I(V_{2,p};Z_2|UU_1U_2X_1X_2YV_{1,c}V_{2,c}V_{1,p})\notag\\
	&\overset{(a)}\le I(X_1X_2;Y) - I(V_{1,c};Z_1|UU_1U_2X_1X_2Y)\notag\\
	&\quad- I(V_{2,c};Z_2|UU_1U_2X_1X_2YV_{1,c}) \notag\\
	&= I(X_1;Y) + I(X_2;Y|X_1)- I(V_{1,c};Z_1|UU_1U_2X_1X_2Y)\notag\\
	&\quad- I(V_{2,c};Z_2|UU_1U_2X_1X_2YV_{1,c}) \notag\\
	&\overset{(b)} \le I(X_1;Y)- I(V_{1,c};Z_1|UU_1U_2X_1X_2Y)\notag\\
	&\quad- I(V_{2,c};Z_2|UU_1U_2X_1X_2YV_{1,c}),
\end{align}
where $(a)$ follows from the nonnegativity of mutual information, and $(b)$ follows from that $I(X_2;UU_1U_2X_1Z_1Y) = 0$, and 
\begin{align}\label{R1SINGLEBOUNDR2equal0case2}
	R_1 & \le  I(U_1;Z_{2}|UU_{2}X_{2}) - I(V_{1,c};X_1Z_1|UU_1U_2X_{2}Z_{2}) \notag\\
	&\quad+ I(X_1X_{2};YV_{1,c}V_{2,c}|UU_1U_2)\notag\\
	&\quad- I(V_{1,p};Z_1|UU_1U_2X_1X_2YV_{1,c}V_{2,c}) \notag\\
	&\quad
	- I(V_{2,p};Z_2|UU_1U_2X_1X_2YV_{1,c}V_{2,c}V_{1,p}) \notag\\
	&\overset{(a)} \le I(U_1;Z_{2}|UU_{2}X_{2}) - I(V_{1,c};X_1Z_1|UU_1U_2X_{2}Z_{2}) \notag\\
	&\quad+ I(X_1X_{2};YV_{1,c}V_{2,c}|UU_1U_2)\notag\\
	& \overset{(b)} \le I(U_1;Z_{2}|UU_{2}X_{2}) - I(V_{1,c};X_1Z_1|UU_1U_2X_{2}Z_{2}) \notag\\
	&\quad+ I(X_1;YV_{1,c}V_{2,c}|UU_1U_2X_{2}),
\end{align}
where $(a)$ follows from the nonnegativity of mutual information, and $(b)$ follows from 
\begin{align}
	&I(X_{2};YV_{1,c}V_{2,c}|UU_1U_2)\notag\\
	&\le I(X_{2};UU_1U_2X_1Z_1YV_{1,c}V_{2,c})\notag\\
	&\overset{(a)} = I(X_{2};V_{1,c}V_{2,c}|UU_1U_2X_1Z_1Y)\notag\\
	&\overset{(b)} = I(X_{2};V_{2,c}|UU_1U_2X_1Z_1Y)\notag\\
	& \overset{(c)}\le H(V_{2,c}|UU_1U_2X_1Z_1)- H(V_{2,c}|UU_1U_2X_1Z_1YX_2Z_2)\notag\\
	& \overset{(d)}\le H(V_{2,c}|UU_1U_2X_1Z_1)- H(V_{2,c}|UU_1U_2X_1Z_1X_2Z_2)\notag\\
	& = I(V_{2,c};X_2Z_2|UU_1U_2X_1Z_1)\notag\\
	&\overset{(e)} = 0,
\end{align}
where $(a)$ follows from that $I(X_2;UU_1U_2X_1Z_1Y) = 0$, $(b)$ follows from the Markov chain $X_2-UU_1U_2X_1Z_1YV_{2,c} - V_{1,c}$, $(c)$ follows from that conditioning reduces entropy, $(d)$ follows from the Markov chain $Y-UU_1U_2X_1Z_1X_2Z_2-V_{2,c}$, and $(e)$ follows from that $I(X_2;UU_1U_2X_1Z_1Y) = 0$. 
Let $\mathcal{I}_{\text{R-D}}(R_1,R_2 = 0,D_1,D_2)$ denote the union of (\ref{R1SINGLEBOUNDR2equal0case1}), (\ref{R1SINGLEBOUNDR2equal0case2}), (\ref{equ:ourResConsA}), (\ref{equ:ourResConsD}), $R_2=0$, and the sensing distortion constraints. The results in Theorem~\ref{innerBound:CD} is achievable if one can prove that $\mathcal{I}_{\text{R-D}}(R_1,R_2 = 0,D_1,D_2)$ is achievable. In fact,  $\mathcal{I}_{\text{R-D}}(R_1,R_2 = 0,D_1,D_2)$ can be obtained by the similar achievability scheme as that in Appendix~\ref{achievabilityProposedScheme}. The key difference is that we need to take $R_{2,c}=R_{2,p}=0$, $V_{1,p}=v_{1,p}^*,V_{2,p}=v_{2,p}^*$ almost surely for some specific values $v_{1,p}^*\in\mathcal{V}_{1,p}$, $v_{2,p}^*\in\mathcal{V}_{2,p}$, i.e., transmitter~2 sends no message and two transmitters send no private compressed information. In such a scheme, the common compressed information of both two transmitters, i.e., $(l^*_{B},m^*_{B},$ $l^*_{B+1},m^*_{B+1})$ can be sent by transmitter~1 to the receiver as $I(X_1;Y)>0$, and the corresponding channel uses $N_{\tilde{B}}$ also satisfies the condition that $\frac{N_{\tilde{B}}}{N}$ is a finite number.

\section{Proof of Theorem~\ref{Theorem:CompInner}}\label{appendix:proofOfTheorem4}
By the definition of $\mathcal{I}_{\text{R-D}}^{\text{our},\text{com}}$, it is straightforward to have  
\begin{align}
	\mathcal{I}_{\text{R-D}}^{\text{our},\text{com}}\subseteq\mathcal{I}_{\text{R-D}}^{\text{our}}.
\end{align}
We proceed to prove that $\mathcal{I}_{\text{R-D}}^{\text{aw}}  \subseteq \mathcal{I}_{\text{R-D}}^{\text{our},\text{com}}$ by comparing the rate bounds before Fourier-Motzkin elimination. For national consistence, we use variables $U$ and $UU_1U_2$ to replace $U_0$ and $\underline{U}$, respectively, in the Ahmadipour--Wigger scheme throughout this paper. Let 
\begin{subequations}\label{equ:innerAssumepVariableCommon}
	\begin{align}
		&P_{V_{1,c}|UU_1U_2X_1Z_1}=P_{V_1|UU_2X_1Z_1}, \label{equ:innerAssumep_1}\\ 
		&P_{V_{2,c}|UU_1U_2X_2Z_2}=P_{V_2|UU_1X_2Z_2}, \label{equ:innerAssumep_2}
	\end{align} 
\end{subequations}
and 
\begin{align}
	&V_{1,p}=v_{1,p}^*,V_{2,p}=v_{2,p}^* \label{equ:innerAssumep_3}
\end{align}
almost surely for some specific values $v_{1,p}^*\in\mathcal{V}_{1,p}$, $v_{2,p}^*\in\mathcal{V}_{2,p}$.

Our rate bounds~\eqref{equ:OurRateBeforFME} become
\begin{subequations}
	\begin{align}
		&R_{v_{k,c}} > I(V_{k,c};X_kZ_k|UU_1U_2),\label{equ:ourRateBoundsBeforeFME_1}\\
		&R_{k,c}  < I(U_k;Z_{\bar{k}}|UU_{\bar{k}}X_{\bar{k}}), \label{equ:ourRateBoundsBeforeFME_2}\\
		&R_{k,c} + R_{v_{k,c}}   < I(U_k;Z_{\bar{k}}|UU_{\bar{k}}X_{\bar{k}}) + I(V_{k,c};X_{\bar{k}}Z_{\bar{k}}|UU_1U_2), \label{equ:ourRateBoundsBeforeFME_3}\\
		&R_{k,p}   < I(X_k;X_{\bar{k}}YV_{1,c}V_{2,c}|UU_1U_2),\label{equ:ourRateBoundsBeforeFME_4}\\
		&R_{1,p}  + R_{2,p}  < I(X_1X_2;YV_{1,c}V_{2,c}|UU_1U_2),\label{equ:ourRateBoundsBeforeFME_5}\\
		&R_{1,c} + R_{v_{1,c}} + R_{2,c} + R_{v_{2,c}} 
		+ R_{1,p} +  R_{2,p}  \notag\\
		&\qquad<  I(X_1X_2;Y) + I(V_{1,c};X_1X_2Y|UU_1U_2) \notag\\
		&\qquad\quad+ I(V_{2,c};X_1X_2YV_{1,c}|UU_1U_2),\label{equ:ourRateBoundsBeforeFME_6}
	\end{align}
\end{subequations}
for $k\in\{1,2\}$, where we have set $R_{v_{k,p}}=0$. The rate bounds before Fourier-Motzkin elimination for $\mathcal{I}_{\text{R-D}}^{\text{aw}}$ in~\cite[Theorem~3]{ahmadipour2023information} are 
\begin{subequations}\label{equ:SOTARateBoundsBeforeFME}
	\begin{align}
		&R_{v_k} > I(V_k;X_kZ_k|UU_1U_2), \label{equ:SOTARateBoundsBeforeFME_1}\\
		&R_{k,c}  + R_{v_{\bar{k}}} < I(U_kV_{\bar{k}};X_{\bar{k}}Z_{\bar{k}}|UU_{\bar{k}}), \label{equ:SOTARateBoundsBeforeFME_2}\\
		&R_{k,c}  + R_{v_1} + R_{v_2} < I(U_kV_{\bar{k}};X_{\bar{k}}Z_{\bar{k}}|UU_{\bar{k}}) \notag\\
		&\qquad + I(V_k;X_{\bar{k}}Z_{\bar{k}}|UU_1U_2), \label{equ:SOTARateBoundsBeforeFME_3}\\
		&R_{k,p}   < I(X_k;YV_{1}V_{2}|UU_1U_2X_{\bar{k}}),\\
		&R_{k,p}  + R_{v_k} < I(X_k;Y|UX_{\bar{k}}) + I(V_1;X_1X_2Y|UU_1U_2)\notag\\
		&\qquad+ I(V_2:X_1X_2YV_1|UU_1U_2), \\
		&R_{1,p}  + R_{2,p} + R_{v_k} < I(X_1X_2;Y|UU_{\bar{k}}) \notag\\
		&\qquad+ I(V_1;X_1X_2Y|UU_1U_2)\notag\\
		&\qquad+ I(V_2:X_1X_2YV_1|UU_1U_2), \\
		&R_{1,p}  + R_{2,p} < I(X_1X_2;YV_1V_2|UU_1U_2),\\
		&R_{1,p}  + R_{2,p} + R_{v_1} + R_{v_2} < I(X_1X_2;Y|U) \notag\\
		&\qquad+ I(V_1;X_1X_2Y|UU_1U_2)+ I(V_2:X_1X_2YV_1|UU_1U_2), \\
		&R_{1,c} + R_{v_{1}} + R_{2,c} + R_{v_{2}} 
		+ R_{1,p} +  R_{2,p} \notag\\
		&< I(X_1X_2;Y)+I(V_1;X_1X_2Y|UU_1U_2)\notag\\
		&\qquad+ I(V_2:X_1X_2YV_1|UU_1U_2),
	\end{align}
\end{subequations}
for $k\in\{1,2\}$.

Given~\eqref{equ:innerAssumepVariableCommon} and~\eqref{equ:innerAssumep_3}, we can find that constraints~\eqref{equ:ourRateBoundsBeforeFME_1},~\eqref{equ:ourRateBoundsBeforeFME_4},~\eqref{equ:ourRateBoundsBeforeFME_5},~\eqref{equ:ourRateBoundsBeforeFME_6} in our rate bounds are also contained in~\eqref{equ:SOTARateBoundsBeforeFME}. Moreover, constraints~\eqref{equ:ourRateBoundsBeforeFME_2} and~\eqref{equ:ourRateBoundsBeforeFME_3} must be satisfied for the rate tuple~\eqref{equ:SOTARateBoundsBeforeFME} due to constraints~\eqref{equ:SOTARateBoundsBeforeFME_1},~\eqref{equ:SOTARateBoundsBeforeFME_2}, and~\eqref{equ:SOTARateBoundsBeforeFME_3}. Therefore, our rate bounds always contain the existing bounds~\eqref{equ:SOTARateBoundsBeforeFME}.
Moreover, we note that both our proposed scheme and the Ahmadipour--Wigger scheme use the same estimation functions. Combining the above analysis, one can obtain that 
\begin{align}
	\mathcal{I}_{\text{R-D}}^{\text{aw}}  \subseteq
	\mathcal{I}_{\text{R-D}}^{\text{our},\text{com}}. 
\end{align}

\section{Converse Proof of Theorem~\ref{ourOuterBound}}\label{appendix:proofOuterBound}
The proofs of rate bounds~\eqref{equ:outerDependenceBound}, dependence balance constraint~\eqref{dependenceBalance}, and genie-aided state estimator~\eqref{estimator:genie-aided} are the same as those in~\cite{kobayashi2019joint} and are omitted. The proof of sensing constraints~\eqref{sensingConsA} is given as follows. 
For $k=1$, we have  
\begin{align}
	I(S_{T_1}^n;X_1^nZ_1^n) \ge I(S_{T_1}^n;A_1^{n}),
\end{align}
where $A_1^{n}$ denotes the estimated parameter sequence of transmitter~1.
The inequality holds due to the Markov chain $S_{T_1}^n - (X_1^n,Z_1^n) - A_1^{n}$.
For term $I(S_{T_1}^n;A_1^{n})$, we have 
\begin{align}\label{inequ:distour}
	I(S_{T_1}^n;A_1^{n}) &= H(S_{T_1}^n) - H(S_{T_1}^n|A_1^{n}) \notag\\
	& \overset{(a)}= \sum_{i=1}^n H(S_{T_1,i}) - H(S_{T_1,i}|A_1^{n},S_{T_1}^{i-1}) \notag\\
	& \overset{(b)} \ge\sum_{i=1}^n H(S_{T_1,i}) - H(S_{T_1,i}|A_{1,i})\notag\\
	& = \sum_{i=1}^n I(S_{T_1,i};A_{1,i}) \notag\\
	& \overset{(c)}\ge \sum_{i=1}^n f_{1,\text{R-D}}(\mathbb{E}[d_1(S_{T_1,i};A_{1,i})])\notag\\
	& = n \bigg(\frac{1}{n}\sum_{i=1}^n f_{1,\text{R-D}}(\mathbb{E}[d_1(S_{T_1,i};A_{1,i})]) \bigg)\notag\\
	& \overset{(d)}\ge nf_{1,\text{R-D}}\bigg(\frac{1}{n}\sum_{i=1}^n\mathbb{E}[d_1(S_{T_1,i};A_{1,i})] \bigg)\notag\\
	& \overset{(e)}= nf_{1,\text{R-D}}(\mathbb{E}[d_1(S_{T_1}^n;A_{1}^n)])\notag\\
	& \overset{(f)}= nf_{1,\text{R-D}}(D_1),
\end{align}
where $(a)$ follows from the fact that the sensing parameter sequence $S_{T_1}^n$ is i.i.d.; $(b)$ follows because conditioning reduces entropy, $(c)$ follows from the definition of rate-distortion function, $(d)$ follows from the convexity of rate-distortion function and Jensen's inequality, $(e)$ follows from the definition (\ref{definitionDistortion}) of distortion for blocks of length $n$, $(f)$ follows from the fact that rate-distortion function $f_{1,\text{R-D}}(\cdot)$ is a nonincreasing function of $D_1$ and sensing distortion constraint for $S_{T_1}$.

For term $I(S_{T_1}^n;X_1^nZ_1^n)$, we have
\begin{align}
	&I(S_{T_1}^n;X_1^nZ_1^n) \notag\\
	& \overset{(a)}\le I(S_{T_1}^n;X_1^nZ_1^n|W_1) \notag\\ 
	& = H(X_1^nZ_1^n|W_1) - H(X_1^nZ_1^n|W_1S_{T_1}^n) \notag\\
	& = \sum_{i=1}^n H(X_{1,i}Z_{1,i}|W_1X_1^{i-1}Z_{1}^{i-1})\notag\\
	&\qquad- H(X_{1,i}Z_{1,i}|W_1S_{T_1}^nX_1^{i-1}Z_{1}^{i-1})\notag\\
	& \overset{(b)}=\sum_{i=1}^n H(Z_{1,i}|W_1X_{1,i}X_1^{i-1}Z_{1}^{i-1})\notag\\
	&\qquad- H(Z_{1,i}|W_1S_{T_1}^nX_{1,i}X_1^{i-1}Z_{1}^{i-1})\notag\\
	& \overset{(c)} \le \sum_{i=1}^n H(Z_{1,i}|X_{1,i})- H(Z_{1,i}|W_1S_{T_1}^nX_{1,i}X_{2,i}X_1^{i-1}Z_{1}^{i-1})\notag\\
	& \overset{(d)} = \sum_{i=1}^n H(Z_{1,i}|X_{1,i})- H(Z_{1,i}|S_{T_1,i}X_{1,i}X_{2,i})\notag\\
	& = \sum_{i=1}^n I(X_{2,i}S_{T_1,i};Z_{1,i}|X_{1,i})\notag\\
	& \overset{(e)}= nI(X_{2,Q}S_{T_1,Q};Z_{1,Q}|X_{1,Q}Q)\notag\\
	& \overset{(f)}= nI(X_{2}S_{T_1};Z_{1}|X_{1}Q),
\end{align}
where $(a)$ follows from the facts that $W_1$ and $S_{T_1}^n$ are independent and conditioning reduces entropy, $(b)$ follows from the fact that $X_{1,i}$ is a function of $(W_1,Z_1^{i-1})$, $(c)$ follows from that conditioning reduces entropy, $(d)$ follows from the Markov chain $(W_1\{S_{T_1,l}\}_{l\neq i}X_1^{i-1}Z_{1}^{i-1}) - (S_{T_1,i}X_{1,i}X_{2,i}) - Z_{1,i}$, $(e)$ follows from we define a random variable $Q$ that is uniformly distributed over $[1:n]$, and $(f)$ follows by defining $X_1=X_{1,Q},X_2=X_{2,Q},S_{T_1}=S_{T_1,Q},Z_1=Z_{1,Q}$.
Combining the above results, we obtain that 
\begin{align}
	I(X_{2}S_{T_1};Z_{1}|X_{1}Q) \ge f_{1,\text{R-D}}(D_1).
\end{align}
For $k=2$, we can obtain the similar result.

\section{Proof of Example \ref{example:private}}\label{appendix:proofExamplePrivate}
Given $Z_1 = X_1\oplus S_2$ and $Z_2 = X_2\oplus S_1$, we have
\begin{subequations}
	\begin{align}
		I(U_1;X_{2}Z_{2}|UU_{2}) = I(U_1;X_{2}S_1|UU_{2}) = 0, 
	\end{align}
	\begin{align}
		I(U_2;X_{1}Z_{1}|UU_{1}) = I(U_2;X_{1}S_2|UU_{1}) = 0, 
	\end{align}
\end{subequations}
since $X_1U_1-U-U_2X_2$ forms a Markov chain and $S_1,S_2$ are independent of $UU_1U_2X_1X_2$. 

Then, consider the inequality constraints~(22d) for $k=2$ in $\mathcal{I}_{\text{R-D}}^{\text{aw}}$ presented in~\cite[Theorem~3]{ahmadipour2023information}, 
\begin{align}
	I(U_2;X_{1}Z_{1}|UU_{1}) +  I(V_2;&X_{1}Z_{1}|UU_1U_2) \notag\\
	&\ge  I(V_{2};X_{2}Z_{2}|UU_1U_2). 
\end{align}
Given $I(U_2;X_{1}Z_{1}|UU_{1}) = 0$, we have
\begin{align}
	0 &\ge I(V_{2};X_{2}Z_{2}|UU_1U_2) - I(V_2;X_{1}Z_{1}|UU_1U_2)\notag\\
	&=  - H(V_2|UU_1U_2X_2Z_2) + H(V_2|UU_1U_2X_1Z_1)\notag\\
	&\overset{(a)} = H(V_2|UU_1U_2X_1Z_1)- H(V_2|UU_1U_2X_1Z_1X_2Z_2)\notag\\
	& = I(V_2;X_2Z_2|UU_1U_2X_1Z_1) \ge 0,
\end{align}
where $(a)$ follows from that $V_2-UU_1U_2X_2Z_2-X_1Z_1$ forms a Markov chain. We can obtain a similar result for $k=1$ as 
\begin{align}
	I(V_{1};X_{1}Z_{1}|UU_1&U_2) - I(V_1;X_{2}Z_{2}|UU_1U_2) \notag\\
	&= I(V_1;X_1Z_1|UU_1U_2X_2Z_2) = 0.
\end{align}

Then, considering the first term in function $\min(\cdot)$ for the single rate bound on $R_2$, i.e., inequality constraint~(22a) for $k=2$, in $\mathcal{I}_{\text{R-D}}^{\text{aw}}$ presented in~\cite[Theorem~3]{ahmadipour2023information}, we have 
\begin{align}
	R_2^{\text{aw}}& \le  I(U_2;X_{1}Z_{1}|UU_{1}) + I(V_2;X_{1}Z_{1}|UU_1U_2) \notag\\
	&\qquad- I(V_{2};X_{2}Z_{2}|UU_1U_2) + I(X_2;Y|UX_{1})\notag\\
	& \qquad +I(V_2;X_1X_2Y|UU_1U_2) + I(V_{1};X_1X_2YV_2|UU_1U_2) \notag\\
	&\qquad- I(V_{2};X_{2}Z_{2}|UU_1U_2) \notag\\
	&  = I(X_2;Y|UX_{1})+I(V_2;X_1X_2Y|UU_1U_2) \notag\\
	&\qquad+ I(V_{1};X_1X_2YV_2|UU_1U_2)  - I(V_{2};X_{2}Z_{2}|UU_1U_2) \notag\\
	& = I(X_2;Y|UX_{1})+I(V_2;X_1X_2YV_1|UU_1U_2) \notag\\
	&\qquad+ I(V_{1};X_1X_2Y|UU_1U_2)  - I(V_{2};X_{2}Z_{2}|UU_1U_2) \notag\\
	& \overset{(a)} = I(X_2;Y|UX_{1}) + I(V_{1};X_1X_2Y|UU_1U_2) \notag\\
	&\qquad- I(V_2;Z_2|UU_1U_2X_1X_2YV_1)\notag\\
	&\le I(X_2;Y|UX_{1}) + I(V_{1};X_1X_2Y|UU_1U_2)\notag\\
	& \overset{(b)}= H(Y_1Y_2|UX_1) - H(Y_1Y_2|UX_1X_2)  \notag\\
	&\qquad+ I(V_{1};X_1X_2Y_1Y_2|UU_1U_2)\notag\\
	& \overset{(c)}= H(S_1) + H(Y_2|UX_1Y_1) - H(S_1) - H(S_2)  \notag\\
	&\qquad + I(V_{1};X_1X_2S_1S_2|UU_1U_2)\notag\\
	&\overset{(d)}\le 1 - 0.5 +  I(V_{1};X_1X_2S_1S_2|UU_1U_2)\notag\\
	& \overset{(e)}= 0.5 + I(V_{1};X_1X_2Z_1Z_2|UU_1U_2)\notag\\
	& \overset{(f)}= 0.5 + I(V_{1};X_2Z_2|UU_1U_2)\notag\\
	& = 0.5 + H(X_2Z_2|UU_1U_2) - H(X_2Z_2|UU_1U_2V_1) \notag\\
	& \overset{(g)} \le 0.5 + H(X_2Z_2|UU_1U_2) - H(X_2Z_2|UU_1U_2X_1Z_1V_1) \notag\\
	& \overset{(h)} \le 0.5 + H(X_2Z_2|UU_1U_2) - H(X_2Z_2|UU_1U_2X_1Z_1) \notag\\
	& = 0.5 + I(X_1Z_1;X_2Z_2|UU_1U_2)\notag\\
	& = 0.5 + I(X_1S_2;X_2S_1|UU_1U_2)\notag\\
	& \overset{(i)}= 0.5,
\end{align}
where $(a)$ follows from the Markov chain $V_2-UU_1U_2X_2Z_2 - X_1YV_1$, $(b)$ follows from that $Y=(Y_1,Y_2)$, $(c)$ follows from that $Y_1=X_1\oplus S_1$, $Y_2=X_2\oplus S_2$, $(d)$ follows from that the entropy of binary variable is no more than 1 and $H(S_2) = 0.5$, $(e)$ follows from that $Z_1 = X_1\oplus S_2,Z_2=X_2\oplus S_1$,
$(f)$ follows from that $I(V_1;X_1Z_1|UU_1U_2X_2Z_2) = 0$, $(g)$ follows from that conditioning reduces entropy, $(h)$ follows from the Markov chain $V_1-UU_1U_2X_1Z_1-X_2Z_2$, and $(i)$ follows from that given $UU_1U_2$, $X_1S_2$ and $X_2S_1$ are mutually independent. 

\section{Proof of Example \ref{example:common}}\label{appendix:proofExampleCommon}
Considering the inequality constraints~(22d) and~(22f) for $k=1$ in $\mathcal{I}_{\text{R-D}}^{\text{aw}}$ presented in~\cite[Theorem~3]{ahmadipour2023information},
\begin{subequations}
	\begin{align}
		&I(U_1;X_{2}Z_{2}|UU_{2}) +  I(V_1;X_{2}Z_{2}|UU_1U_2) \notag\\
		&\qquad\qquad\qquad\qquad\qquad\ge  I(V_{1};X_{1}Z_{1}|UU_1U_2), \label{proof:example2_1},\\
		&I(X_1;Y|UX_{2}) + I(V_1;X_1X_2Y|UU_1U_2) \notag\\
		&\qquad+ I(V_{2};X_1X_2YV_1|UU_1U_2) \ge I(V_{1};X_{1}Z_{1}|UU_1U_2),\label{proof:example2_2}
	\end{align}
\end{subequations}
we note that the sum of the right hand side of (\ref{proof:example2_1}) and (\ref{proof:example2_2}) is no more than the sum of the left hand side, i.e., 
\begin{align}\label{proof:example2_sum}
	&I(U_1;X_{2}Z_{2}|UU_{2}) +  I(V_1;X_{2}Z_{2}|UU_1U_2)+ I(X_1;Y|UX_{2}) \notag\\
	&\quad+ 
	I(V_1;X_1X_2Y|UU_1U_2) + I(V_{2};X_1X_2YV_1|UU_1U_2) \notag\\
	&\qquad\ge I(V_{1};X_{1}Z_{1}|UU_1U_2) + I(V_{1};X_{1}Z_{1}|UU_1U_2).
\end{align}
Given $Y = X_1 \oplus S_1 \oplus N, X_2 \oplus S_2$, $Z_1 = (X_1\oplus S_1, X_1 \oplus S_2)$, $Z_2 = X_1\oplus B$, one can prove that
\begin{align}
	I(X_1;Y|UX_{2}) \overset{(a)}= 0,
\end{align}
where $(a)$ follows from that $H(N) = 1$. Thus, (\ref{proof:example2_sum}) becomes
\begin{align}\label{proof:example2_sum_update}
	&I(U_1;X_2Z_{2}|UU_{2}) +  I(V_1;X_{2}Z_{2}|UU_1U_2)\notag\\
	&\quad+  
	I(V_1;X_1X_2YV_2|UU_1U_2) + I(V_{2};X_1X_2Y|UU_1U_2)\notag\\
	&\qquad \ge I(V_{1};X_{1}Z_{1}|UU_1U_2) + I(V_{1};X_{1}Z_{1}|UU_1U_2).
\end{align}
Based on the Markov chains $X_2-UU_2-U_1$, $X_2Z_2-UU_1U_2X_1Z_1-V_1$ and $X_2YV_2-UU_1U_2X_1Z_1-V_1$, inequality (\ref{proof:example2_sum_update}) can be further written as
\begin{align}\label{proof:example2_sum_update_new}
	&I(U_1;Z_{2}|UU_{2}X_{2}) + I(V_{2};X_1X_2Y|UU_1U_2) \notag\\
	& \ge I(V_1;X_1Z_1|UU_1U_2X_2Z_2) + I(V_1;Z_1|UU_1U_2X_1X_2YV_2).
\end{align}
For the right hand side of (\ref{proof:example2_sum_update_new}), we have
\begin{align}\label{example2:rate-distortionRelax1}
	&I(V_1;X_1Z_1|UU_1U_2X_2Z_2) \notag\\
	&\ge I(V_1;Z_1|UU_1U_2X_1X_2Z_2)\notag\\
	&\overset{(a)}= I(V_1;S_1S_2|UU_1U_2X_1X_2,X_1\oplus B)\notag\\
	&\ge I(V_1;S_1|UU_1U_2X_1X_2,X_1\oplus B)\notag\\
	&\overset{(b)}\ge H(S_1) - H(S_1|U_1X_2,X_1\oplus B,V_1)\notag\\
	& = I(S_1;U_1X_2V_1Z_2)\notag\\
	&\overset{(c)}\ge I(S_1;\hat{S}_{T_2})
\end{align}
where $(a)$ follows from that $Z_1 = (X_1\oplus S_1, X_1 \oplus S_2)$ and $Z_2 = X_1\oplus B$, $(b)$ follows from that $S_1$ is independent of $UU_1U_2X_1X_2,X_1\oplus B$ and conditioning reduces entropy, $(c)$ follows from the Markov chain $S_1-U_1X_2V_1Z_2-\hat{S}_{T_2}$ based on the parameter estimator for $\mathcal{I}_{\text{R-D}}^{\text{aw}}$ in~\cite[Theorem~3]{ahmadipour2023information},
and 
\begin{align}\label{example2:rate-distortionRelax2}
	&I(V_1;Z_1|UU_1U_2X_1X_2YV_2) \notag\\
	&\overset{(a)}= I(V_1;S_1S_2|UU_1U_2X_1X_2,S_1\oplus N,S_2V_2)\notag\\
	& \overset{(b)}= H(S_1) - H(S_1|UU_1U_2X_1X_2,S_1\oplus N,S_2V_1V_2)\notag\\
	&\overset{(c)} =  H(S_1) - H(S_1|UU_1U_2X_1X_2,S_1\oplus N,S_2V_1V_2 B)\notag\\
	& = H(S_1) - H(S_1|UU_1U_2X_1X_2,S_1\oplus N,S_2V_1V_2 B,X_1\oplus B)\notag\\
	&\overset{(d)}\ge H(S_1) - H(S_1|U_1X_2V_1,X_1\oplus B)\notag\\
	& = I(S_1;U_1X_2V_1Z_2)\notag\\
	&\overset{(e)}\ge I(S_1;\hat{S}_{T_2})
\end{align}
where $(a)$ follows from that $Y=(X_1 \oplus S_1\oplus N,X_2\oplus S_2)$ and $Z_1 = (X_1\oplus S_1, X_1 \oplus S_2)$, $(b)$ follows from $S_1$ is independent of $UU_1U_2X_1X_2,S_1\oplus N,S_2B$ as $H(N)=1$ and Markov chain $S_1-UU_1U_2X_2Z_2-V_2$ with $Z_2 = X_1\oplus B$, $(c)$ follows from that $S_1$ is independent of $B$ given $UU_1U_2X_1X_2,S_1\oplus N,S_2V_1V_2$, $(d)$ follows from that conditioning reduces entropy, and $(e)$ follows from the Markov chain $S_1-U_1X_2V_1Z_2-\hat{S}_{T_2}$ based on the parameter estimator for $\mathcal{I}_{\text{R-D}}^{\text{aw}}$ in~\cite[Theorem~3]{ahmadipour2023information}.

For the left hand side of (\ref{proof:example2_sum_update_new}), we have
\begin{align}
	&I(U_1;Z_{2}|UU_{2}X_{2}) + I(V_{2};X_1X_2Y|UU_1U_2) \notag\\
	&\overset{(a)}= I(U_1;Z_{2}|UU_{2}X_{2}) + I(V_{2};X_1X_2|UU_1U_2) \notag\\
	&\qquad+ I(V_2;S_1\oplus N,S_2|UU_1U_2X_1X_2)\notag\\
	& \overset{(b)}\le I(U_1;Z_{2}|UU_{2}X_{2}) + I(V_{2};X_1X_2|UU_1U_2) \notag\\
	&\qquad+ H(S_1\oplus N,S_2|UU_1U_2X_1X_2B) \notag\\
	&\qquad- H(S_1\oplus N,S_2|UU_1U_2X_1X_2V_2B)\notag\\
	&= I(U_1;Z_{2}|UU_{2}X_{2}) + I(V_{2};X_1X_2|UU_1U_2) \notag\\
	&\qquad+ I(V_2;S_1\oplus N,S_2|UU_1U_2X_1X_2B)\notag\\
	&\overset{(c)} = I(U_1;Z_{2}|UU_{2}X_{2}) + I(V_{2};X_1X_2|UU_1U_2)\notag\\
	& = I(U_1;Z_{2}|UU_{2}X_{2}) + H(X_1|UU_1U_2) - H(X_1|UU_1U_2V_2) \notag\\
	&\qquad+ I(V_{2};X_2|UU_1U_2X_1)\notag\\
	& \overset{(d)}\le I(U_1;Z_{2}|UU_{2}X_{2}) + H(X_1|UU_1U_2) \notag\\
	&\qquad- H(X_1|UU_1U_2X_2Z_2V_2) + I(V_{2};X_2|UU_1U_2X_1)\notag\\
	& \overset{(e)}= I(U_1;Z_{2}|UU_{2}X_{2}) + I(X_1;X_2Z_2|UU_1U_2) \notag\\
	&\qquad+ I(V_{2};X_2|UU_1U_2X_1)\notag\\
	& \overset{(f)}= I(U_1;Z_{2}|UU_{2}X_{2}) + I(X_1;Z_2|UU_1U_2X_2) \notag\\
	&\qquad+ I(V_{2};X_2|UU_1U_2X_1)\notag\\
	& = H(Z_{2}|UU_{2}X_{2}) - H(Z_2|UU_1U_2X_1X_2)\notag\\
	&\qquad+ I(V_{2};X_2|UU_1U_2X_1)\notag\\
	&\overset{(g)}\le H(X_1\oplus B) - H(X_1\oplus B|X_1) + I(V_{2};X_2|UU_1U_2X_1)\notag\\
	&\overset{(h)}\le 0.5 + I(V_{2};X_2|UU_1U_2X_1),
\end{align}
where $(a)$ follows from that $Y=(X_1 \oplus S_1 \oplus N, X_2 \oplus S_2)$, $(b)$ follows from that $S_1\oplus N,S_2$ are independent of $B$ given $UU_1U_2X_1X_2$ and conditioning reduces entropy, $(c)$ follows from the Markov chain $S_1\oplus N,S_2-UU_1U_2X_2Z_2-V_2$ with $Z_2 = X_1\oplus B$, $(d)$ follows from that conditioning reduces entropy, $(e)$ follows from the Markov chain $X_1-UU_1U_2X_2Z_2-V_2$, $(f)$ follows from the Markov chain $X_1-UU_1U_2-X_2$, $(g)$ follows from $Z_2 = X_1\oplus B$ and conditioning reduces entropy, $(h)$ follows from that the entropy of binary variable is no more than 1 and $H(B) = 0.5$.
Moreover, based on the inequality constraint~(22d) for $k=2$ in $\mathcal{I}_{\text{R-D}}^{\text{aw}}$ presented in~\cite[Theorem~3]{ahmadipour2023information},
one can know that 
\begin{align}
	0& \overset{(a)}= I(U_2;X_{1}Z_{1}|UU_{2})\notag\\
	&\overset{(b)}\ge I(V_{2};X_{2}Z_{2}|UU_1U_2) - I(V_2;X_{1}Z_{1}|UU_1U_2)\notag\\
	&\overset{(c)} = I(V_{2};X_{2}Z_{2}|UU_1U_2) - I(V_2;X_{1}|UU_1U_2) \notag\\
	&\qquad- I(V_2;S_1S_2|UU_1U_2X_{1})\notag\\
	&\overset{(d)} \ge  I(V_{2};X_{2}Z_{2}|UU_1U_2) - I(V_2;X_{1}|UU_1U_2) \notag\\
	&\qquad- H(S_1S_2|UU_1U_2X_{1}X_2B) \notag\\
	&\qquad + H(S_1S_2|UU_1U_2X_{1}X_2V_2B)\notag\\
	& =  I(V_{2};X_{2}Z_{2}|UU_1U_2) - I(V_2;X_{1}|UU_1U_2) \notag\\
	&\qquad- I(V_2;S_1S_2|UU_1U_2X_1X_2B)\notag\\
	& \overset{(e)}= I(V_{2};X_{2}Z_{2}|UU_1U_2) - I(V_2;X_{1}|UU_1U_2)\notag\\
	& = H(V_2|UU_1U_2X_1) - H(V_2|UU_1U_2X_{2}Z_{2})\notag\\
	& \overset{(f)}= H(V_2|UU_1U_2X_1) - H(V_2|UU_1U_2X_{2}Z_{2}X_1)\notag\\
	& = I(V_2;X_2Z_2|UU_1U_2X_1) \notag\\
	&\ge I(V_2;X_2|UU_1U_2X_1),
\end{align}
where $(a)$ follows from the Markov chain $X_1U_1-U-U_2X_2$ and $Z_1 = (X_1\oplus S_1, X_1 \oplus S_2)$, $(b)$ follows from the inequality constraint~(22d) for $k=2$ in $\mathcal{I}_{\text{R-D}}^{\text{aw}}$ presented in~\cite[Theorem~3]{ahmadipour2023information}, $(c)$ follows from that $Z_1 = (X_1\oplus S_1, X_1 \oplus S_2)$, $(d)$ follows from that $S_1S_2$ is independent of $B$ given $UU_1U_2X_{1}X_2$ and conditioning reduces entropy, $(e)$ follows from the Markov chain $S_1S_2-UU_1U_2X_2Z_2-V_2$ and $Z_2 = X_1\oplus B$, $(f)$ follows from the Markov chain $X_1 -UU_1U_2X_{2}Z_{2}- V_2$.

Therefore, the inequality (\ref{proof:example2_sum_update_new}) means that 
\begin{align}
	I(S_1;\hat{S}_{T_2}) \le 0.25 < 0.5 = H(S_1),
\end{align}
i.e., the mutual information $I(S_1;\hat{S}_{T_2})$ is strictly smaller than the entropy $H(S_1)$, which shows that transmitter~2 cannot achieve zero distortion based on the rate-disotrtion theory \cite{el2011network}.

\section{Proof of Strict Inclusion for Inner Bounds in Example \ref{example:general}}\label{appendix:proofExampleGeneral}
We first shown that tuple $(0.918563,0,0.3)$ is not in $\mathcal{I}_{\text{R-D}}^{\text{our},\text{com}}$ by proving that $R_1^{\text{our},\text{com}}<0.918563$ holds for $\mathcal{I}_{\text{R-D}}^{\text{our},\text{com}}$ and tuple $(0,0,0.13072)$ is not in $\mathcal{I}_{\text{R-D}}^{\text{aw}}$ by proving that $D_2^{\text{aw}}>0.13072$ holds for $\mathcal{I}_{\text{R-D}}^{\text{aw}}$. Then, we show that tuple $(0.11697,0,0.1783)$ is not in $\mathcal{I}_{\text{R-D}}^{\text{aw}}$ by proving that $R_1^{\text{aw}}<0.11697$ when $D_2^{\text{aw}}= 0.1783$.

\subsection{Proof of $R_1^{\text{our},\text{com}}<0.918563$}
Due to the fact that $\min\{a,b\}\le a$ and $V_{1,p}=v^*_{1,p},V_{2,p}=v^*_{2,p}$ almost surely for some specific values $v^*_{1,p}\in\mathcal{V}_{1,p},v^*_{2,p}\in\mathcal{V}_{2,p}$ in $\mathcal{I}_{\text{R-D}}^{\text{our},\text{com}}$, we obtain a relaxed single-user bound for $R_1$ in $\mathcal{I}_{\text{R-D}}^{\text{our},\text{com}}$ based on (\ref{equ:ourResA}) as
\begin{align}\label{equ:RelaxedBoundSingleUser}
	&R_1^{\text{our},\text{com}}\le I(U_1;Z_{2}|UU_{2}X_2) - I(V_{1,c};X_1Z_1|UU_1U_2X_{2}Z_{2})  \notag\\
	&\quad+ I(X_1;X_{2}YV_{1,c}V_{2,c}|UU_1U_2)\notag\\
	& \overset{(a)} =  I(U_1;Z_{2}|UU_{2}X_2) + I(V_{1,c};X_{2}Z_{2}|UU_1U_2) \notag\\
	&\quad- I(V_{1,c};X_{1}Z_{1}|UU_1U_2)   + I(X_1;YV_{1,c}V_{2,c}|UU_1U_2X_{2})\notag\\
	&  = I(U_1;Z_{2}|UU_{2}X_2)  + I(V_{1,c};X_{2}Z_{2}|UU_1U_2) \notag\\
	&\quad- I(V_{1,c};X_{1}Z_{1}|UU_1U_2) + I(X_1;Y|UU_1U_2X_{2}) \notag\\
	&\quad+ I(X_1;V_{2,c}|UU_1U_2X_2Y)  + I(X_1;V_{1,c}|UU_1U_2X_2YV_{2,c}),
\end{align}
where $(a)$ follows from the Markov chains $X_2Z_2 - UU_1U_2X_1Z_1 - V_{1,c}$ and $X_1-UU_1U_2-X_2$.

Now we proceed to derive upper bounds for the terms in (\ref{equ:RelaxedBoundSingleUser}). We have
\begin{align}\label{equ:APPENDIXBUSE1}
	I(U_1;Z_{2}|UU_{2}X_{2}) &\overset{(a)}= I(U_1;B\cdot X_1,X_2\oplus S_1|UU_{2}X_{2})\notag\\
	& = I(U_1;B\cdot X_1|UU_{2}X_{2}) \notag\\
	&\qquad+ I(U_1;X_2\oplus S_1|UU_2X_2,B\cdot X_1)\notag\\
	& \overset{(b)}= I(U_1;B\cdot X_1|UU_{2}X_{2}),
\end{align}
where $(a)$ follows from the fact that $Z_2 = (BX_1, X_2 \oplus S_1)$ in the example, $(b)$ follows from the fact that $S_1$ is independent of $U,U_1,U_2,X_1,X_2,B,B\cdot X_1$. 

For term $I(V_{1,c};X_{2}Z_{2}|UU_1U_2)$, we have
\begin{align}\label{equ:APPENDIXBUSE2}
	&I(V_{1,c};X_{2}Z_{2}|UU_1U_2)\notag\\
	&= H(X_{2}Z_{2}|UU_1U_2) - H(X_{2}Z_{2}|UU_1U_2V_{1,c}) \notag\\
	&\overset{(a)}\le H(X_{2}Z_{2}|UU_1U_2) - H(X_{2}Z_{2}|UU_1U_2X_1Z_1V_{1,c}) \notag\\
	&\overset{(b)}=H(X_{2}Z_{2}|UU_1U_2) - H(X_{2}Z_{2}|UU_1U_2X_1Z_1) \notag\\
	& = I(X_1Z_1;X_2Z_2|UU_1U_2)\notag\\
	& = I(X_1;X_2Z_2|UU_1U_2) + I(Z_1;X_2Z_2|UU_1U_2X_1)\notag\\
	& \overset{(c)}= I(X_1;X_2Z_2|UU_1U_2) + I(N;X_2Z_2|UU_1U_2X_1)\notag\\
	&\overset{(d)}= I(X_1;X_2Z_2|UU_1U_2) \notag\\
	&\overset{(e)}= I(X_1;B\cdot X_1, X_2\oplus S_1|UU_1U_2X_2)\notag\\
	&\overset{(f)}= I(X_1;B\cdot X_1|UU_1U_2X_2),
\end{align}
where $(a)$ follows from that conditioning reduces entropy, $(b)$ follows from the Markov chain $X_2Z_2-UU_1U_2X_1Z_1-V_{1,c}$, $(c)$ follows from that $Z_1 = X_1\oplus N$ in the example, $(d)$ follows from that $N$ is independent of $UU_1U_2X_1X_2Z_2$ as $Z_2 = (BX_1,X_2\oplus S_1)$, $(e)$ follows from the Markov chain $X_1U_1-U-U_2X_2$ and $Z_2 = (BX_1,X_2\oplus S_1)$, $(f)$ follows from that $S_1$ is independent of $X_1$ given $UU_1U_2X_2,B\cdot X_1$.

For term $I(X_1;Y|UU_1U_2X_{2})$, we have
\begin{align}\label{equ:APPENDIXBUSE3}
	&I(X_1;Y|UU_1U_2X_{2}) \notag\\
	&\overset{(a)}=  I(X_1;X_1\oplus S_1, X_2\oplus S_2|UU_1U_2X_{2}) \notag\\
	& = I(X_1;X_1\oplus S_1|UU_1U_2X_{2}) \notag\\
	&\qquad+ I(X_1;X_2\oplus S_2|UU_1U_2X_{2},X_1\oplus S_1)\notag\\
	&\overset{(b)}= I(X_1;X_1\oplus S_1|UU_1U_2X_{2}) \notag\\
	& = H(X_1\oplus S_1|UU_1U_2X_{2}) - H(X_1\oplus S_1|UU_1U_2X_1X_{2})\notag\\
	&\overset{(c)}\le H(X_1\oplus S_1) - H(S_1)\notag\\
	&\overset{(d)}\le 1 - H(S_1) \notag\\
	& \overset{(e)}< 0.205,
\end{align}
where $(a)$ follows from the fact that $Y = (Y_1,Y_2) = (X_1\oplus S_1,X_2\oplus S_2)$ in the example, $(b)$ follows from the fact that $S_2$ is independent of $UU_1U_2X_1X_2S_1$, $(c)$ follows from that conditioning reduces entropy and $S_1$ is independent of $UU_1U_2X_1X_2$, $(d)$ follows from that the entropy of binary random variable is no more than 1, $(e)$ follows from that $P_{S_1}(1) = 0.24$ in the example.

For term $I(X_1;V_{2,c}|UU_1U_2X_2Y)$, we have
\begin{align}\label{equ:APPENDIXBUSE4}
	&I(X_1;V_{2,c}|UU_1U_2X_2Y) \notag\\
	&\le I(V_{2,c};X_1X_2Y|UU_1U_2)\notag\\
	& \overset{(a)} = I(V_{2,c};X_1X_2,X_1\oplus S_1,X_2\oplus S_2|UU_1U_2)\notag\\
	& = I(V_{2,c};X_1X_2S_1S_2|UU_1U_2)\notag\\
	& = I(V_{2,c};X_1X_2S_1|UU_1U_2) + I(V_{2,c};S_2|UU_1U_2X_1X_2S_1)\notag\\
	& \overset{(b)} \le I(V_{2,c};X_1X_2S_1|UU_1U_2) + H(S_2|UU_1U_2X_1X_2S_1B) \notag\\
	&\qquad- H(S_2|UU_1U_2X_1X_2S_1V_{2,c}B)\notag\\
	& \overset{(c)}= I(V_{2,c};X_1X_2S_1|UU_1U_2) \notag\\
	&\qquad+ I(S_2;V_{2,c}|UU_1U_2X_1X_2S_1BZ_2)\notag\\
	& \overset{(d)}=I(V_{2,c};X_1X_2S_1|UU_1U_2) \notag\\
	& = I(V_{2,c};X_1|UU_1U_2) + I(V_{2,c};X_2S_1|UU_1U_2X_1)\notag\\
	& = H(X_1|UU_1U_2) - H(X_1|UU_1U_2V_{2,c}) \notag\\
	&\qquad+ I(V_{2,c};X_2S_1|UU_1U_2X_1)\notag\\
	& \overset{(e)} \le H(X_1|UU_1U_2) - H(X_1|UU_1U_2X_2Z_2V_{2,c}) \notag\\
	&\qquad+ I(V_{2,c};X_2S_1|UU_1U_2X_1)\notag\\
	& \overset{(f)}= H(X_1|UU_1U_2) - H(X_1|UU_1U_2X_2Z_2)  \notag\\
	&\qquad+ I(V_{2,c};X_2S_1|UU_1U_2X_1)\notag\\
	& \overset{(g)}= H(X_1|UU_1U_2) - H(X_1|UU_1U_2X_2,B\cdot X_1,X_2\oplus S_1) \notag\\
	&\qquad+ I(V_{2,c};X_2S_1|UU_1U_2X_1)\notag\\
	& \overset{(h)} =H(X_1|UU_1U_2) - H(X_1|UU_1U_2,B\cdot X_1) \notag\\
	&\qquad+ I(V_{2,c};X_2S_1|UU_1U_2X_1)\notag\\
	& = I(B\cdot X_1;X_1|UU_1U_2) + I(V_{2,c};X_2S_1|UU_1U_2X_1),
\end{align}
where $(a)$ follows from that $Y = (Y_1,Y_2) = (X_1\oplus S_1,X_2\oplus S_2)$ in the example, $(b)$ follows from that $S_2$ is independent of $UU_1U_2X_1X_2S_1B$ and conditioning reduces entropy, $(c)$ follows from that $Z_2 = (B\cdot X_1,X_2\oplus S_1)$ in the example, $(d)$ follows from the Markov chain $S_2-UU_1U_2X_2Z_2-V_{2,c}$, $(e)$ follows from that conditioning reduces entropy, $(f)$ follows from the Markov chain $X_1-UU_1U_2X_2Z_2-V_{2,c}$, $(g)$ follows from that $Z_2 = (B\cdot X_1,X_2\oplus S_1)$, $(h)$ follows from the Markov chain $X_2,S_1-UU_1U_2,B\cdot X_1-X_1$.

For term $I(X_1;V_{1,c}|UU_1U_2X_2YV_{2,c})- I(V_{1,c};X_{1}Z_{1}|U$ $U_1U_2)$, we have
\begin{align}\label{equ:APPENDIXBUSE5}
	&I(X_1;V_{1,c}|UU_1U_2X_2YV_{2,c})- I(V_{1,c};X_{1}Z_{1}|UU_1U_2)\notag\\
	& = H(V_{1,c}|UU_1U_2X_2YV_{2,c}) - H(V_{1,c}|UU_1U_2X_1X_2YV_{2,c})\notag\\
	&\qquad- H(V_{1,c}|UU_1U_2) + H(V_{1,c}|UU_1U_2X_{1}Z_{1})\notag\\
	& \overset{(a)} \le H(V_{1,c}|UU_1U_2)- H(V_{1,c}|UU_1U_2X_1X_2YV_{2,c})\notag\\
	&\qquad- H(V_{1,c}|UU_1U_2) + H(V_{1,c}|UU_1U_2X_{1}Z_{1})\notag\\
	& = - H(V_{1,c}|UU_1U_2X_1X_2YV_{2,c})+ H(V_{1,c}|UU_1U_2X_{1}Z_{1})\notag\\
	&\overset{(b)}=- H(V_{1,c}|UU_1U_2X_1X_2YV_{2,c})\notag\\
	&\qquad+ H(V_{1,c}|UU_1U_2X_{1}Z_{1}X_2YV_{2,c})\notag\\
	& = - I(V_{1,c};Z_1|UU_1U_2X_1X_2YV_{2,c})\notag\\
	&\le 0,
\end{align}
where $(a)$ follows from that conditioning reduces entropy, $(b)$ follows from the Markov chain $X_2YV_{2,c} - UU_1U_2X_{1}Z_{1} - V_{1,c}$.

Combining (\ref{equ:RelaxedBoundSingleUser}), (\ref{equ:APPENDIXBUSE1}), (\ref{equ:APPENDIXBUSE2}), (\ref{equ:APPENDIXBUSE3}), (\ref{equ:APPENDIXBUSE4}), and (\ref{equ:APPENDIXBUSE5}), we have 
\begin{align}
	&R_1^{\text{our},\text{com}} < I(U_1;B\cdot X_1|UU_{2}X_{2}) + I(X_1;B\cdot X_1|UU_1U_2X_2) \notag\\
	&\qquad+I(B\cdot X_1;X_1|UU_1U_2) + I(V_{2,c};X_2S_1|UU_1U_2X_1) \notag\\
	&\qquad+ 0.205.
\end{align}

Next, we show that $I(V_{2,c};X_2S_1|UU_1U_2X_1) = 0$. Given $Z_1 = X_1\oplus N$, we have 
\begin{align}
	I(U_2;Z_1|UU_1X_1) &= I(U_2;N|UU_1X_1) \overset{(a)}= 0,
\end{align}
where $(a)$ follows from that $N$ is independent of $UU_1U_2X_1$. Considering (\ref{equ:ourResConsA}) for $k=2$, we have 
\begin{align}\label{equ:APPENDIXBUSE6}
	0 &\ge  I(V_{2,c};X_2Z_2|UU_1U_2X_1Z_1)\notag\\
	& = H(V_{2,c}|UU_1U_2X_1Z_1) - H(V_{2,c}|UU_1U_2X_1Z_1X_2Z_2)\notag\\
	& \overset{(a)} =  H(V_{2,c}|UU_1U_2X_1Z_1) - H(V_{2,c}|UU_1U_2X_2Z_2)\notag\\
	& = H(V_{2,c}|UU_1U_2X_1Z_1) - H(V_{2,c}|UU_1U_2) \notag\\
	&\qquad+ H(V_{2,c}|UU_1U_2)- H(V_{2,c}|UU_1U_2X_2Z_2)\notag\\
	& = I(V_{2,c};X_2Z_2|UU_1U_2) - I(V_{2,c};X_1Z_1|UU_1U_2),\notag\\
	& \overset{(b)} = I(V_{2,c};X_2Z_2|UU_1U_2) - I(V_{2,c};X_1,X_1\oplus N|UU_1U_2),\notag\\
	& = I(V_{2,c};X_2Z_2|UU_1U_2) - I(V_{2,c};X_1|UU_1U_2) \notag\\
	&\qquad- I(V_{2,c};N|UU_1U_2X_1)\notag\\
	& \overset{(c)}= I(V_{2,c};X_2Z_2|UU_1U_2) - I(V_{2,c};X_1|UU_1U_2) \notag\\
	& \overset{(d)}= I(V_{2,c};X_2Z_2|UU_1U_2X_1)\notag\\
	&\overset{(e)} = I(V_{2,c};X_2,B\cdot X_1,X_2\oplus S_1|UU_1U_2X_1)\notag\\
	& = I(V_{2,c};X_2,B\cdot X_1,S_1|UU_1U_2X_1)\notag\\
	&\ge I(V_{2,c};X_2S_1|UU_1U_2X_1),
\end{align}
where $(a)$ follows from the Markov chain $X_1Z_1 - UU_1U_2X_2Z_2 - V_{2,c}$, $(b)$ follows from that $Z_1 = X_1\oplus N$ in the example, $(c)$ follows from that $N$ is independent of $V_{2,c}$ given $UU_1U_2X_1$, $(d)$ follows from the Markov chain $X_1-UU_1U_2X_2Z_2-V_{2,c}$, $(e)$ follows from that $Z_2 = (BX_1, X_2\oplus S_1)$ in the example. 

Given (\ref{equ:APPENDIXBUSE6}), we have
\begin{align}\label{upperboundR1_proveExample}
	&R_1^{\text{our},\text{com}}\notag\\
	& \le I(U_1;B\cdot X_1|UU_{2}X_{2}) + I(X_1;B\cdot X_1|UU_1U_2X_2) \notag\\
	&\qquad +I(B\cdot X_1;X_1|UU_1U_2) + 0.205\notag\\
	& = H(B\cdot X_1|UU_{2}X_{2}) - H(B\cdot X_1|UU_1U_{2}X_1X_{2})\notag\\
	& \qquad + H(B\cdot X_1|UU_1U_2) - H(B\cdot X_1|UU_1U_2X_1) + 0.205\notag\\
	&\overset{(a)}\le H(B\cdot X_1) - H(B\cdot X_1|UU_1U_{2}X_1X_{2})+ H(B\cdot X_1)\notag\\
	&\qquad - H(B\cdot X_1|UU_1U_2X_1) + 0.205\notag\\
	&\overset{(b)} = H(B\cdot X_1) - H(B\cdot X_1|X_1)+ H(B\cdot X_1) \notag\\
	&\qquad- H(B\cdot X_1|X_1) + 0.205\notag\\
	& = 2I(X_1;B\cdot X_1)+ 0.205\notag\\
	&\overset{(c)} < 2 * 0.322 + 0.205\notag\\
	& = 0.849 < 0.918563,
\end{align}
where $(a)$ follows from that conditioning reduces entropy, $(b)$ follows from the Markov chain $UU_1U_2X_2-X_1-B\cdot X_1$, $(c)$ follows from $P_B = 0.5$ and $X$ is a binary random variable. 

\subsection{Proof of $D_2^{\text{aw}}>0.13072$}
Considering the inequality constraints~(22d) and~(22f) for $k=1$ in $\mathcal{I}_{\text{R-D}}^{\text{aw}}$ presented in~\cite[Theorem~3]{ahmadipour2023information},
\begin{subequations}
	\begin{align}
		&I(U_1;X_{2}Z_{2}|UU_{2}) +  I(V_1;X_{2}Z_{2}|UU_1U_2) \notag\\
		&\qquad\qquad\qquad\qquad\qquad\ge  I(V_{1};X_{1}Z_{1}|UU_1U_2), \label{proof:exampleGeneral_1}\\
		& I(X_1;Y|UX_{2}) + I(V_1;X_1X_2Y|UU_1U_2) \notag\\
		&\qquad+ I(V_{2};X_1X_2YV_1|UU_1U_2) \ge I(V_{1};X_{1}Z_{1}|UU_1U_2),\label{proof:exampleGeneral_2}
	\end{align}
\end{subequations}
we note that the sum of the right hand side of (\ref{proof:exampleGeneral_1}) and (\ref{proof:exampleGeneral_2}) is no more than the sum of the left hand side, i.e., 
\begin{align}\label{proof:exampleGeneral_sum}
	&I(U_1;X_{2}Z_{2}|UU_{2}) +  I(V_1;X_{2}Z_{2}|UU_1U_2)+ I(X_1;Y|UX_{2}) \notag\\
	&\qquad+ 
	I(V_1;X_1X_2Y|UU_1U_2) + I(V_{2};X_1X_2YV_1|UU_1U_2) \notag\\
	&\qquad\qquad \ge I(V_{1};X_{1}Z_{1}|UU_1U_2) + I(V_{1};X_{1}Z_{1}|UU_1U_2), 
\end{align}
which can be equivalently transformed to
\begin{align}\label{proof:exampleGeneral_sumUpdate}
	&I(U_1;Z_{2}|UU_{2}X_{2}) + I(X_1;Y|UX_{2}) + I(V_{2};X_1X_2Y|UU_1U_2) \notag\\
	& \ge I(V_{1};X_{1}Z_{1}|UU_1U_2X_2Z_2) + I(V_{1};Z_{1}|UU_1U_2X_1X_2YV_2) 
\end{align}
based on the Markov chains $X_2Z_2YV_2-UU_1U_2X_1Z_1-V_1$ and $X_1-UU_1U_2-X_2$.
By the similar procedures as those in (\ref{equ:APPENDIXBUSE1}), (\ref{equ:APPENDIXBUSE3}),  (\ref{equ:APPENDIXBUSE4}), (\ref{equ:APPENDIXBUSE6}), and (\ref{upperboundR1_proveExample}), we have 
\begin{align}
	&I(U_1;Z_2|UU_2X_2) + I(X_1;Y|UX_{2})+ I(V_{2};X_1X_2Y|UU_1U_2)\notag\\
	& < I(U_1;B\cdot X_1|UU_2X_2) + 0.205 +  I(B\cdot X_1;X_1|UU_1U_2)\notag\\
	& \le H(B\cdot X_1) -H(B\cdot X_1|X_1) + 0.205\notag\\
	& = I(X_1;B\cdot X_1) + 0.205 \notag\\
	& < 0.322 + 0.205\notag\\
	& = 0.527 < 0.6.
\end{align}
For the left hand side of (\ref{proof:exampleGeneral_sumUpdate}), by the similar procedures as those in (\ref{example2:rate-distortionRelax1}) and (\ref{example2:rate-distortionRelax2}), we have
\begin{align}\label{ine:distortionRelaxExample1}
	&I(V_1;X_{1}Z_{1}|UU_1U_2X_{2}Z_{2}) \notag\\
	&\overset{(a)}= I(V_1;X_1,X_1\oplus N|UU_1U_2X_2Z_2)\notag\\
	& \ge I(V_1;N|UU_1U_2X_1X_2Z_2)\notag\\
	& = H(N|UU_1U_2X_1X_2Z_2) - H(N|UU_1U_2X_1X_2Z_2V_1)\notag\\
	& \overset{(b)}= H(N) - H(N|UU_1U_2X_1X_2Z_2V_1)\notag\\
	& \overset{(c)} \ge H(N) - H(N|U_1X_2Z_2V_1)\notag\\
	& = I(N;U_1X_2Z_2V_1)\notag\\
	& \overset{(d)}\ge I(N;\hat{S}_{T_2})
\end{align}
where $(a)$ follows from that $Z_1 = X_1 \oplus N$ in the example, $(b)$ follows from that $N$ is independent of $UU_1U_2X_1X_2Z_2$ as $Z_2 = (B\cdot X_1,X_2\oplus S_1)$ in the example, $(c)$ follows from that conditioning reduces entropy, $(d)$ follows from that $N-U_1X_2Z_2V_1-\hat{S}_{T_2}$ forms a Markov chain, and
\begin{align}\label{ine:distortionRelaxExample2}
	&I(V_1;Z_1|UU_1U_2X_1X_2YV_2)\notag\\
	&\overset{(a)}= I(V_1;N|UU_1U_2X_1X_2YV_2) \notag\\
	& = H(N|UU_1U_2X_1X_2YV_2)  - H(N|UU_1U_2X_1X_2YV_1V_2)\notag\\
	& \overset{(b)}= H(N) - H(N|UU_1U_2X_1X_2YV_1V_2)\notag\\
	& \overset{(c)}= H(N) - H(N|UU_1U_2X_1X_2YV_1V_2B)\notag\\
	& \overset{(d)}= H(N) - H(N|UU_1U_2X_1X_2YV_1V_2B,B\cdot X_1,X_2\oplus S_1)\notag\\
	&= H(N) - H(N|UU_1U_2X_1X_2YV_1V_2BZ_2)\notag\\
	& \overset{(e)}\ge H(N) - H(N|U_1X_2Z_2V_1) \notag\\
	& = I(N;U_1X_2Z_2V_1)\notag\\
	& \overset{(f)}\ge I(N;\hat{S}_{T_2}),
\end{align}
where $(a)$ follows from that $Z_1 = X_1 \oplus N$ in the example, $(b)$ follows from that $N$ is independent of $UU_1U_2X_1X_2YB$ as $Y = (X_1\oplus S_1,X_2\oplus S_2)$ in the example and $N-UU_1U_2X_2Z_2-V_2$ forms a Markov chain as $Z_2 = (BX_1,X_2\oplus S_1)$, $(c)$ follows from that $N$ is independent of $B$ given $UU_1U_2X_1X_2YV_1V_2$ for the numerical example, $(d)$ follows from that given $B,X_1,X_2,Y$, one can exactly know $B\cdot X_1$ and $X_2\oplus S_1$, $(e)$ follows from that conditioning reduces entropy, $(f)$ follows from that $N-U_1X_2Z_2V_1-\hat{S}_{T_2}$ forms a Markov chain.

Combining the above results, one can obtain that 
\begin{align}
	I(N;\hat{S}_{T_2}) < 0.3
\end{align}
holds for $\mathcal{I}_{\text{R-D}}^{\text{aw}}$. According to the rate-distortion theory \cite{el2011network}, one can know that the distortion between $N$ and $\hat{S}_{T_2}$ must be strictly larger than 0.138\footnote{When the distortion is equal to 0.138, the mutual information is $I(N;\hat{S}_{T_2}) = 0.3023 > 0.3$.}, i.e., 
\begin{align}
	D_2^{\text{aw}} > 0.138 > 0.13072.
\end{align}

\subsection{Proof of $R_1^{\text{aw}}<0.11697$ when $D_2^{\text{aw}}= 0.1783$}
Since $\min\{a,b\}\le a$, we can obtain a relaxed single-user bound for $R_1$ by considering the inequality constraint~(22a) for $k=1$ in $\mathcal{I}_{\text{R-D}}^{\text{aw}}$ presented in~\cite[Theorem~3]{ahmadipour2023information} as
\begin{align}
	&R_1^{\text{aw}} \le I(U_1;X_{2}Z_{2}|UU_{2}) + I(V_1;X_{2}Z_{2}|UU_1U_2) \notag\\
	&\qquad- I(V_{1};X_{1}Z_{1}|UU_1U_2) + I(X_1;Y|UX_{2}) \notag\\
	&\qquad+ I(V_1;X_1X_2Y|UU_1U_2) + I(V_{2};X_1X_2YV_1|UU_1U_2) \notag\\
	&\qquad- I(V_{1};X_{1}Z_{1}|UU_1U_2)\notag\\
	&\overset{(a)} = I(U_1;X_{2}Z_{2}|UU_{2}) +I(X_1;Y|UX_{2}) \notag\\
	&\qquad+ I(V_{2};X_1X_2Y|UU_1U_2) - I(V_{1};X_{1}Z_{1}|UU_1U_2X_{2}Z_{2}) \notag\\
	&\qquad- I(V_{1};Z_{1}|UU_1U_2X_1X_2YV_2) \notag\\
	& \overset{(b)}\le I(X_1;B\cdot X_1) + H(X_1\oplus S_1) - H(S_1)\notag\\
	&\qquad - I(V_{1};X_{1}Z_{1}|UU_1U_2X_{2}Z_{2}) \notag\\
	&\qquad- I(V_{1};Z_{1}|UU_1U_2X_1X_2YV_2) \notag\\
	& \overset{(c)}\le I(X_1;B\cdot X_1) + H(X_1\oplus S_1) - H(S_1)\notag\\
	&\qquad- I(N;\hat{S}_{T_2}) - I(N;\hat{S}_{T_2}),
\end{align}
where $(a)$ follows from the Markov chain $X_2Z_2YV_2-UU_1U_2X_1Z_1-V_1$, $(b)$ follows the similar procedures as those in (\ref{equ:APPENDIXBUSE1}), (\ref{equ:APPENDIXBUSE3}),  (\ref{equ:APPENDIXBUSE4}), (\ref{equ:APPENDIXBUSE6}), (\ref{upperboundR1_proveExample}), $(c)$ follows based on (\ref{ine:distortionRelaxExample1}) and (\ref{ine:distortionRelaxExample2}).

Given $P_{S_1}(1) = 0.24$, $P_{S_2}(1) = 0.05$, $P_{N}(1) = 0.3$, $P_{B}(1) = 0.5$ in the example, we have 
\begin{align}
	I(X_1;B\cdot X_1) \le \max_{P_{X_1}}I(X_1;B\cdot X_1)\approx 0.321928094887362,
\end{align}
where the inequality holds with equality if and only if $P_X(1) = 0.4$. We can also know that 
\begin{align}
	H(X_1\oplus S_1) - H(S_1) \le 1 - H(S_1) \approx 0.204959720615478,
\end{align}
where the inequality holds with equality if and only if $P_X(1) = 0.5$. Moreover, when $D_2^{\text{aw}}\approx 0.1783$, there is $I(N;\hat{S}_{T_2}) = 1- H(S_1)$. Combing with these results, we have
\begin{align}
	R_1^{\text{aw}} &< \max_{P_{X_1}}I(X_1;B\cdot X_1) - (1-H(S_1))\notag\\
	&\approx0.321928094887362 - 0.204959720615478\notag\\
	& =0.11697
\end{align}
when $D_2^{\text{aw}}\approx 0.1783$. 

For our results, when choosing 
\begin{subequations}
	\begin{align}
		&X_k = U_k \oplus  \Theta_k = U \oplus \Sigma_k \oplus  \Theta_k , \ k\in\{1,2\}, \\
		&V_{1,c} = \tilde{N}, \ V_{1,p} = V_{2,c} = \phi, \  V_{2,p} = \phi,  
	\end{align}
\end{subequations}
for $\mathcal{I}_{\text{R-D}}^{\text{our},\text{com}}$, one can find that by taking 
\begin{align}
	P_U = 0, P_{\Sigma_1} = 0.4, P_{\Theta_1} = 0, P_{\Sigma_2} = 0, P_{\Theta_2} = 0.5,
\end{align}
and choosing $V_{1,c} = \tilde{N}$ to make that $I(N;\tilde{N}) = \max(H(X_1\oplus S_1) - H(S_1)) = 1 - H(S_1)$, $D_2^{\text{our},\text{com}}\approx 0.1783$ and $R_1^{\text{our},\text{com}}=\max_{P_{X_1}}I(X_1;B\cdot X_1) - (1-H(S_1))\approx0.11697$ can be achieved simultaneously.

\section{Characterization of Outer Bound for Example~\ref{example:general}}\label{appendix:parallelChannelExtension}
The adaptive parallel channel extension of our outer bound is given as follows. Let $\Delta(\mathcal{U})$ denote the set of all distributions of $U$ and $\Delta(\mathcal{U}|\mathcal{V})$ denote the set of all conditional distributions of $U$ given $V$. Then for any mapping $F:\Delta(\mathcal{X}_1\times\mathcal{X}_2)\rightarrow \Delta(\mathcal{Z}_{\text{pc}}|\mathcal{X}_1\times\mathcal{X}_2\times\mathcal{Y}\times\mathcal{Z}_1\times\mathcal{Z}_2)$, the optimal capacity-distortion region $\mathcal{C}(D_1,D_2)$ of ISAC over MAC is contained in $\mathcal{O}_{\text{R-D-PC}}^{\text{our}}$ where 
\begin{subequations}
	\begin{align}
		R_1 &\le  I(X_1;YZ_1Z_2Z_{\text{pc}}|X_2T), \\
		R_2 &\le  I(X_2;YZ_1Z_2Z_{\text{pc}}|X_1T), \\
		R_1 + R_2 &\le  I(X_1X_2;YZ_1Z_2Z_{\text{pc}}|T), \label{equ:sumRateNouse}\\
		R_1 + R_2 &\le  I(X_1X_2;Y), 
	\end{align}
\end{subequations}
with the dependence-balanced constraint 
\begin{align}\label{dependenceBalanceNew}
	I(X_1;X_2|T)\le I(X_1;X_2|Z_1Z_2Z_{\text{pc}}T)
\end{align}
and sensing constraints 
\begin{subequations}
	\begin{align}
		f_{k,\text{R-D}}(D_k) &\le I(S_{T_k}X_{\bar{k}};Z_kZ_{\text{pc}}|X_kQ),k\in\{1,2\}, 
	\end{align}
	\begin{align}
		\mathbb{E}[d_k(S_{T_k},\hat{S}_{T_k}(X_1,X_2,Z_1,Z_2,Z_{\text{pc}}))]\le D_k,\  k\in\{1,2\},
	\end{align}
\end{subequations}
where the rate-distortion function $f_{k,\text{R-D}}(D_k)$ is defined in~\eqref{definitionRateDistortionFunctions}, and $\hat{S}_{T_k}(X_1,X_2,Z_1,Z_2,Z_{\text{pc}}))$ is a genie-aided estimator given as 
\begin{align}\label{estimator:ParallelChannelExtension}
	&\hat{s}^*_{T_k}(x_1,x_2,z_1,z_2,z_{\text{pc}}) = \arg\min_{s'_{T_k}\in\hat{\mathcal{S}}_{T_k}}\sum_{s_{T_k}\in\mathcal{S}_{T_k}}\notag\\
	&\qquad\quad P_{S_{T_k}|X_1X_2Z_1Z_2Z_{\text{pc}}}(s_{T_k}|x_1x_2z_1z_2z_{\text{pc}})d_k(s_{T_k},s'_{T_k}).
\end{align} 
The random variables $QTX_1X_2SS_{T_1}S_{T_2}YZ_1Z_2Z_{\text{pc}}\hat{S}_{T_1}\hat{S}_{T_2}$ have the joint distribution 
\begin{align}
	&P_{QT}P_{X_1X_2|T}P_{SS_{T_1}S_{T_2}}P_{YZ_1Z_2|X_1X_2S}P^{+}_{Z_{\text{pc}}|QTX_1X_2SYZ_1Z_2}\notag\\
	&P_{\hat{S}_{T_1}|X_1X_2Z_1Z_2Z_{\text{pc}}}P_{\hat{S}_{T_2}|X_1X_2Z_1Z_2Z_{\text{pc}}},
\end{align}
such that for all $q,t$
\begin{align}
	P^{+}_{Z_{\text{pc}}|QTX_1X_2SYZ_1Z_2}(z_{\text{pc}}|qtx_1x_2syz_1z_2) = F(P_{X_1X_2|T}(x_1x_2|t)).
\end{align}

The idea of choosing the parallel channel $P^{+}_{Z_{\text{pc}}|TX_1X_2YSZ_1Z_2}(z_{\text{pc}}|tx_1x_2syz_1z_2)$ is to reduce the amount of dependence, which makes the characterization more tractable. In other words, one can choose the parallel channel $P^{+}_{Z_{\text{pc}}|TX_1X_2SYZ_1Z_2}(z_{\text{pc}}|tx_1x_2syz_1z_2)$ to make the dependence-balanced constraint more stringent, consequently reducing the set of allowable input distributions. In this example, we consider the same technique used in \cite{tandon2009outer} with the choice of $Z_{\text{pc}}=X_2$. 
Given such a choice, we have 
\begin{align}
	I(X_1;X_2|T) \le I(X_1;X_2|Z_1Z_2Z_{\text{pc}}T) = 0,
\end{align}
which implies only distributions of the type 
\begin{align}
	P_{X_1X_2|T} = P_{X_1|T}P_{X_2|T}
\end{align}
are allowed. The corresponding result is given by 
\begin{subequations}
	\begin{align}
		&\mathcal{O}_{\text{R-D-PC}}^{\text{our},X_2} = \bigg\{ (R_1,R_2,D_1,D_2) :\notag\\
		&R_1 \le  I(X_1;YZ_1Z_2|X_2T) \\
		&R_2 \le  H(X_2|T), \\
		&R_1 + R_2 \le  I(X_1X_2;Y), \\
		&f_{1,\text{R-D}}(D_1) \le I(S_{T_1}X_{2};Z_1X_2|X_1Q), \\
		&f_{2,\text{R-D}}(D_2) \le I(S_{T_2}X_{1};Z_2|X_2Q), \\
		&\mathbb{E}[d_k(S_{T_k},\hat{S}_{T_k}(X_1,X_2,Z_1,Z_2))]\le D_k,\  k\in\{1,2\}, \\
		&\bigg\},\notag
	\end{align}
\end{subequations}
where the sum rate constraint (\ref{equ:sumRateNouse}) is redundant \cite{tandon2009outer} and the region is evaluated over the set of input distributions of the form $P_{QTX_1X_2} = P_{QT}P_{X_1|T}P_{X_2|T}$.

We consider the composite function $\omega(t) = \frac{1-\sqrt{|1-2t|}}{2}$ for $0\le t\le 1$. We refer to the entropy function as which is defined as 
\begin{align}
	h^{(k)}(t_1,\cdots,t_k) = - \sum_{i=1}^kt_i\log (t_i), 
\end{align}
for $t_i\ge 0, i \in[1:k]$, and $\sum_{i=1}^kt_i=1$. Specifically, we denote $h^{(2)}(t,1-t)$ simply as $h(t)$. Note that for composite function $\omega(t) = \frac{1-\sqrt{|1-2t|}}{2}$ for $0\le t\le 1$, the following property holds:
\begin{align}
	\omega(2t(1-t)) = \min(t,1-t).
\end{align}
As a consequence, the following holds:
\begin{align}\label{PropertyCompositeFun}
	h(\omega(2t(1-t))) = h(t). 
\end{align}
Now we characterize the outer bound result. Let the cardinality of the auxiliary random variables $Q$ and $T$ be fixed and arbitrary, say $\mathcal{Q},\mathcal{T}$. Then, the joint distribution $P_{QT}P_{X_1|T}P_{X_2|T}$ can be described by the following variables 
\begin{subequations}
	\begin{align}
		&\pi_{1t} = P_{X_1|T}(0|t), \ t= 1, ...,|\mathcal{T}|, \\
		&\pi_{2t} = P_{X_2|T}(0|t), \ t= 1, ...,|\mathcal{T}|, \\
		&\kappa_{t} = P_{T}(t) = \sum_{q}P_{QT}(qt),\ t= 1, ...,|\mathcal{T}|, \\
		&\kappa_{qt} = P_{QT}(qt), \  q= 1, ...,|\mathcal{Q}|,\  t= 1, ...,|\mathcal{T}|,
	\end{align}
\end{subequations}
Our outer bound can be characterized in terms of three variables $\alpha_1,\alpha_2,\alpha$, which are functions of $P_{QT}P_{X_1|T}P_{X_2|T}$, and are given as 
\begin{subequations}\label{VarOutIntroduce}
	\begin{align}
		\alpha_1 &= \sum_{t}\kappa_{t}\pi_{1t}(1-\pi_{1t}) = \sum_{t}\kappa_{t}\alpha_{1t},\\
		\alpha_2 &= \sum_{t}\kappa_{t}\pi_{2t}(1-\pi_{2t}) = \sum_{t}\kappa_{t}\alpha_{2t},\\
		\alpha &= \sum_{t}\kappa_{t}\pi_{1t} = \sum_{t}\kappa_{t}\alpha_{t}.
	\end{align}
\end{subequations}
It should be noted that $\alpha_1,\alpha_2$ both lie in the range $[0,\frac{1}{4}]$ as $0\le \pi_{jt}\le 1$ for $j=1,2$, $t=1,...,|\mathcal{T}|$, and $\alpha$ lies in the range $[\alpha_1,1]$.

The upper bounds for terms in $\mathcal{O}_{\text{R-D-PC}}^{\text{our},X_2}$ are given as follows. 
\begin{align}
	R_1&\le I(X_1;YZ_1Z_2|X_2T) \notag\\
	&\overset{(a)}= H(X_1|T)\notag\\
	&= \sum_{t}\kappa_{t}h(\pi_{1t}) \notag\\
	&\overset{(b)}= \sum_{t}\kappa_{t}h(\omega(2\pi_{1t}(1-\pi_{1t})))\notag\\
	&\overset{(c)}= \sum_{t}\kappa_th(\omega(2\alpha_{1t})) \notag\\
	&\overset{(d)}\le h(\omega(2\alpha_1)),
\end{align}
where $(a)$ follows from that $Y=(X_1\oplus S_1,X_2\oplus S_2)$, $Z_1 = X_1 \oplus N$, $Z_2 = (BX_1, X_2 \oplus S_1)$, and $X_1$ is independent of $X_2$ given $T$ for $\mathcal{O}_{\text{R-D-PC}}^{\text{our},X_2}$, $(b)$ follows from (\ref{PropertyCompositeFun}), $(c)$ follows from (\ref{VarOutIntroduce}), $(d)$ follows from the application of Jensen’s inequality. Similarly, we have
\begin{align}
	R_2 \le H(X_2|T)  \le h(\omega(2\alpha_2)),
\end{align}
and 
\begin{align}
	&R_1 + R_2 \notag\\
	&\le I(X_1X_2;Y) \notag\\
	&= H(Y) - H(Y|X_1X_2)\notag\\
	& = h^{(4)}(P_{Y_1Y_2}(0,0),P_{Y_1Y_2}(0,1),P_{Y_1Y_2}(1,0),P_{Y_1Y_2}(1,1)) \notag\\
	&\qquad- H(S_1) - H(S_2)\notag\\
	&\overset{(a)}\le h(P_{Y_1Y_2}(0,0)+P_{Y_1Y_2}(0,1)) + 1 - H(S_1) - H(S_2)\notag\\
	& = h(P_{Y_1}(0)) + 1 - H(S_1) - H(S_2)\notag\\
	& = h\bigg(P_{S_1}(1) + (1 - 2P_{S_1}(1))\alpha\bigg) + 1 - H(S_1) - H(S_2),
\end{align}
where $(a)$ follows from the fact that
\begin{align}
	h^{(4)}(a,b,c,d) &= \frac{1}{2} h^{(4)}(a,b,c,d) + \frac{1}{2} h^{(4)}(b,a,d,c)\notag\\
	&\le h^{(4)}(\frac{a+b}{2},\frac{a+b}{2},\frac{c+d}{2},\frac{c+d}{2})\notag\\
	& = h(a+b) + 1
\end{align}
due to the concavity of the entropy function and the application of Jensen’s inequality.
For the sensing constraints, we have
\begin{align}
	I(S_{T_2}X_{1};Z_2|X_2Q) &\overset{(a)}= H(Z_2|X_2Q) - H(Z_2|X_1X_2Q)\notag\\
	&= H(B\cdot X_1|X_2Q) - H(B\cdot X_1|X_1Q) \notag\\
	& \overset{(b)}\le H(B\cdot X_1) - H(B\cdot X_1|X_1)\notag\\
	& = h\bigg(P_{B}(1)(1-\alpha)\bigg)- (1-\alpha) H(B),
\end{align} 
where $(a)$ follows from that $S_{T_2} = N$, $Z_2 = (BX_1,X_2\oplus S_1)$, $(b)$ follows from that conditioning reduces entropy and $Q$ is independent of channel variables $B,N,S_1,S_2$.

\section*{Acknowledgment}
The authors would like to thank Dr. Jiayu Zhang from the Huawei Techologies Co., Ltd. for his professional
advice on the improved inner bound.

\ifCLASSOPTIONcaptionsoff
  \newpage
\fi



%
\bibliographystyle{IEEEtran}
\bibliography{ISACMAC}

\end{document}